\documentclass[usenatbib]{mn2e}
\usepackage{multirow}
\usepackage{rotating}
\usepackage{colortbl}
\usepackage{color}
\usepackage[fleqn]{amsmath}
\usepackage{amssymb}
\usepackage{amsfonts}
\usepackage{verbatim}
\usepackage{scalefnt}
\usepackage[percent]{overpic}

\newcommand{\comments}[1]{} 
\newcommand\T{\rule{0pt}{2.6ex}}       
\newcommand\B{\rule[-1.2ex]{0pt}{0pt}} 

\title[Growth and activity of BHs in galaxy mergers]{Growth and activity of black holes in galaxy mergers with varying mass ratios}

\author[Pedro R. Capelo et al.]{Pedro R. Capelo,$^{1}$\thanks{E-mail: capelop@umich.edu} Marta Volonteri,$^{1,2}$ Massimo Dotti,$^{3,4}$\newauthor Jillian M. Bellovary,$^{5}$ Lucio Mayer$^{6}$ and Fabio Governato$^{7}$\\
$^1$Department of Astronomy, University of Michigan, Ann Arbor, MI 48109, USA\\
$^2$Institut d'Astrophysique de Paris, 98bis Boulevard Arago, F-75014 Paris, France\\
$^3$Dipartimento di Fisica G. Occhialini, Universit$\grave{a}$ degli Studi di Milano Bicocca, Piazza della Scienza 3, I-20126 Milano, Italy\\
$^4$INFN, Sezione Milano--Bicocca, Piazza della Scienza 3, I-20126 Milano, Italy\\
$^5$Department of Physics and Astronomy, Vanderbilt University, Nashville, TN 37235, USA\\
$^6$Institute for Computational Science, University of Z$\ddot{u}$rich, Winterthurerstrasse 190, CH-8057 Z$\ddot{u}$rich, Switzerland\\
$^7$Department of Astronomy, University of Washington, Box 351580, Seattle, WA 98195, USA}

\begin{document}

\maketitle

\begin{abstract}
We study supermassive black holes (BHs) in merging galaxies, using a suite of hydrodynamical simulations with very high spatial ($\sim$10~pc) and temporal ($\sim$1~Myr) resolution, where we vary the initial mass ratio, the orbital configuration, and the gas fraction. (i) We address the question of when and why, during a merger, increased BH accretion occurs, quantifying gas inflows and BH accretion rates. (ii) We also quantify the relative effectiveness in inducing AGN activity of merger-related versus secular-related causes, by studying different stages of the encounter: the stochastic (or early) stage, the (proper) merger stage, and the remnant (or late) stage. (iii) We assess which galaxy mergers preferentially enhance BH accretion, finding that the initial mass ratio is the most important factor. (iv) We study the evolution of the BH masses, finding that the BH mass contrast tends to decrease in minor mergers and to increase in major mergers. This effect hints at the existence of a preferential range of mass ratios for BHs in the final pairing stages. (v) In both merging and dynamically quiescent galaxies, the gas accreted by the BH is not necessarily the gas with {\it low} angular momentum, but the gas that {\it loses} angular momentum.
\end{abstract}



\section{Introduction}\label{agn2014:sec:Introduction}

Supermassive black holes (BHs) are believed to reside at the centre of most massive galaxies in the local Universe and to obey tight relationships between their mass and several quantities of the host spheroid \citep[e.g.][]{KormendyRichstone95,magorrian1998,fm00,Gebhardt2000,Gultekin2009,McConnell2013}. A small fraction of these systems have been detected as an active galactic nucleus (AGN), which is a consequence of high levels of accretion of gas onto the central BH \citep[e.g.][]{Salpeter64,LB1969}. This gas is believed to originate at large (galactic: from 0 to $\sim$kpc) scales and, through the loss of most of its angular momentum, to flow towards the centre of the massive system, down to sub-pc scales, where it can be accreted \citep[e.g.][]{Jogee_2006} and drive the growth of BHs \citep{2012NewAR..56...93A}. Many possible explanations have been proposed, including minor interactions (and extremely minor mergers) and/or internal (secular) processes, which include instabilities driven by bars and violent gas instabilities at high redshift \citep[e.g.][]{Gabor_Bournaud_2013}. One of the first mechanisms brought forward were the large-scale gravitational torques produced by major galaxy mergers. During these encounters, torques generate large-scale gas inflows that drive the gas down to pc-scale \citep[e.g.][]{Shlosman1989,Shlosman1990,Barnes92,BarnesHernquist96,MihosHernquist1996}.

Observationally, we know that not {\it all} AGN activity is merger driven. By performing detailed morphological and kinematic-neighbour studies of a sample of $\sim$400 AGN hosts and of a control sample of inactive galaxies from the Cosmic Evolution Survey (COSMOS) field at redshifts 0.3~$<$~z~$<$~1.0, \citet{2009ApJ...691..705G} find that the asymmetry distributions and the neighbour counts of the two samples are consistent with each other, suggesting that strong interactions are not more common among AGN than normal galaxies. \citet{Cisternas2011} perform a similar study and reach the same conclusions, by visually inspecting the morphologies of 140 AGN hosts and of 1264 inactive galaxies, finding a lack of strong distortions in more than 85 per cent of the AGN hosts, and similar distortion fractions between active and inactive galaxies.

On the other hand, there are suggestions that mergers increase AGN activity. \citet{Ellison2011} compare a sample of 11060 Sloan Digital Sky Survey (SDSS) galaxies with a close companion to a mass- and redshift-matched control sample of 110600 galaxies with no close companion, and find that the AGN fraction in close pairs of galaxies increases, with decreasing projected separation, by up to $\sim$2.5 at $\lesssim$10~kpc. \citet{Silverman2011} and \citet{Lackner_et_al_2014}, studying instead COSMOS galaxies at $0.25<z<1$, find similar AGN-fraction increases.

Therefore, we need to shift from asking the question ``Is AGN activity merger driven?" to ``Which galaxy mergers enhance AGN activity?". Our goal is to understand {\it if}, {\it when} and {\it how} galaxy mergers trigger AGN activity, depending on the dynamics and thermodynamics of the merger itself.

Theoretically, numerous simulations have considered the triggering of BH accretion in equal-mass galaxy mergers \citep[e.g.][]{DiMatteo2005,hopkins2006,HQ2010}. 1:1 mergers, while triggering the strongest burst of activity, are extremely rare in the Universe. Fewer studies have considered BH fuelling in minor mergers (\citealt{Younger2008}, \citealt{Johansson2009}; see also \citealt{Cox2008} for an extensive parameter study of galaxy mergers, although with no BHs). These studies have generally resolved scales of $\simeq$100~pc and focussed on the evolution of galaxies along observed scaling relations, but not on the dynamics of gaseous inflows. More recent studies of major galactic mergers have improved greatly on resolution \citep[e.g.][]{Kim_et_al_2009,Karl_et_al_2010,Saitoh_et_al_2011,Teyssier_et_al_2010,Chapon_et_al_2013,Renaud_et_al_2014} but were again focused on 1:1 mergers and lacked the presence of central BHs. \citet{Hayward_et_al_2014} included BH accretion and feedback, but focused only on major mergers. High-resolution simulations that trace both dynamics and accretion are needed to accurately estimate gas inflows in nuclear regions in minor galaxy mergers. As an example, the importance of the initial mass ratio of the merging galaxies on the final fate of the BHs has been discussed in \citet{Callegari2009,Callegari2011}, \citet{VW2012,VW2014}.

We present detailed analysis aiming at understanding which BH would be more active (and therefore more visible) during a merger event. We anticipate that the galaxy--galaxy interaction must redistribute the angular momentum of the gas in order to drive consistent inflows triggering any AGN activity. The effectiveness of such angular momentum redistribution depends, as will be demonstrated, on the galaxy mass ratio: in minor mergers, the secondary galaxy is significantly affected by the gravitational torques exerted by the primary, while the primary itself remains basically unperturbed during the whole interaction. Major mergers, on the other hand, can significantly affect both galaxies, triggering major accretion episodes onto both BHs.


\section{Numerical Setup}\label{agn2014:sec:Numerical_setup}

\begin{table} \centering
\vspace{-3.5pt}
\caption[Merger parameters]{Main simulation parameters for our six major mergers (Runs~1--6), four minor mergers (Runs~7--10), and three control runs (Runs~C1--C3). (1) Run number. (2) Initial mass ratio $q_{\textnormal{\tiny \textsc{G}}}$ between the merging galaxies. (3) Initial angle $\theta_1$ between the primary galaxy's angular momentum vector and the overall orbital angular momentum vector, in radians. (4) Initial angle $\theta_2$ between the secondary galaxy's angular momentum vector and the overall orbital angular momentum vector, in radians. (5) Initial eccentricity $e$ of the orbit. (6) First pericentric distance $R_{\rm peri}$ between the two galaxies, as a fraction of the virial radius of $G_1$. (7) Initial separation $R_{\rm init}$ between the two galaxies, divided by the sum of the initial virial radii of the merging galaxies. (8) Gas fraction in the galactic disc. (9) BH feedback efficiency $\epsilon_f$ (except for Runs~C1 and C2, where accretion has been shut off).
\label{agn2014:tab:merger_params}}
\vspace{5pt}
{\small
\begin{tabular*}{0.475\textwidth}{ccccccccc}
\hline
Run & $q_{\textnormal{\tiny \textsc{G}}}$ & $\theta_1$ & $\theta_2$ & $e$ & $R_{\rm peri}$ & $R_{\rm init}$ & gas & $\epsilon_f$ \T \B \\
\hline
01 & 1:1 & 0 & 0 & 1 & 0.2 & 1 & 0.3 & 0.001 \T \B \\
02 & 1:2 & 0 & 0 & 1 & 0.2 & 1 & 0.3 & 0.001 \B \\
03 & 1:2 & $\pi/4$ & 0 & 1 & 0.2 & 1 & 0.3 & 0.001 \B \\
04 & 1:2 & $\pi$ & 0 & 1 & 0.2 & 1 & 0.3 & 0.001 \B \\
05 & 1:2 & 0 & $\pi$ & 1 & 0.2 & 1 & 0.3 & 0.001 \B \\
06 & 1:2 & 0 & 0 & 1 & 0.2 & 1 & 0.6 & 0.001 \B \\
\hline
07 & 1:4 & 0 & 0 & 1 & 0.2 & 1 & 0.3 & 0.001 \T \B \\
08 & 1:4 & $\pi/4$ & 0 & 1 & 0.2 & 1 & 0.3 & 0.001 \B \\
09 & 1:6 & 0 & 0 & 1 & 0.2 & 1 & 0.3 & 0.001 \B \\
10 & 1:10 & 0 & 0 & 1 & 0.2 & 1 & 0.3 & 0.001 \B \\
\hline
C1 & 1:2 & 0 & 0 & 1 & 0.2 & 1 & 0.3 & No acc. \T \B \\
C2 & 1:2 & 0 & 0 & 1 & 0.2 & 1 & 0.6 & No acc. \B \\
C3 & 1:2 & 0 & 0 & 1 & 0.2 & 1 & 0.3 & 0.005 \B \\
\hline
\end{tabular*}
\vspace{5pt}
}
\end{table}

In this section, we describe the numerical setup of our merger simulations, which include encounters of two disc galaxies set at $z=3$ (near the peak of the cosmic merger rate), with different mass ratios, orbital configuration, and gas fractions. For the remainder of the paper, we define `major' (`minor') mergers those encounters with an initial mass ratio $q_{\textnormal{\tiny \textsc{G}}} \equiv M_2/M_1 > 0.25$ ($q_{\textnormal{\tiny \textsc{G}}} \le 0.25$; see \citealt{Mayer2013} for a discussion on the boundary between major and minor mergers), where $M_1$ and $M_2$ are the initial virial masses of the primary ($G_1$) and secondary ($G_2$) galaxy, respectively, and $M_1 \ge M_2$. We also define `low-gas-fraction' (`high-gas-fraction') mergers those encounters where the fraction of total (baryonic) mass of the the galactic discs in gaseous form is 30 (60) per cent. Additionally, when we specify a merger by its Run number, we refer to column 1 in Table~\ref{agn2014:tab:merger_params}.


\subsection{Orbital configuration}\label{agn2014:sec:Orbital_configuration}

\begin{table*} \centering
\vspace{-3.5pt}
\caption[Galactic simulation parameters]{Main galactic parameters at the beginning of the simulation. (1) Galaxy (primary -- G1 or secondary -- G2) and merger. (2)~Virial mass. (3)~Stellar bulge mass. (4)~Stellar disc mass. (5)~Gas disc mass. (6)~Disc scale radius. (7)~BH mass. (8)~DM particle mass. (9)~DM particle softening length. The disc mass is the sum of the stellar disc mass and the gas disc mass. The stellar bulge scale radius and the disc scale height are always equal to $0.2\,r_{\rm disc}$ and $0.1\,r_{\rm disc}$, respectively. All other parameters are the same for all galaxies and all mergers: gas and stellar particle mass ($4.6 \times 10^{3}$ and $3.3 \times 10^{3}\; {\rm M}_{\odot}$, respectively) and softening (20 and 10~pc, respectively); BH softening (5~pc); DM halo spin and concentration parameters ($\lambda=0.04$ and $c_{\rm vir}=3$, respectively); and redshift ($z=3$).
\label{agn2014:tab:galactic_params}}
\vspace{5pt}
{\small
\begin{tabular*}{0.96\textwidth}{cccccccccc}
\hline
Galaxy & $M_{\rm vir}$ & $M_{\rm stell.\,bulge}$ & $M_{\rm stell.\,disc}$ & $M_{\rm gas.\,disc}$ & $r_{\rm disc}$ & $M_{\rm BH}$ & $M_{\rm DM\,part.}$ & $\epsilon_{\rm DM\,part.}$ \T \B \\
{[}Merger] & [$10^{11} \hbox{M}_{\odot}$] & [$10^{9} \hbox{M}_{\odot}$] & [$10^{9} \hbox{M}_{\odot}$] & [$10^{9} \hbox{M}_{\odot}$] & [kpc] & [$10^{6} \hbox{M}_{\odot}$] & [$10^{5} \hbox{M}_{\odot}$] & [pc] \B \\
\hline
G1 [1:1, 1:2, 1:4 low-gas-frac] & $2.21$ & $1.77$ & $6.18$ & $2.65$ & 1.13 & $3.53$ & $1.1$ & 30 & \T \B \\
G1 [1:2 high-gas-frac]              & $2.21$ & $1.77$ & $3.53$ & $5.30$ & 1.13 & $3.53$ & $1.1$ & 30 & \B \\
G1 [1:6]                                    & $2.21$ & $1.77$ & $6.18$ & $2.65$ & 1.13 & $3.53$ & $0.8$ & 27 & \B \\
G1 [1:10]                                  & $2.21$ & $1.77$ & $6.18$ & $2.65$ & 1.13 & $3.53$ & $0.5$ & 23 & \B \\
\hline
G2 [1:2 low-gas-frac]               & $1.10$ & $0.88$ & $3.09$ & $1.32$ & 0.90 & $1.77$ & $1.1$ & 30 & \T \B \\
G2 [1:2 high-gas-frac]              & $1.10$ & $0.88$ & $1.77$ & $2.65$ & 0.90 & $1.77$ & $1.1$ & 30 & \B \\
G2 [1:4]                                    & $0.55$ & $0.44$ & $1.54$ & $0.66$ & 0.71 & $0.88$ & $1.1$ & 30 & \B \\
G2 [1:6]                                    & $0.37$ & $0.29$ & $1.03$ & $0.44$ & 0.62 & $0.59$ & $0.8$ & 27 & \B \\
G2 [1:10]                                  & $0.22$ & $0.18$ & $0.62$ & $0.26$ & 0.53 & $0.35$ & $0.5$ & 23 & \B \\
\hline
\end{tabular*}
\vspace{5pt}
}
\end{table*}

In accordance with \citet{VW2014}, and to avoid the effects of using different global orbital parameters, we set the galaxies of all encounters to initially follow parabolic orbits (eccentricity $e=1$), found to be the most common orbits in cosmological simulations of galaxy formation \citep{Benson05}. We set the initial separation $R_{\rm init}$ to be equal to the sum of the two initial virial radii, and the first pericentric distance $R_{\rm peri}$ to be equal to 20 per cent \citep{Khochfar2006} of the initial virial radius of $G_1$, defined in Section~\ref{agn2014:sec:Galaxies}. We also vary the orbital configuration of each galaxy, by changing the angle $\theta$ between the individual galactic angular momentum vector and the global orbital angular momentum vector, to consider coplanar, prograde--prograde (Runs~1--2, 6--7, 9--10, and C1--C3), retrograde--prograde (Run~4), and prograde--retrograde (Run~5) mergers, and inclined-primary mergers (Runs~3 and 8). In Table~\ref{agn2014:tab:merger_params}, we list the global orbital parameters of all simulations of the suite in columns~5--7, together with their orbital configuration in columns 3--4. We note that the effect of using different global orbital parameters (e.g. more or less radial orbits) might be very important (as hinted by, e.g., \citealt{Callegari2011}; see also \citealt{DiMatteo_et_al_2008}), but we chose to focus on more `internal' galactic parameters, such as galactic mass, gas fraction, and internal disc rotation.


\subsection{Galaxies}\label{agn2014:sec:Galaxies}

All galaxies are composite systems of dark matter (DM), gas, stars, and a central BH (described in Section~\ref{agn2014:sec:Black_holes}). Most of this description follows \citet{Springel_White_1999} and \citet{springel2005b}. Most values in this section were chosen for consistency with previous work \citep{Callegari2009,Callegari2011,VW2012,VW2014}. See Table~\ref{agn2014:tab:galactic_params} for a complete list.

DM is described by a spherical Navarro--Frenk--White \citep{NFW1996} density profile up to the virial radius, and by an exponentially decaying profile outside the virial radius \citep{Springel_White_1999}. The DM halo spin and concentration parameters are initialized to $\lambda = 0.04$ \citep{Vitvitska_et_al_2002} and $c_{\rm vir}=3$, respectively. This value of the concentration parameter is only slightly lower than what expected from recent DM-only simulations of $z=3$ systems \citep[$c \lesssim 4$ for the range of masses considered in this paper; e.g.][]{Dutton_Maccio_2014,Diemer_Kravtsov_2014}.

The baryonic component is comprised of a stellar bulge and a mixed stellar and gaseous disc. The disc is described by an exponential surface density profile and by an isothermal sheet \citep{Spitzer_1942,Camm_1950}, with a total mass equal to 4 per cent of the virial mass of the galaxy. The disc scale radius $r_{\rm disc}$ is then determined by imposing conservation of specific angular momentum of the material that forms the disc, whereas the disc scale height $z_{\rm disc}$ is set to be 10 per cent of $r_{\rm disc}$. The fraction of total (baryonic) mass of the disc in gaseous form is 30 (60) per cent in the low- (high-) gas-fraction simulations (see Table~\ref{agn2014:tab:merger_params}, column 8), consistent with the range of molecular gas fractions in high-redshift galaxies from \citet{Tacconi2010}. The bulge, making up for 0.8 per cent of the virial mass of the galaxy, is described by a spherical \citet{Hernquist1990} density profile with a scale radius equal to 20 per cent of the disc scale radius. In each merger, $G_1$ has a virial mass of $2.21 \times 10^{11}\; {\rm M}_{\odot}$ \citep[consistent with][]{adelberger2005b} and, consequently, a stellar bulge mass of $1.77 \times 10^{9}\; {\rm M}_{\odot}$, a disc mass of $8.83 \times 10^{9}\; {\rm M}_{\odot}$, and a disc scale radius of 1.13~kpc. All other galaxies have their quantities scaled according to $q_{\textnormal{\tiny \textsc{G}}}$, which varies from 0.1 to 1 (see Table~\ref{agn2014:tab:merger_params}, column 2).

Stellar and gas particles initially have the same particle mass ($3.3 \times 10^{3}$ and $4.6 \times 10^{3}\; {\rm M}_{\odot}$, respectively) and softening length (10 and 20~pc, respectively) in all the encounters of the suite. In order to limit excursions of BHs from the centre of each galaxy, we impose the DM particles to have a mass smaller than 15 per cent of that of the smaller BH in each merger. For this reason, the mass and softening length of DM particles in the 1:1, 1:2, and 1:4 mergers were set to $1.1 \times 10^{5}\; {\rm M}_{\odot}$ and 30~pc, respectively. In the other encounters, on the other hand, because of the much lower mass of the secondary BH, DM particle masses and softening lengths were lowered accordingly (1:6 merger: $8 \times 10^{4}\; {\rm M}_{\odot}$ and 27~pc; 1:10 merger: $5 \times 10^{4}\; {\rm M}_{\odot}$ and 23~pc). The total initial number of particles varies between 8 and 13 million, depending on the merger. Overall, the whole suite used $1.3 \times 10^8$ particles and the total equivalent time amounted to 29~Gyr of evolution. All mergers were simulated using 256 processors, and each merger required on average $\sim$10$^5$ processor-hours.

All galaxies are initialized with solar metallicity and a uniform stellar population with an age of 2~Gyr and a Kroupa initial mass function \citep{Kroupa_IMF,Raiteri_et_al_1996}. Before the proper merger simulation, we `relax' each galaxy, i.e. evolve it in isolation for 0.1~Gyr, gradually increasing the star formation (SF) efficiency from $c^*=0.005$ to 0.015. This is done to avoid unphysical bursts of supernovae at the beginning of our simulations, due to the fact that at the onset of a simulation, there has not been any effective feedback to heat the gas and prevent it from cooling and forming stars.

We performed all our simulations using the N-body smoothed particle hydrodynamics (SPH) code {\scshape gasoline} \citep{gasoline}, an extension of the pure gravity tree code {\scshape pkdgrav} \citep{stadel01}. The version of {\scshape gasoline} we use includes explicit line cooling for atomic hydrogen and helium, and metals \citep[][]{Shen_Wadsley_Stinson_2010}, as well as a physically motivated prescription for SF, supernova feedback and stellar winds \citep{Stinson2006}. In this prescription, stars are allowed to form if the parent gas particle is colder than 6000~K and denser than 100~a.m.u. cm$^{-3}$, and supernovae release $10^{51}$ erg into the surrounding gas, according to the blast-wave formalism of \citet{Stinson2006}. The minimum gas temperature is set at 500~K, to ensure that the Jeans mass is resolved.


\subsection{Black Holes}\label{agn2014:sec:Black_holes}

After each galaxy has been initialized, we place a single BH at its centre, with a mass proportional to the mass of the stellar bulge, according to the relation $M_{\rm BH}=2 \times 10^{-3} M_{\rm Bulge}$ (\citealt{MarconiHunt2003}; see Table~\ref{agn2014:tab:galactic_params}, column 7). For this reason, the initial galactic mass ratio of the merger ($q_{\textnormal{\tiny \textsc{G}}}$; see Table~\ref{agn2014:tab:merger_params}, column 2) is also equal to the initial BH mass ratio. After the relaxation period (see Section~\ref{agn2014:sec:Galaxies}), the BH masses are re-initialized to their initial values. BHs are implemented as sink particles \citep{Bellovary10} which accrete surrounding gas according to a Bondi--Littleton--Hoyle (hereafter, Bondi) accretion formula:

\begin{equation}
\dot{M}_{\rm BH} = \frac{4 \pi \alpha G^2 M_{\rm BH}^2 \rho}{(c_s^2+v^2)^{3/2}},
\end{equation}

\noindent where $c_s$ is the local speed of sound, $\rho$ is the local gas density, $v$ is the relative velocity of the BH with respect to the gas, and $G$ is the gravitational constant. In order to realistically model accretion from an inhomogeneous mix of hot and cold gas particles around the BH, the accretion rate is computed as the sum of the Bondi accretion rate of each individual gas particle near the BH, rather than simply averaging the gas quantities over all the neighbouring particles. This method allows the accretion rate to be weighted more heavily by nearby, cold, dense gas particles (and less by more distant, hot ones) rather than treating them all equally. The particles that contributed the most to the accretion are favoured for mass removal: this way, BHs accrete from particles that are nearby, cold, and dense, rather than simply the nearest particle. As opposed to previous work \citep[e.g.][]{Bellovary10,VW2012,VW2014}, in which $\alpha=1$, the boost factor $\alpha$ in our accretion calculations is equal to 3. We also allow for mildly super-Eddington accretion, limiting the accretion rate to $\alpha \dot{M}_{\rm BH-Edd}$, where

\begin{equation}
\dot{M}_{\rm BH-Edd} = \frac{4 \pi G M_{\rm BH} m_p}{\epsilon_r \sigma_T c},
\end{equation}

\noindent where $m_p$ is the proton mass, $\sigma_T$ is the Thomson cross section, $\epsilon_r=0.1$ is the radiative efficiency, and $c$ is the speed of light in vacuum. A fixed fraction $\epsilon_r$ of the accretion energy rate $\dot{M}_{\rm BH}c^2$ is emitted as radiation. A fraction $\epsilon_f$ of this BH luminosity is injected, in the form of thermal energy, in the nearest gas particle. The softening length of all BHs in all mergers is 5~pc and BH properties (mass and accretion rate) are evaluated every 0.1~Myr. In order to evaluate the effects of BH physics, we also ran additional simulations (`control runs', described in Section~\ref{agn2014:sec:Control_runs}), where we either shut off the BH accretion altogether or increased the BH feedback efficiency (see Table~\ref{agn2014:tab:merger_params}, column 9).


\section{Accretion and growth of black holes}\label{agn2014:sec:BH_accretion_and_growth}

\begin{figure}
\centering
\vspace{2.5pt}
\includegraphics[width=0.99\columnwidth,angle=0]{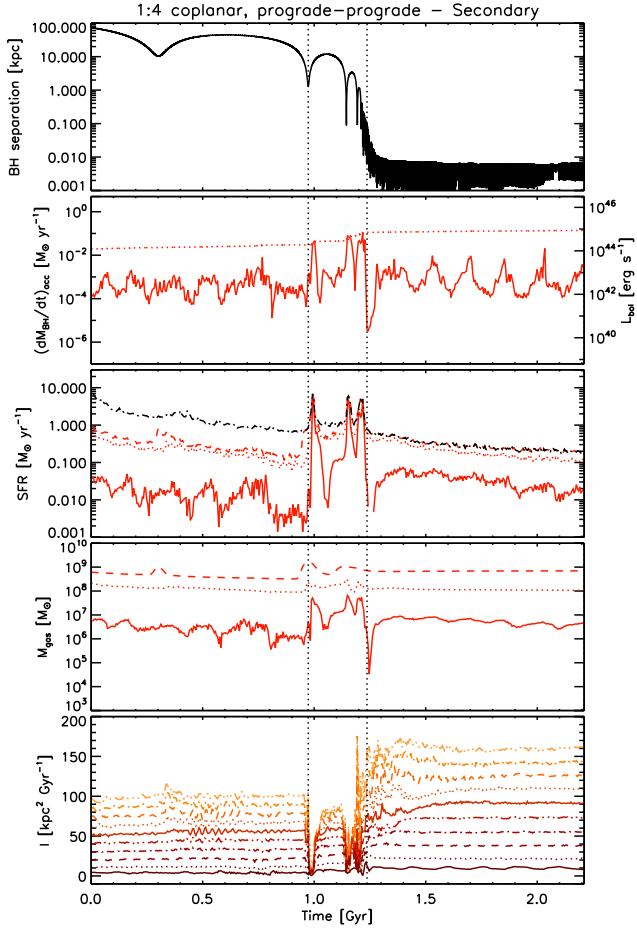}
\vspace{-5pt}
\caption[Temporal evolution of the 1:4 coplanar, prograde--prograde simulation -- Secondary BH]{Temporal evolution of the 1:4 coplanar, prograde--prograde merger -- Main quantities of and around the secondary BH. In all panels, the vertical, dotted, black lines show the separation between the stochastic, the merger, and the remnant stage. In this and in the following figures: BH accretion rate and SFR are averaged over 5~Myr; gas mass and specific angular momentum are shown every 5~Myr; BH mass and separation are shown every 0.1~Myr. {First panel}: separation between the two BHs. {Second panel}: BH accretion rate (solid) and BH Eddington accretion rate (dotted). {Third panel}: global SFR across both galaxies (dash-dotted, black), and SFR in concentric spheres around the BH: 0--0.1 (solid, red), 0--1 (dotted, red), and 0--10 (dashed, red) kpc. {Fourth panel}: gas mass in concentric spheres around the local centre of mass near the BH: 0--0.1 (solid), 0--1 (dotted), and 0--10 (dashed) kpc. {Fifth panel}: gas specific angular momentum magnitude in concentric shells around the local centre of mass near the BH: 0--100 (solid), 100--200 (dotted), 200--300 (dashed), 300--400 (dash-dotted), 400--500 (dash-triple-dotted), 500--600 (solid), 600--700 (dotted), 700--800 (dashed), 800--900 (dash-dotted), and 900-1000 (dash-triple-dotted) pc. The colour varies from dark to light orange as the radius of the shell increases.}
\label{agn2014:fig:m4_hr_gf0_3_BHeff0_001_phi000000_five_panels_secondary}
\end{figure}

\begin{figure}
\centering
\vspace{2.5pt}
\includegraphics[width=0.99\columnwidth,angle=0]{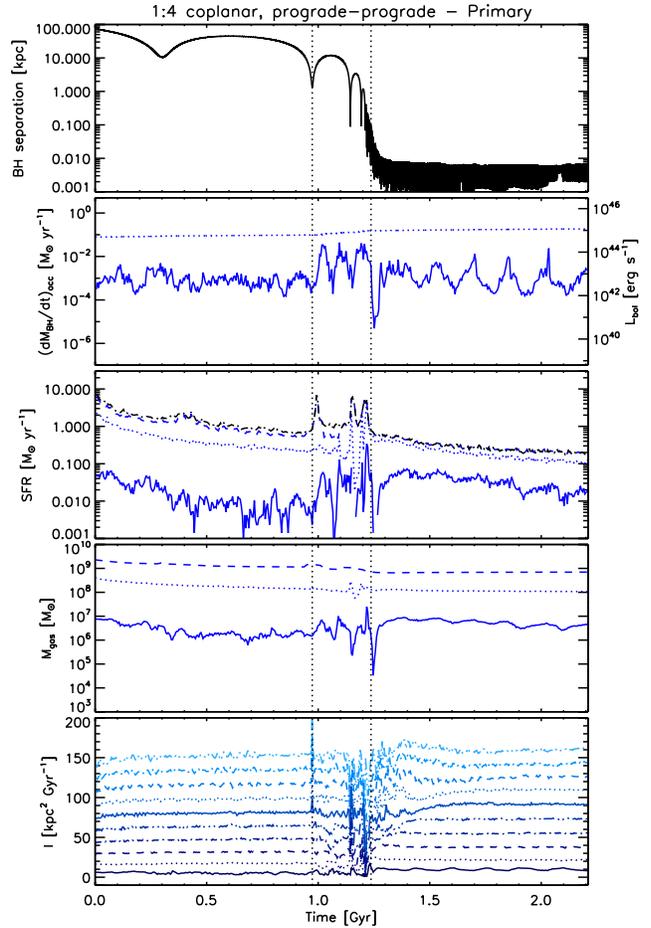}
\vspace{-5pt}
\caption[Temporal evolution of the 1:4 coplanar, prograde--prograde simulation -- Primary BH]{Temporal evolution of the 1:4 coplanar, prograde--prograde merger -- Main quantities of and around the primary BH. Same as Fig.~\ref{agn2014:fig:m4_hr_gf0_3_BHeff0_001_phi000000_five_panels_secondary}. BH accretion, non-total SFR and gas mass are shown in blue. In the fifth panel, the colour of the specific-angular-momentum curve of each gas shell varies from dark to light blue as the radius of the shell increases.}
\label{agn2014:fig:m4_hr_gf0_3_BHeff0_001_phi000000_five_panels_primary}
\end{figure}

In this section, we describe the physical processes influencing the accretion onto BHs before, during, and after the `proper' merger event. We assess the importance of the initial mass ratio and orbital configuration, to understand which mergers preferentially trigger AGN activity, and of the gas fraction in the galactic discs. We divide the history of the encounter into three distinct stages, defined in the next section: the \textit{stochastic} (or early) \textit{stage}; the (proper) \textit{merger stage}; and the \textit{remnant} (or late) \textit{stage}. By assessing the relative effectiveness in triggering AGN activity amongst these stages, it is possible to understand the importance of merger-related versus secular-related causes for enhanced BH accretion.


\subsection{The 1:4 coplanar, prograde--prograde merger}\label{agn2014:sec:1to4_merger}

In this section, we describe in detail a representative merger of our suite. We chose the 1:4 coplanar, prograde--prograde merger (hereafter, the `default merger'; Run~7) because the mass ratio $q_{\textnormal{\tiny \textsc{G}}}=0.25$ is the median of all the mass ratios we considered. Moreover, this ratio is usually chosen as a boundary between major and minor mergers \citep[see][]{Mayer2013}. In Section~\ref{agn2014:sec:Dependence_on_mass_ratio}, we will highlight the differences between this merger, the other minor ($q_{\textnormal{\tiny \textsc{G}}} \le 0.25$) mergers, and the major ($q_{\textnormal{\tiny \textsc{G}}} > 0.25$) mergers.

\begin{figure*}
\centering
\vspace{5pt}
\begin{overpic}[width=0.51\columnwidth,angle=0]{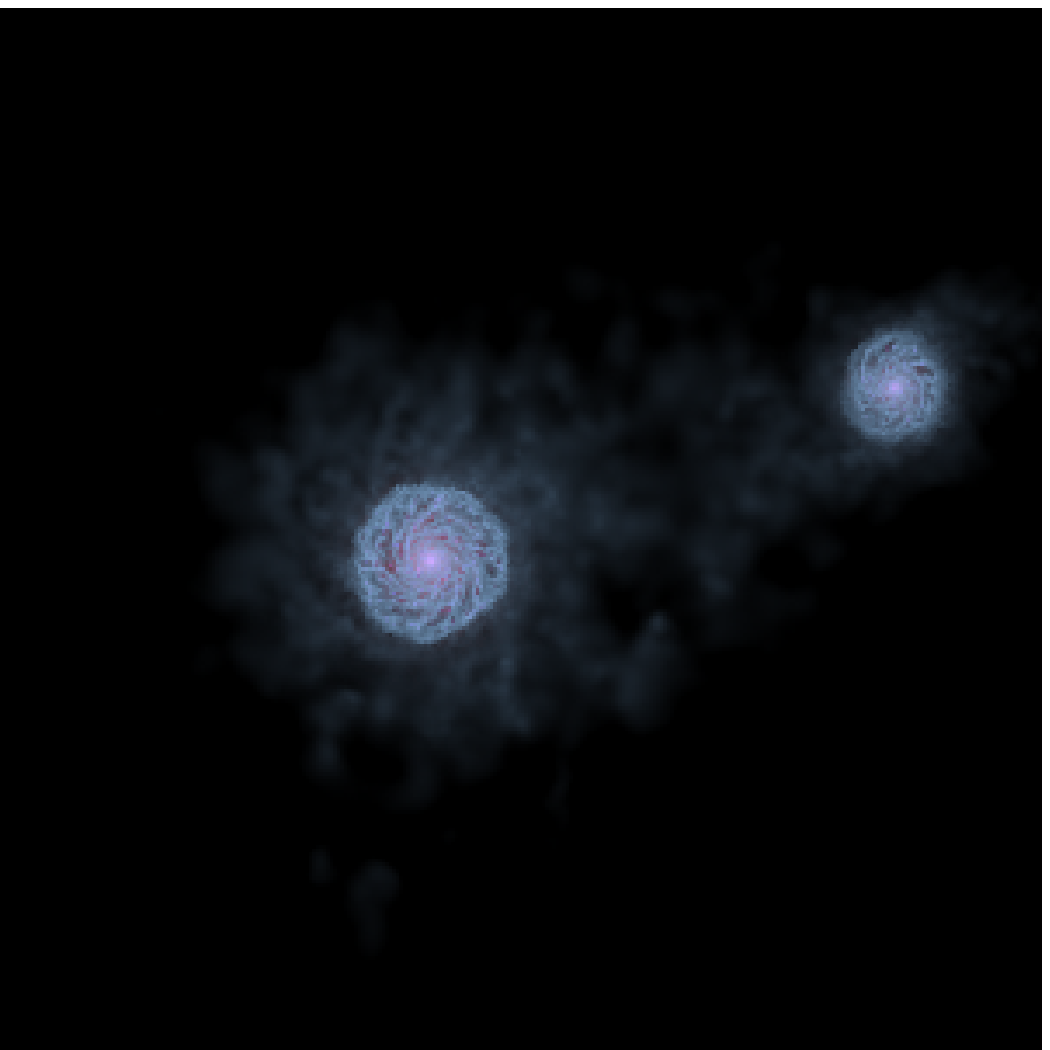}
\put (2,93) {\textcolor{white}{$1$}}
\end{overpic}
\begin{overpic}[width=0.51\columnwidth,angle=0]{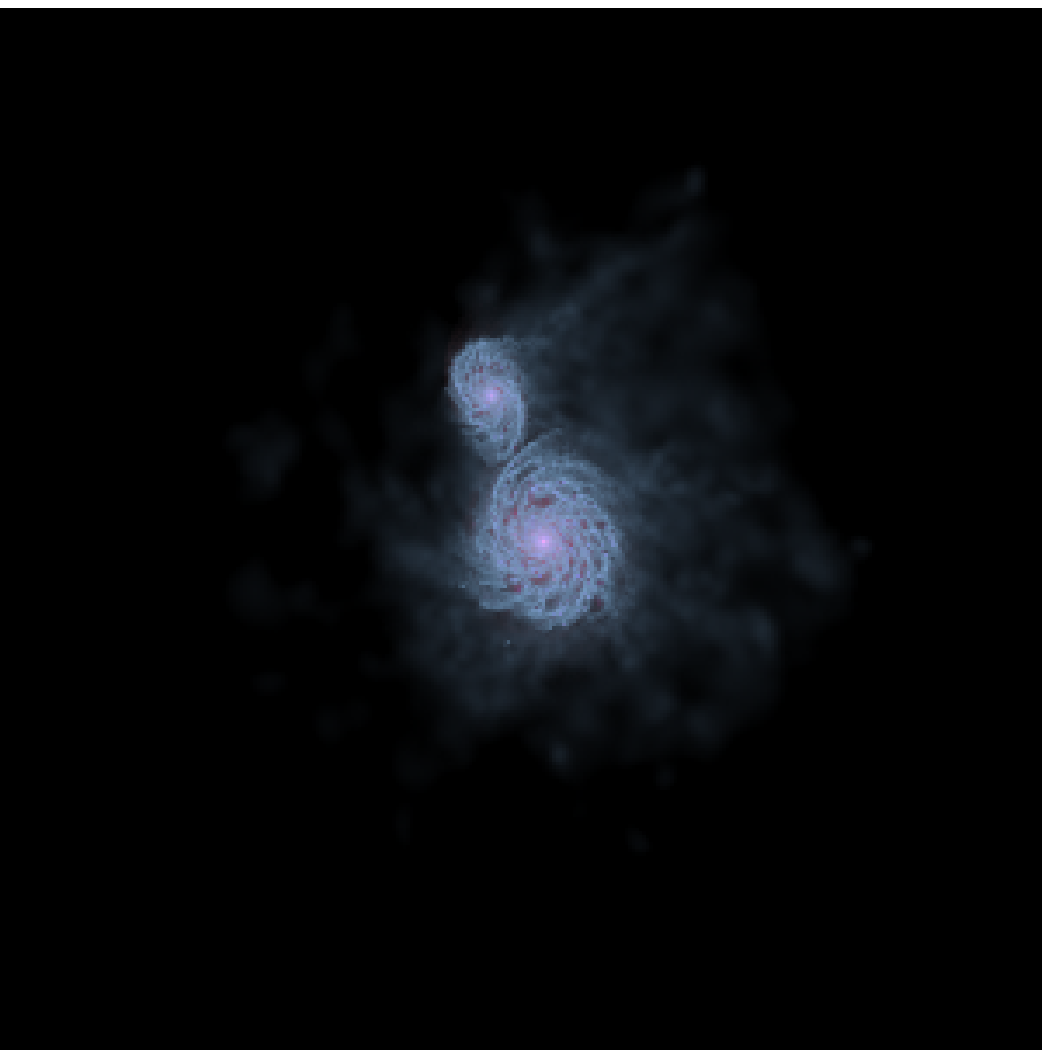}
\put (2,93) {\textcolor{white}{$2$}}
\put (26,93) {\textcolor{white}{First pericentre}}
\end{overpic}
\begin{overpic}[width=0.51\columnwidth,angle=0]{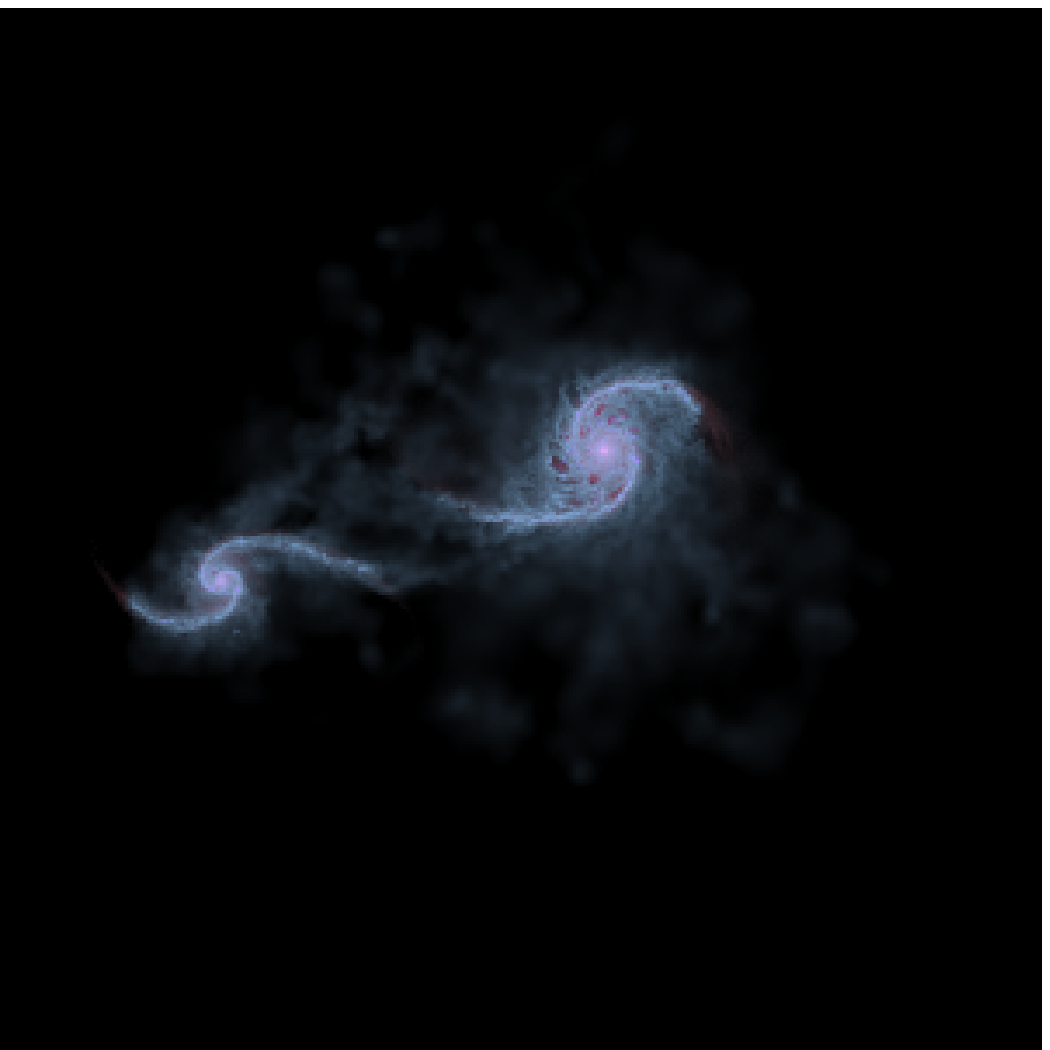}
\put (2,93) {\textcolor{white}{$3$}}
\end{overpic}
\begin{overpic}[width=0.51\columnwidth,angle=0]{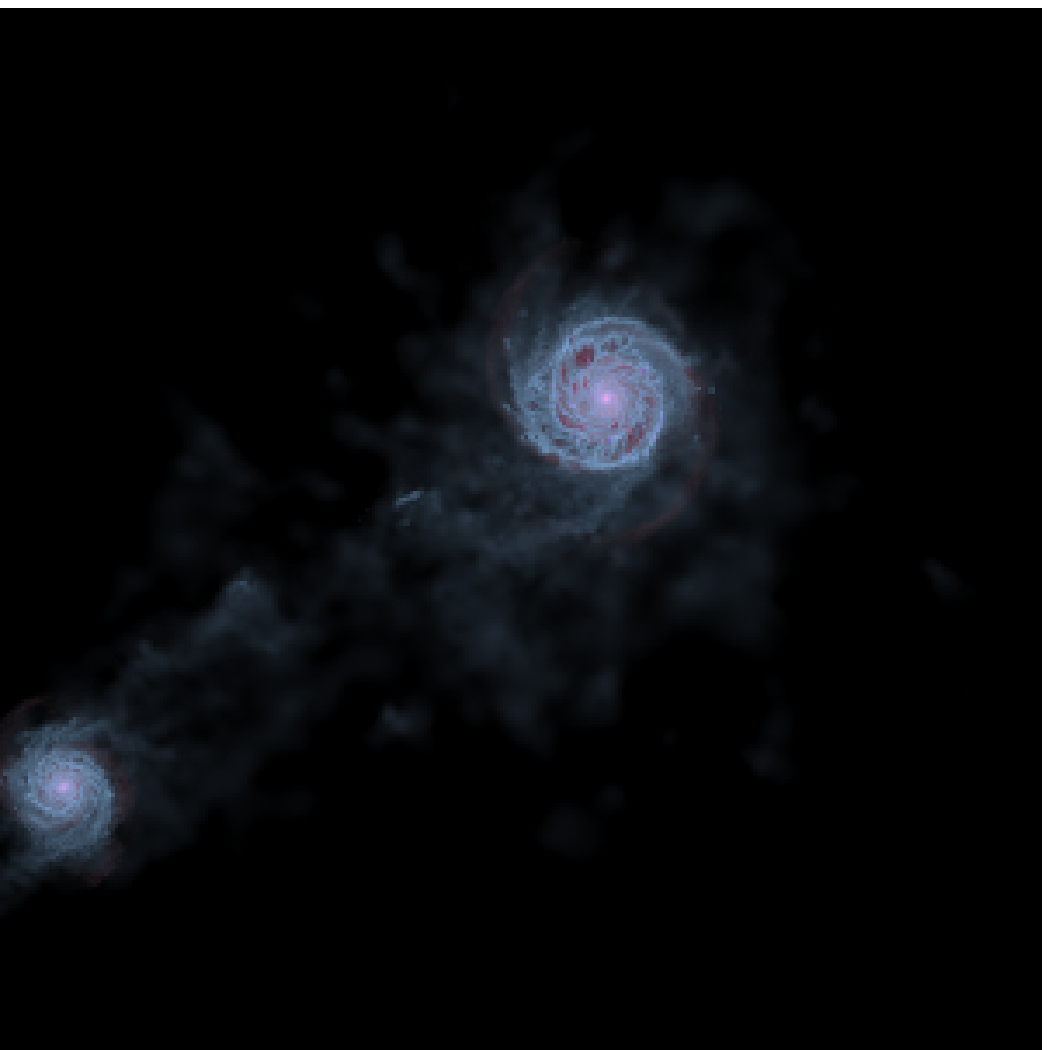}
\put (2,93) {\textcolor{white}{$4$}}
\put (26,93) {\textcolor{white}{First apocentre}}
\end{overpic}
\vskip 0.6mm
\begin{overpic}[width=0.51\columnwidth,angle=0]{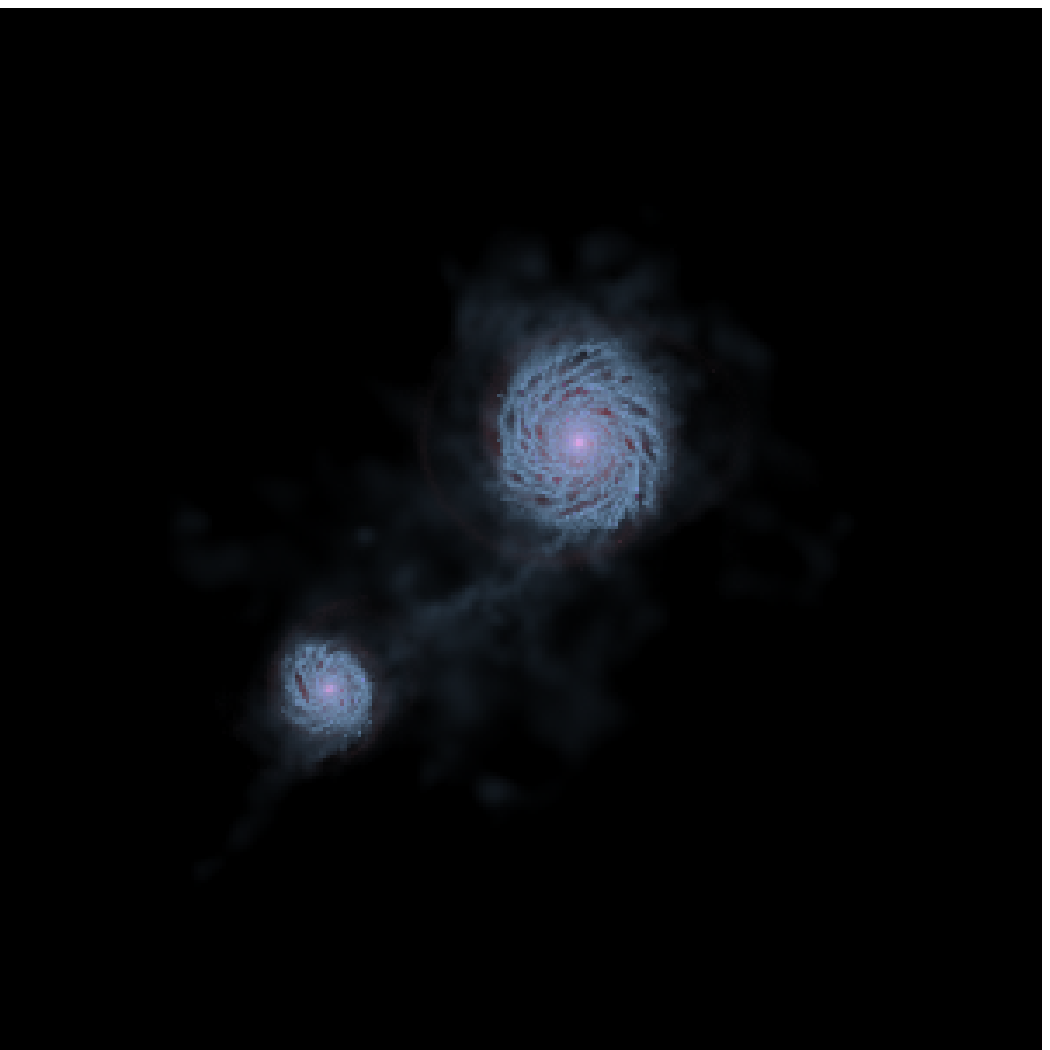}
\put (2,93) {\textcolor{white}{$5$}}
\end{overpic}
\begin{overpic}[width=0.51\columnwidth,angle=0]{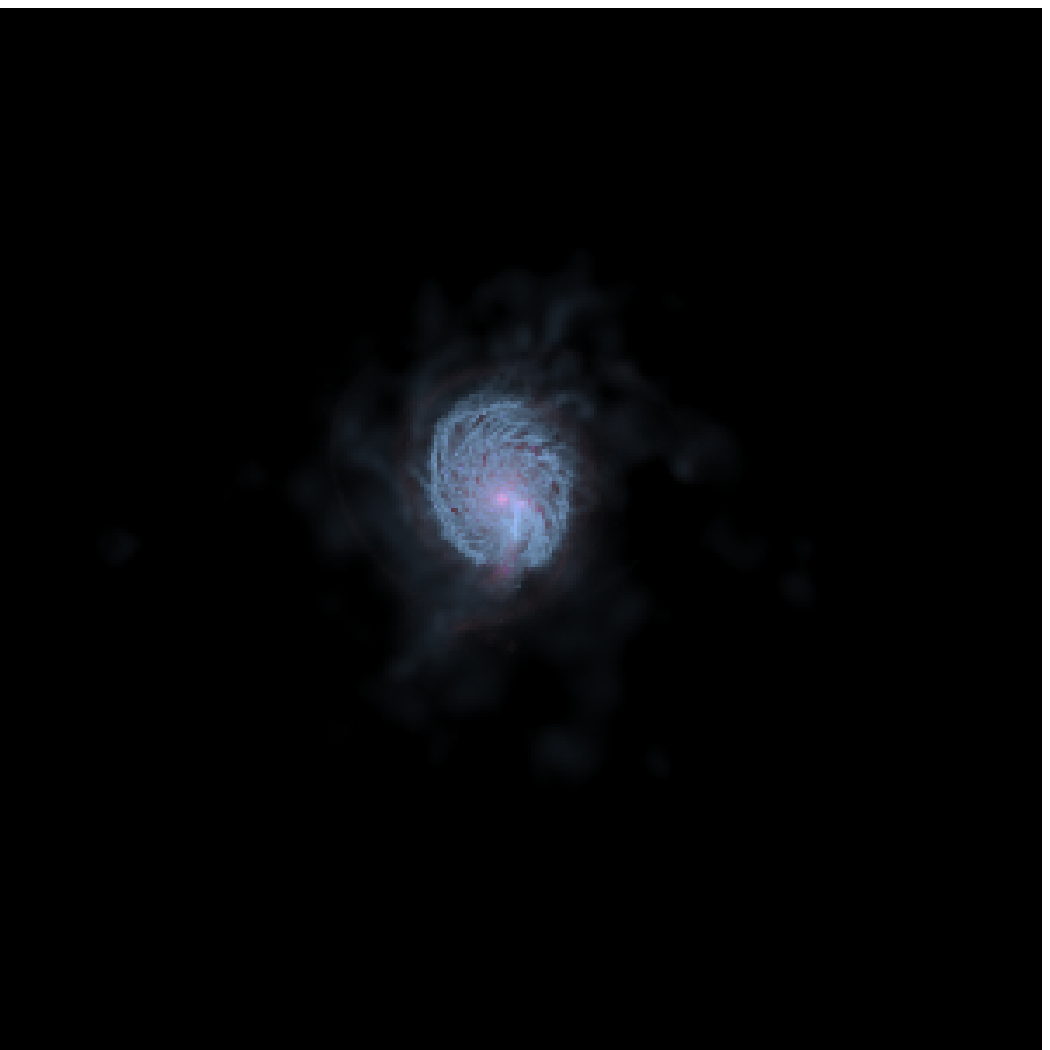}
\put (2,93) {\textcolor{white}{$6$}}
\put (24,93) {\textcolor{white}{Second pericentre}}
\put (8,4) {\textcolor{white}{End of the stochastic stage}}
\end{overpic}
\begin{overpic}[width=0.51\columnwidth,angle=0]{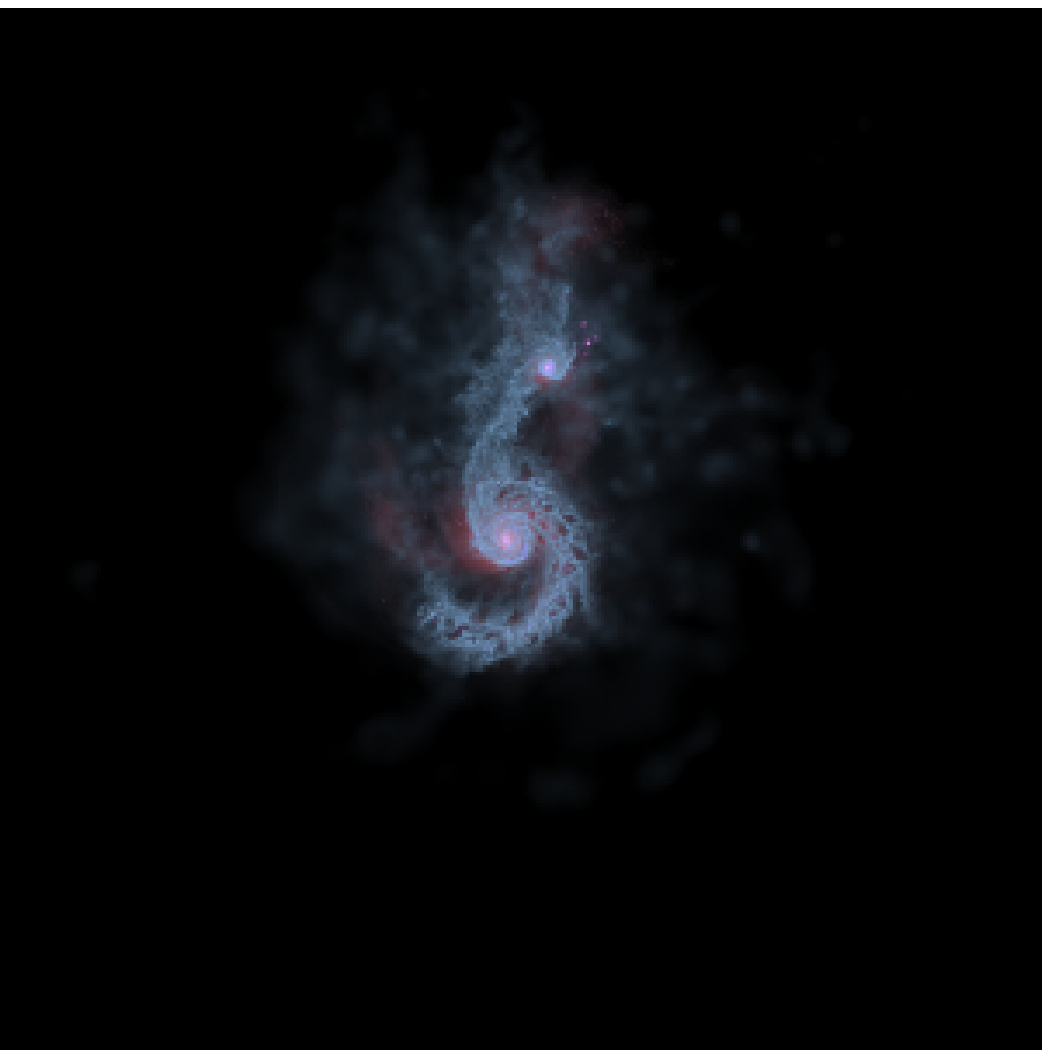}
\put (2,93) {\textcolor{white}{$7$}}
\put (24,93) {\textcolor{white}{Second apocentre}}
\end{overpic}
\begin{overpic}[width=0.51\columnwidth,angle=0]{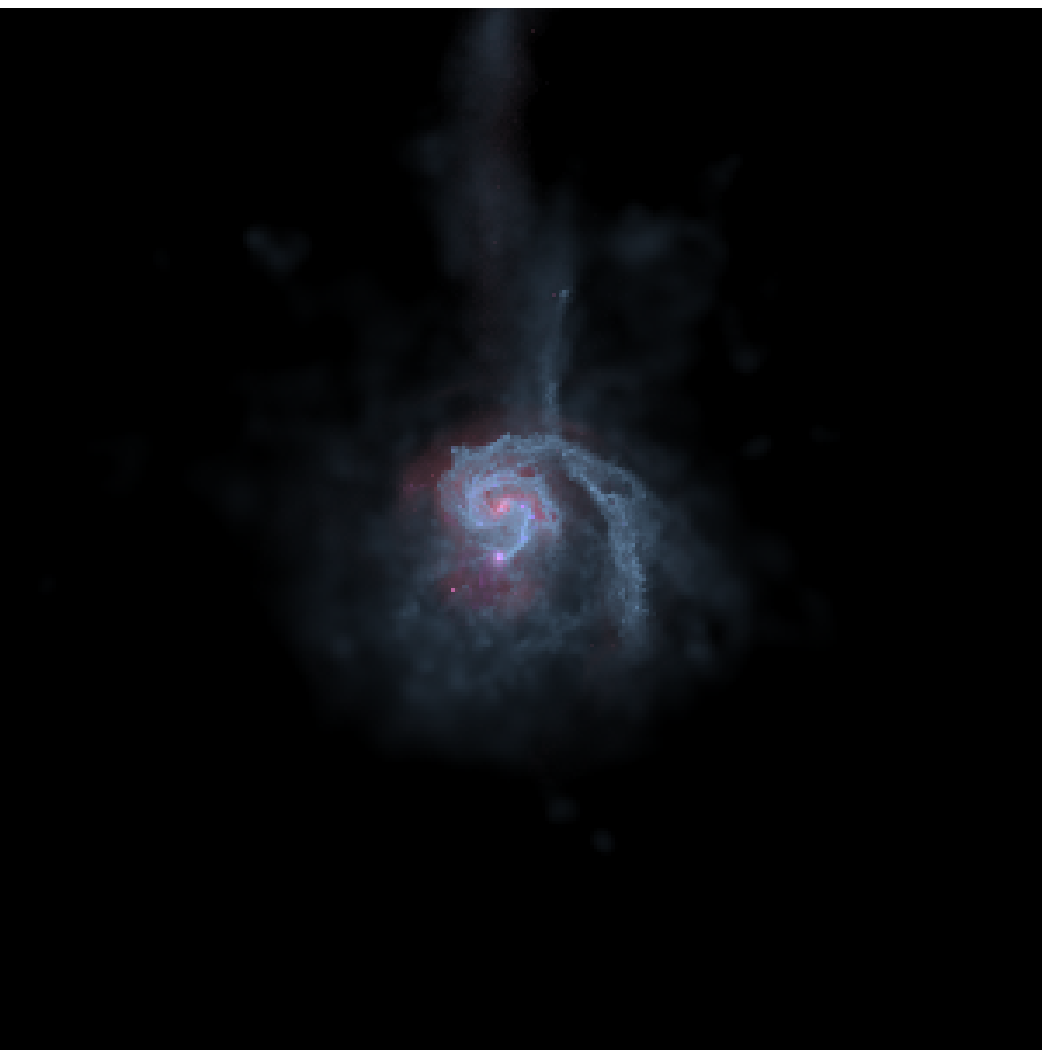}
\put (2,93) {\textcolor{white}{$8$}}
\put (26,93) {\textcolor{white}{Third apocentre}}
\end{overpic}
\vskip 0.6mm
\hspace{2.8pt}\begin{overpic}[width=0.51\columnwidth,angle=0]{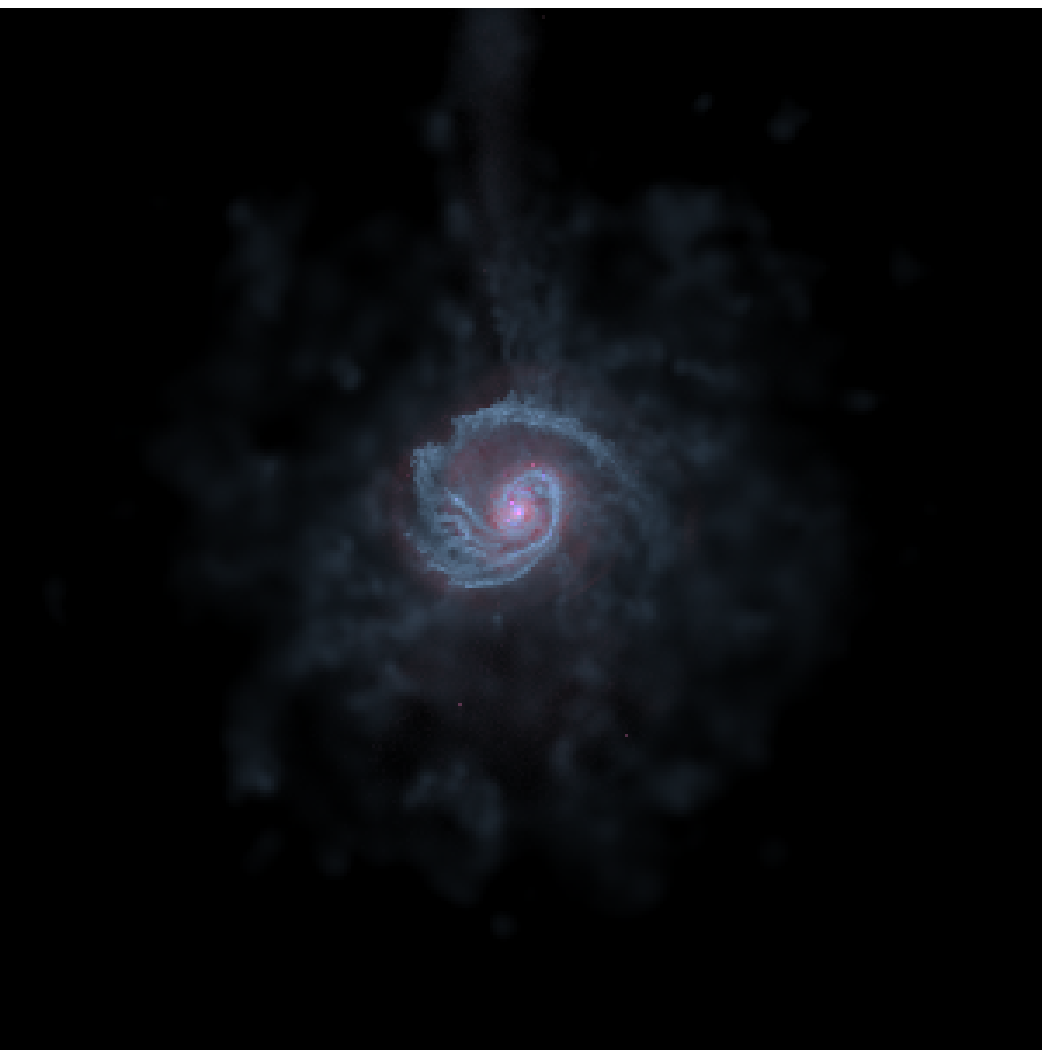}
\put (2,93) {\textcolor{white}{$9$}}
\put (12.5,4) {\textcolor{white}{End of the merger stage}}
\end{overpic}
\begin{overpic}[width=0.51\columnwidth,angle=0]{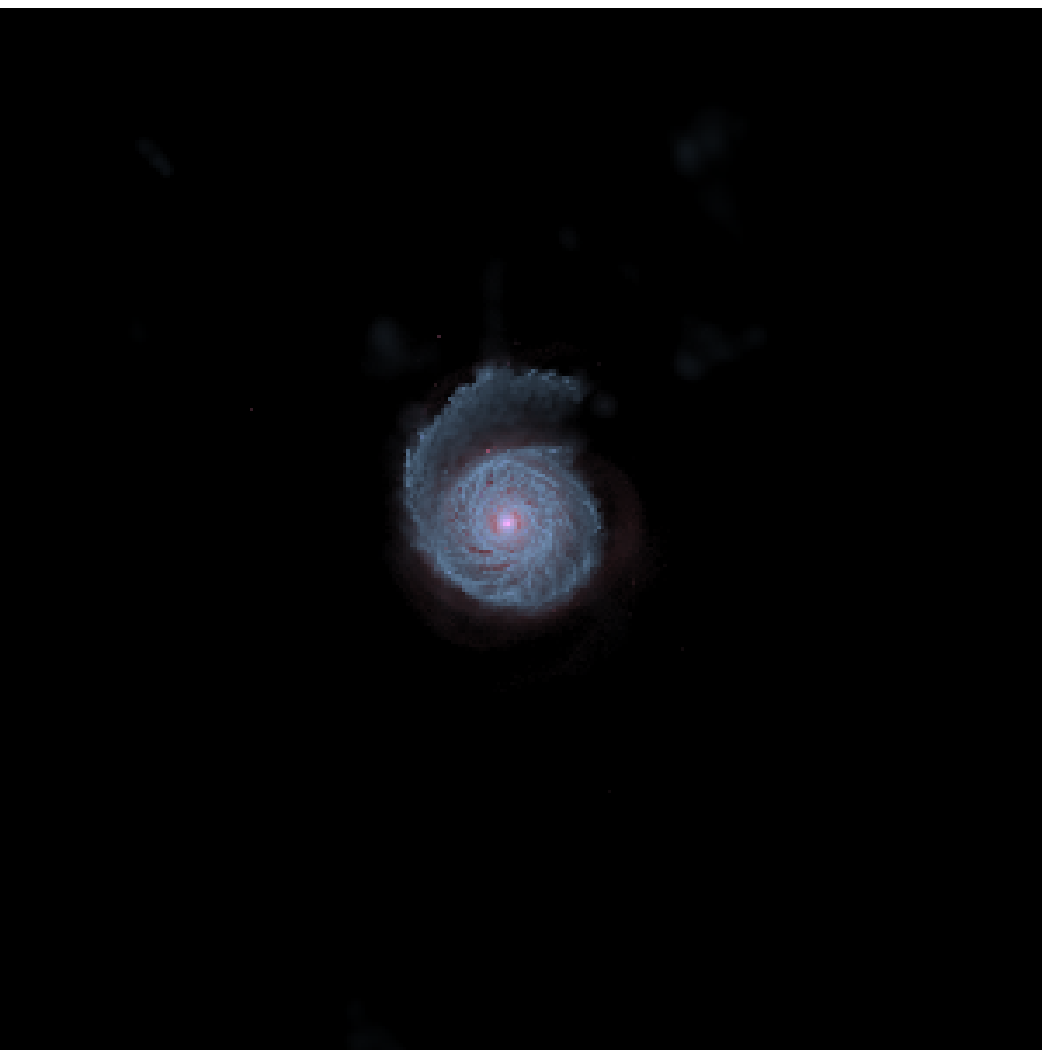}
\put (2,93) {\textcolor{white}{$10$}}
\end{overpic}
\begin{overpic}[width=0.51\columnwidth,angle=0]{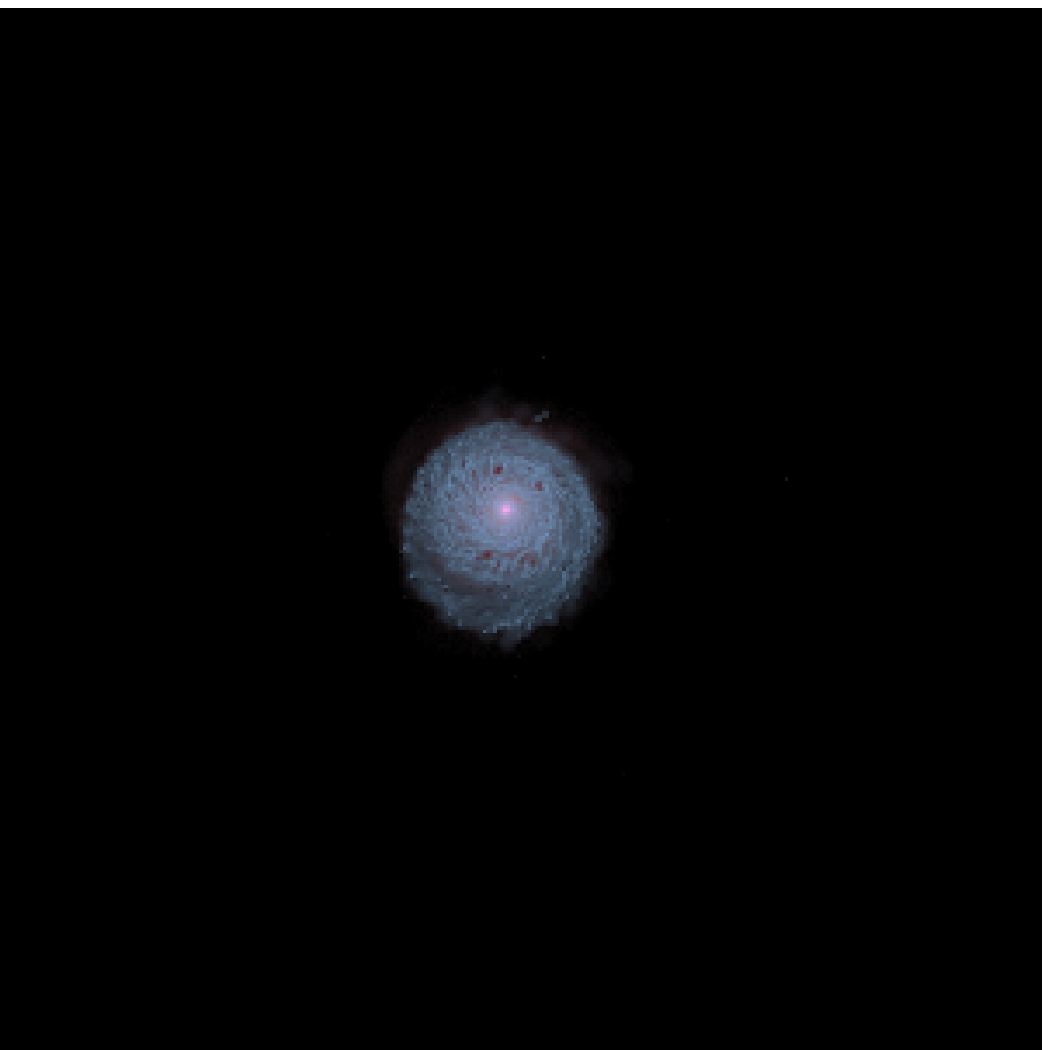}
\put (2,93) {\textcolor{white}{$11$}}
\end{overpic}
\begin{overpic}[width=0.51\columnwidth,angle=0]{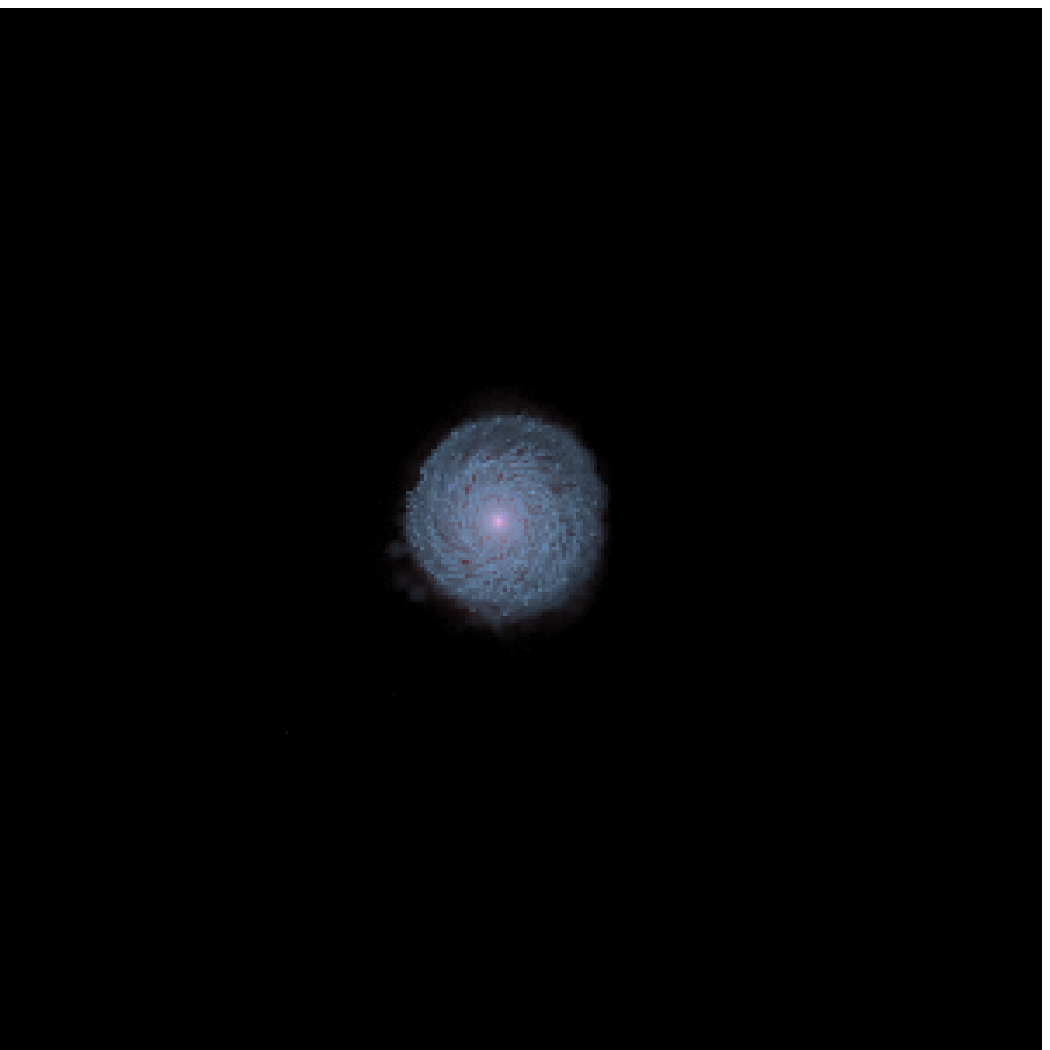}
\put (2,93) {\textcolor{white}{$12$}}
\put (10.5,4) {\textcolor{white}{End of the remnant stage}}
\end{overpic}
\vspace{5pt}
\caption[Stellar and gas density snapshots]{Stellar (red) and gas (blue) density snapshots (viewed face-on) at representative times of the 1:4 coplanar, prograde--prograde merger: (1) 0.20, (2) 0.30 (first pericentric passage), (3) 0.39, (4) 0.61 (first apocentric passage), (5) 0.88, (6) 0.97 (second pericentric passage -- end of the stochastic stage), (7) 1.05 (second apocentric passage), (8) 1.17 (third apocentric passage), (9) 1.24 (end of the merger stage), (10) 1.56, (11) 1.89, and (12) 2.21~Gyr (end of the remnant stage), respectively. We have run the simulations long enough to capture the re-establishment of quiescence after the merger: note how the galaxy in the final snapshot is a normal-looking disc galaxy. The primary (secondary) galaxy starts the parabolic orbit on the left (right) of the first snapshot, moving right- (left-) wards. In order to make the gas more visible, gas density was over-emphasized with respect to stellar density. Each image's size is $70 \times 70$~kpc.}
\label{agn2014:fig:stellar_and_gas_density_snapshots}
\end{figure*}

In Figs~\ref{agn2014:fig:m4_hr_gf0_3_BHeff0_001_phi000000_five_panels_secondary} and \ref{agn2014:fig:m4_hr_gf0_3_BHeff0_001_phi000000_five_panels_primary}, we show in detail the evolution of the default merger, for the main quantities of and around the secondary and primary BH, respectively, whereas, in Fig.~\ref{agn2014:fig:stellar_and_gas_density_snapshots}, we show a more qualitative view of the history of this encounter, through stellar and gas density snapshots for twelve representative times (described below).

In the first panel of Figs~\ref{agn2014:fig:m4_hr_gf0_3_BHeff0_001_phi000000_five_panels_secondary} and \ref{agn2014:fig:m4_hr_gf0_3_BHeff0_001_phi000000_five_panels_primary}, we show the BH separation. The two BHs, embedded in their host galaxies, start at a distance of $\sim$74~kpc (the sum of the initial virial radii of the two galaxies) in an initially parabolic orbit. After 0.3~Gyr, the two galaxies undergo their first pericentric passage, with the two BHs briefly finding themselves at a distance of $\sim$10~kpc (see also panel~2 of Fig.~\ref{agn2014:fig:stellar_and_gas_density_snapshots}). The interaction between the two galaxies has already perturbed the parabolic orbit, causing the two galaxies to be bound to each other and to undergo subsequent pericentric passages. After another 0.7~Gyr, the two galaxies have their second pericentric passage, this time at a much shorter distance ($\sim$1~kpc; see also panel~6 of Fig.~\ref{agn2014:fig:stellar_and_gas_density_snapshots}), and, $\sim$300~Myr (and several pericentric passages) later, a remnant galaxy has formed and the two BHs find themselves at a mutual distance below 10~pc (comparable to the softening lengths of the stellar particles and of the BHs). Since our set-up cannot follow the dynamics of BHs on pc scales, where the main uncertainties on the BH dynamics lie, we have not included a condition for the two sink particles to merge. We caution the reader that this introduces an uncertainty in the estimation of BH accretion during the late stages of the encounter. However, the time-scales for BH merging are still very uncertain and could be even longer than 1~Gyr \citep[the `final parsec problem'; see, e.g.,][]{BBR1980,milosavljevic2001}. Therefore, assuming a specific time-scale for the merging would be highly arbitrary.

In the second panel, we show the BH accretion rate and the BH bolometric luminosity. BH accretion is low, usually well below the Eddington level, except during the 0.3~Gyr following the second pericentric passage, when there are a few BH luminosity peaks, in some cases (e.g. in the secondary BH) reaching Eddington level. These peaks happen during or shortly after the pericentric passages. In the final stages of the encounter, the BH accretion rate becomes quasi-periodic, with a period of $\sim$150~Myr and a change in magnitude of more than two orders of magnitude. Correspondingly, a somewhat spherical cavity in the central gas region has formed, with its radius oscillating between $\sim$60 and $\sim$140~pc with the same temporal period of the BH accretion. When the cavity reaches its maximum radius, the BH accretion is at its minimum, and vice versa. We believe that this is a clear case of BH self-regulation (`breathing'), in which the BHs follow periodic stages of feeding and feedback.

In the third panel of Figs~\ref{agn2014:fig:m4_hr_gf0_3_BHeff0_001_phi000000_five_panels_secondary} and \ref{agn2014:fig:m4_hr_gf0_3_BHeff0_001_phi000000_five_panels_primary}, we show the SF rate (SFR) for three spherical regions centred around the BH (of radii 0.1, 1, and 10~kpc, respectively), and the total SFR of the entire system. SFR is evaluated every 1~Myr, but here we show its average over the same time intervals as those of gas mass and specific angular momentum, which are evaluated every 5~Myr. Central SFR ($<$100~pc) around the BH follows a similar behaviour to that of BH accretion rate, staying at low levels at all times except during the $\sim$300 Myr that follow the second pericentric passage. During this time, central SFR around the secondary BH can increase by more than three orders of magnitude from its previous levels and account for almost the totality of the SFR in the system. The increase in SFR around the primary BH is much more modest, but in both cases it happens at the same time of the BH accretion rate increase. During the final stage, when the two BHs are at a mutual distance of $\lesssim$10~pc, central SFR is higher than during the first stage. Also, SFR around the primary BH is more `centralized': the SFR in the central kpc comprises most of the SFR of the inner 10~kpc, as opposed to during the first stage. The link between BH accretion and SF is at the same time simple (both processes feed off the same reservoir of gas) and complex (the exact correlation between them is still highly debated). In a separate paper (Volonteri et al. 2014, submitted), we present a detailed study on this topic.

In the fourth panel, we show the amount of gas mass in three spherical regions (of radii 0.1, 1, and 10~kpc, respectively) around the local centre of mass\footnote{We calculate the position of the local centre of mass near a BH iteratively, starting from the position of the BH itself. We first calculate the centre of mass of a 100-pc spherical region centred on the BH. We then perform the same computation, using the newly calculated centre of mass as the centre of the new 100-pc spherical region. We continue until the fractional difference between the positions of the `new' and the `old' local centre of mass is less than $10^{-4}$. Since the resulting typical distance between the local centre of mass near a BH and the BH itself varies between 0.01 and 0.02~kpc, depending on the merger, we will use interchangeably `around the local centre of mass near a BH' and `around a BH'.} near each BH. The central gas mass ($<$100~pc) around the primary BH stays approximately constant, while that around the secondary BH increases significantly during the $\sim$300 Myr that follow the second pericentric passage.

In the fifth panel, we show the magnitude of the gas specific angular momentum ($l$), for ten spherical shells (equally spaced in radius, from 0 to 1~kpc) centred around the local centre of mass near each BH. The $l$-curves around the secondary BH do not vary with time up until the second pericentric passage, indicating that there are no large gas inflows or outflows during this stage. The almost equal difference in magnitude from shell to shell is simply due to the different distance from the centre, since specific angular momentum in these galaxies is a linear function of radius. Around the second pericentric passage, all $l$-curves suddenly drop to almost zero. This, and the fact that the central gas mass around the secondary BH increases by more than an order of magnitude at the same time, clearly suggests the presence of a relatively large-scale gas inflow. In the $\sim$300~Myr that follow the second pericentric passage, the $l$-curves around the secondary BH undergo dramatic oscillations, following the pericentric passages, until they return to relatively constant values again, after the formation of the galactic remnant. The behaviour around the primary BH is different. The primary galaxy is not significantly affected by the presence of the companion, and this can be seen very well from the fact that the $l$-curves are essentially constant for the entire duration of the merger.

A division of the merger history into three different stages appears clear from looking at Fig.~\ref{agn2014:fig:m4_hr_gf0_3_BHeff0_001_phi000000_five_panels_secondary} (the main quantities of and around the secondary BH) and from the description above. There exists an initial stage, which we call \textit{stochastic} (or early) \textit{stage}, where the above-mentioned central ($<$1~kpc) quantities are not affected by merger dynamics (see also panels 1--6 of Fig.~\ref{agn2014:fig:stellar_and_gas_density_snapshots}). During this stage, BH luminosities and SFR remain relatively low, and the specific angular momentum of the gas in the central shells remains relatively constant with time. BH accretion is not triggered by merger dynamics but is random \citep{2006ApJS..166....1H}. Note also that this stage includes the first pericentric passage, which appears to not be dynamically important and is not able to induce any global (bar) instabilities because of the presence of the central bulges \citep{MihosHernquist1996}, despite having a clear effect at large distances (see panels~2 and 3 of Fig.~\ref{agn2014:fig:stellar_and_gas_density_snapshots}). The slight increase of the 10-kpc SFR around the two BHs and that of the specific angular momentum of the gas in the outer 100-pc shells around the secondary BH are simply due to the fact that, during this passage, the two BHs find themselves at a distance of $\sim$10~kpc.

The stochastic stage is followed by the (proper) \textit{merger stage} (see also panels 6--9 of Fig.~\ref{agn2014:fig:stellar_and_gas_density_snapshots}), commencing around the time of the second pericentric passage. During this stage, all the relevant quantities are significantly affected by merger dynamics. Merger-induced tidal torques cause the gas to lose angular momentum and flow inwards, creating bursts of central SF and increased BH accretion. At the same time, the mutual distance between the BHs drops from tens of kpc to tens of pc, and a galactic remnant has started to form.

\begin{figure}
\centering
\vspace{2.5pt}
\includegraphics[width=0.99\columnwidth,angle=0]{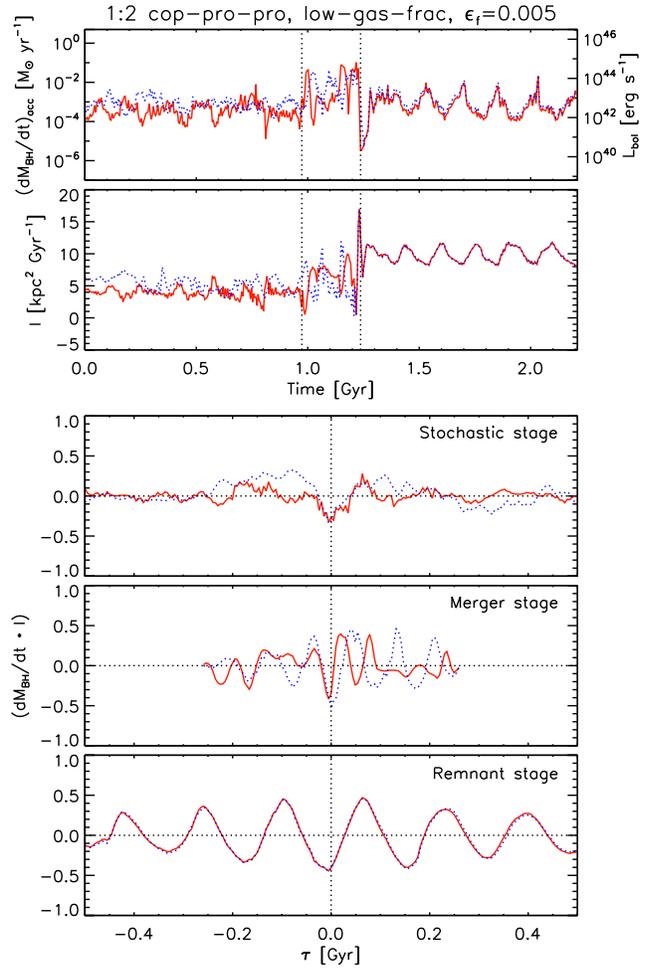}
\vspace{-5pt}
\caption[Cross-correlation for the 1:4 coplanar, prograde--prograde merger]{Cross-correlation between BH accretion rate and central gas specific angular momentum, for the 1:4 coplanar, prograde--prograde merger. All dotted, blue (solid, red) lines refer to the primary (secondary) BH. In the top two panels, the vertical, dotted, black lines show the separation between the stochastic, the merger, and the remnant stage. In the bottom three panels, the horizontal (vertical), dotted, black line denotes zero correlation (delay). {First panel}: BH accretion rate, as a function of time. {Second panel}: central specific angular momentum of the gas (in the innermost 100-pc sphere around the local centre of mass near the BH), as a function of time. {Third panel}: cross-correlation signal between BH accretion rate and central gas specific angular momentum, as a function of lag-time, for the stochastic stage. {Fourth panel}: same as the third panel, but for the merger stage. {Fifth panel}: same as the third panel, but for the remnant stage.}
\label{agn2014:fig:m4_hr_gf0_3_BHeff0_001_phi000000_cross_correlation_Mdot-l}
\end{figure}

Finally, the history of the encounter ends with the \textit{remnant} (or late) \textit{stage}. During this stage (see also panels 9--12 of Fig.~\ref{agn2014:fig:stellar_and_gas_density_snapshots}), the two galaxies have already started to coalesce into one remnant galaxy and, in many cases, the two BHs have reached a separation below the stellar softening length of our simulations. In terms of behaviours of the relevant quantities, this stage is similar to the early stage, with the obvious addition that now some quantities are similar or almost identical. The accretion onto each BH, for example, is now comparable, since the two BHs have access to the same gas reservoir, and their masses are comparable (see also Section~\ref{agn2014:sec:Dependence_on_mass_ratio} for the evolution of the BH mass ratio).

In order to have meaningful comparisons amongst the different stages of the merger (and amongst different mergers, for the next sections), the point in time where we divide the three stages of the encounter should be as unarbitrary as possible, and should be consistent from merger to merger. A fixed time cannot obviously be used, since different mergers have very different merger times, due to the vastly different dynamical-friction time-scales and, more generally, to the different dynamics of each merger.

We divided the stochastic and the merger stage at the second pericentric passage. At that moment in time, in all mergers, there is a clear increase in secondary BH accretion, concurrent to a peak of SF and a drop in gas specific angular momentum around the secondary BH, in all central shells.

The division between the merger and remnant stages is not as straightforward. The behaviour of BH accretion and SF varies from merger to merger. Even though, generally speaking, BH accretion rates are much higher during the merger stage than during the other two stages, there are some cases where the BH accretion rate during the stochastic stage (e.g. one of the BHs in Run~1) or the remnant stage (e.g. the secondary BH in Run~2) are comparable to those in the merger stage. The only common behaviour of all encounters is the fact that the $l$-curves of the gas in the central shells around the secondary BH eventually become flat again, meaning that the dynamically-violent stage of the merger has ceased. We therefore define\footnote{A similar definition could have been used to define the time at which the stochastic stage ends, but we simplify the analysis using the second pericentric passage time, since the results are almost identical.} the time at which the remnant stage starts as the first time after the second pericentric passage when $\Delta l/\Delta t \le 0.3\; l$, over time increments of 0.05~Gyr, where $l$ in this case is the magnitude of the gas specific angular momentum in a spherical region of radius 1~kpc centred around the local centre of mass near the secondary BH. Incidentally, we note that, in many cases, the merger stage coincides with the time when an `angular momentum flip' has occurred: the polar angle of the angular momentum of the gas in the central shells around the secondary BH sharply changes by 180 degrees during the second pericentric passage and by another 180 degrees around the time when the $l$-curves become flat again (see Fig.~\ref{agn2014:fig:m4_hr_gf0_3_BHeff0_001_phi000000_angular_momentum_panels} in the Appendix). Since this physical phenomenon is used here only as a complementary method to divide the history of the encounter into three stages, but does not seem to be directly relevant for BH accretion, we postpone its detailed study to a future work, where we aim to study in more detail the dynamics of the mergers. We note that these definitions for the beginning and end of the merger stage would be very difficult to be detected observationally. On one hand, one could detect a system with a significantly unrelaxed (molecular) gas dynamics and at most be able to say that such system is in its merger stage, without knowing the exact time during that stage. On the other hand, such observational probes would require an exquisite angular resolution, which could be eventually possible using the Atacama Large Millimeter/submillimeter Array (ALMA) in its extended configuration.

\begin{figure}
\centering
\vspace{3.0pt}
\includegraphics[width=0.99\columnwidth,angle=0]{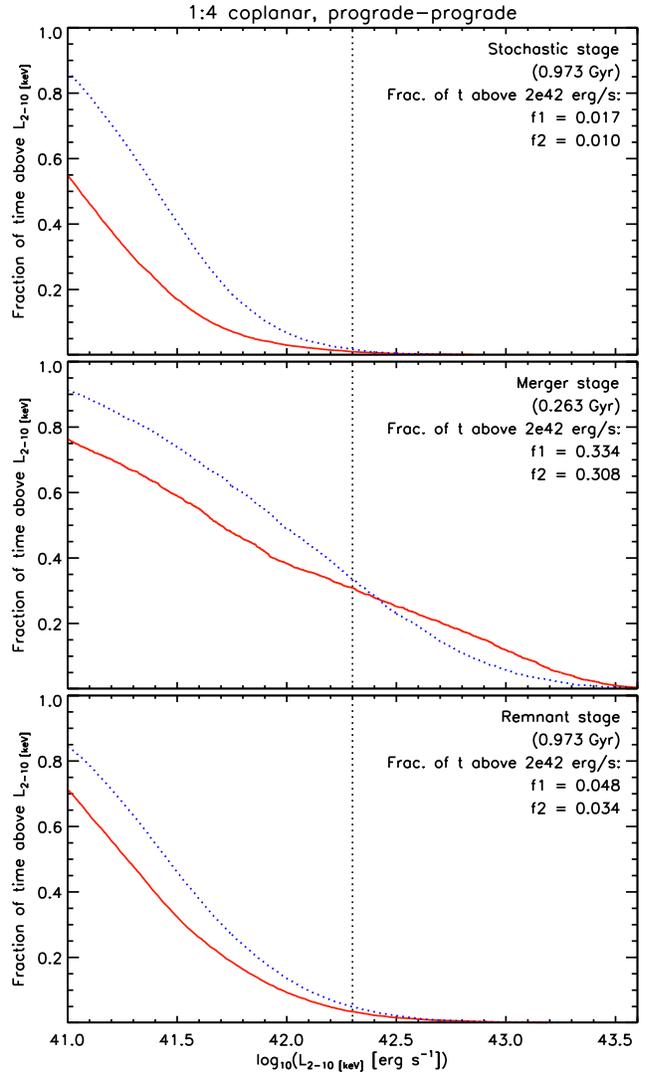}
\vspace{-5pt}
\caption[Fractional cumulative time above a given luminosity, for the 1:4 coplanar, prograde--prograde merger]{Fractional cumulative time above a given luminosity, for the primary (f1) and secondary (f2) BH, in the 1:4 coplanar, prograde--prograde merger. For each stage of the merger, we show the cumulative time, divided by the time-length of the stage, for which a BH accretes above a given level, as a function of the inferred hard-X-ray luminosity. The vertical, dotted, black line shows the typical AGN activity threshold ($L_{2-10\, {\rm keV}} \simeq 2 \times 10^{42}$~erg~s$^{-1}$). {Top panel}: fractional cumulative time for the stochastic stage, for the primary (dotted, blue) and for the secondary (solid, red) BH. {Middle panel}: same as the top panel, but for the merger stage. {Bottom panel}: same as the top panel, but for the remnant stage.}
\label{agn2014:fig:m4_hr_gf0_3_BHeff0_001_phi000000_integral_agn_linlog}
\end{figure}

Finally, the length of the remnant stage clearly depends on how long we ran the simulations. In order to have a meaningful comparison, we impose the length of this stage to be equal to the length of the stochastic stage of each encounter (except for the 1:1 merger, which we ran only for 1.5 Gyr).


\subsubsection{Black hole accretion and gas angular momentum}\label{agn2014:sec:BH_accretion_and_gas_angular_momentum}

In summary, the basic history of the encounter is simple: as far as central quantities (BH accretion rate, central SFR, etc.) are concerned, the two galaxies behave almost as isolated systems during the stochastic stage, until they pass each other for the second time. For a few hundred Myr, during the merger stage, merger dynamics trigger loss of gas angular momentum (and gas inflows), linked to bursts of high SFR and BH accretion rate. Finally, the encounter ends with a remnant galaxy, where BH accretion rate and SFR return to levels comparable to those of the stochastic stage.

It is important to note that the BH accretion rate during the first and last stage, although lower than that during the second stage, is not negligible. Gas gets accreted onto the two BHs during every stage of the merger. In this section, we show that the direct cause for this accretion is the same at all times.

BH accretion is obviously a gas-limited process: without gas in the vicinity of the BH, there would be no accretion \citep{Vito_et_al_2014}. However, the presence of nearby gas is only a necessary, but not sufficient, condition. In fact, a more important condition for accretion is that the specific angular momentum of the central gas needs to be low \citep[see, e.g.,][]{Jogee_2006}. This is obviously the case during the merger stage, for example during the second pericentric passage, when $l$ is at its lowest levels.

With the high spatial and temporal resolution of our simulations, we can show that what matters most is not the amount of specific angular momentum of the gas, but its temporal gradient. The gas that gets accreted is the gas that loses angular momentum. Almost any decrease of specific angular momentum (local minima) of the central gas, regardless of the value it had before, is enough to cause an inflow of gas towards the centre and enhance accretion onto the BH. In other words, any local minima of specific angular momentum can cause an increase in BH accretion. In the merger stage, global torques cause global loss of angular momentum (see the large decrease of specific angular momentum for all ten 100-pc shells during the second pericentric passage in Fig.~\ref{agn2014:fig:m4_hr_gf0_3_BHeff0_001_phi000000_five_panels_secondary}). During the stochastic and remnant stages, this loss of angular momentum is instead caused by random concentrations of gas falling towards the centre.

For all stages of the encounter, we can quantify the link between the local minima of the specific angular momentum and the local maxima of the BH accretion rate, by calculating the cross-correlation function between the BH accretion rate and the central ($<$100~pc) gas specific angular momentum. A cross-correlation analysis quantifies the degree to which two functions (of time, in this particular case) are correlated, by providing the correlation strength of the two functions shifted against one another in lag-time, $\tau$. The presence of a clear, high-amplitude peak (trough) in the cross-correlation function indicates the existence of a strong correlation (anti-correlation) between the two functions, and the lag-time at which such peak (or trough) occurs gives us the delay between the two quantities. In Fig.~\ref{agn2014:fig:m4_hr_gf0_3_BHeff0_001_phi000000_cross_correlation_Mdot-l}, we show this analysis for both BHs of the default merger. During all three stages of the encounter, BH accretion rate and central gas specific angular momentum are strongly anti-correlated, with a lag-time consistent with zero. This shows that, on average, at every stage of the encounter, BH accretion increases when the specific angular momentum of the central gas has a negative temporal gradient\footnote{This result holds -- to different degrees -- for all mergers in our suite, except for the remnant stages of the 1:2 coplanar, retrograde--prograde merger (both BHs) and of the 1:4 inclined-primary merger (secondary BH).}.

\begin{figure}
\centering
\vspace{3.0pt}
\includegraphics[width=0.99\columnwidth,angle=0]{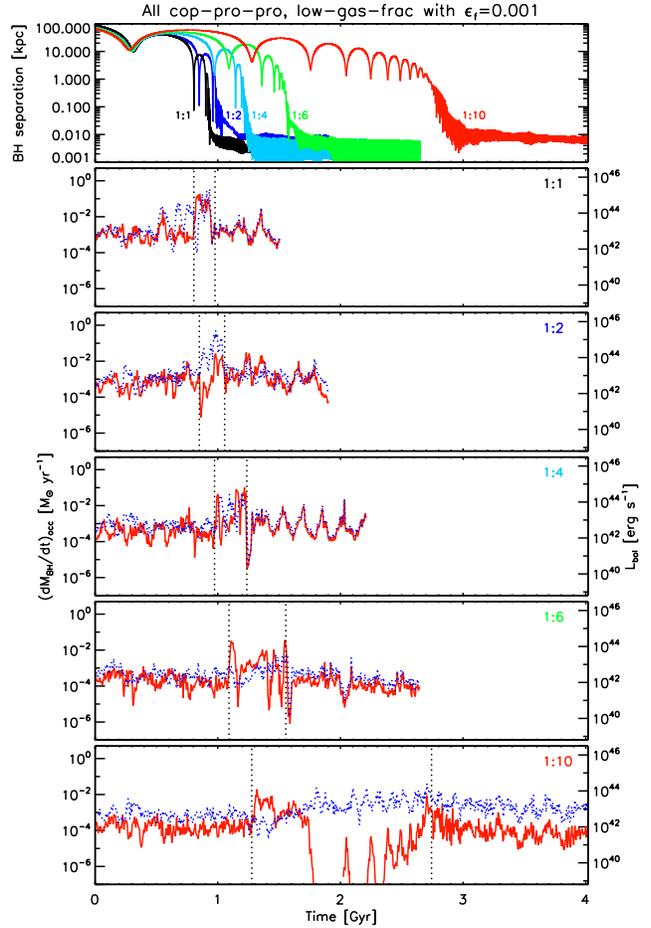}
\vspace{-5pt}
\caption[BH accretion rates for all coplanar, prograde--prograde mergers with low gas fraction and $\epsilon_f=0.001$]{BH accretion rates for all coplanar, prograde--prograde mergers with low gas fraction and $\epsilon_f=0.001$. In panels 2--6, the vertical, dotted, black lines show the separation between the stochastic, the merger, and the remnant stage. {First panel}: separation between the two BHs, for the 1:1 (black), 1:2 (blue), 1:4 (cyan), 1:6 (green), and 1:10 (red) merger. {Second panel}: BH accretion rate for the primary (dotted, blue) and secondary (solid, red) BH of the 1:1 merger. {Third panel}: same as the second panel, but for the 1:2 merger. {Fourth panel}: same as the second panel, but for the 1:4 merger. {Fifth panel}: same as the second panel, but for the 1:6 merger. {Sixth panel}: same as the second panel, but for the 1:10 merger.}
\label{agn2014:fig:Mdot_comparison_allcoplanarlowgasfraction_sixpanels}
\end{figure}


\subsubsection{AGN activity}\label{agn2014:sec:AGN_activity}

\begin{figure}
\centering
\vspace{3.0pt}
\includegraphics[width=0.99\columnwidth,angle=0]{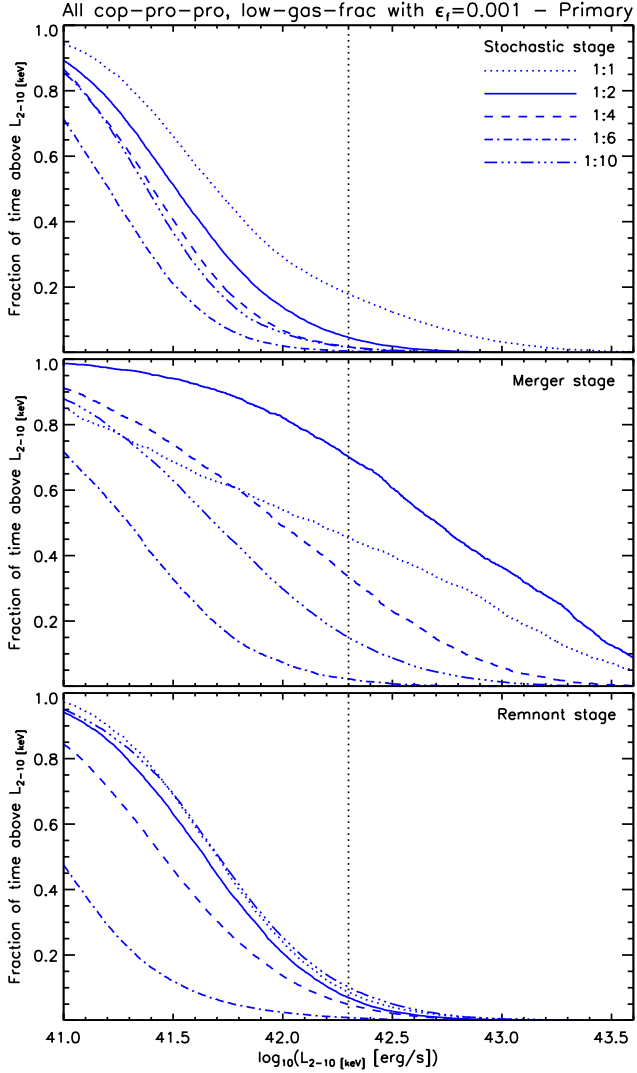}
\vspace{-5pt}
\caption[Fractional cumulative time above a given luminosity -- Coplanar, prograde--prograde mergers with low gas fraction and $\epsilon_f=0.001$ -- Primary BH]{Fractional cumulative time above a given luminosity -- Coplanar, prograde--prograde mergers with low gas fraction and $\epsilon_f=0.001$ -- Primary BH. Same as Fig.~\ref{agn2014:fig:m4_hr_gf0_3_BHeff0_001_phi000000_integral_agn_linlog}. {Top panel}: fractional cumulative time for the stochastic stage, for the primary BH of the 1:1 (dotted), 1:2 (solid), 1:4 (dashed), 1:6 (dash-dotted), and 1:10 (dash-triple-dotted) merger. {Middle panel}: same as the top panel, but for the merger stage. {Bottom panel}: same as the top panel, but for the remnant stage.}
\label{agn2014:fig:integral_agn_comparison_allcoplanar_primary_linlog}
\end{figure}

\begin{figure}
\centering
\vspace{3.0pt}
\includegraphics[width=0.99\columnwidth,angle=0]{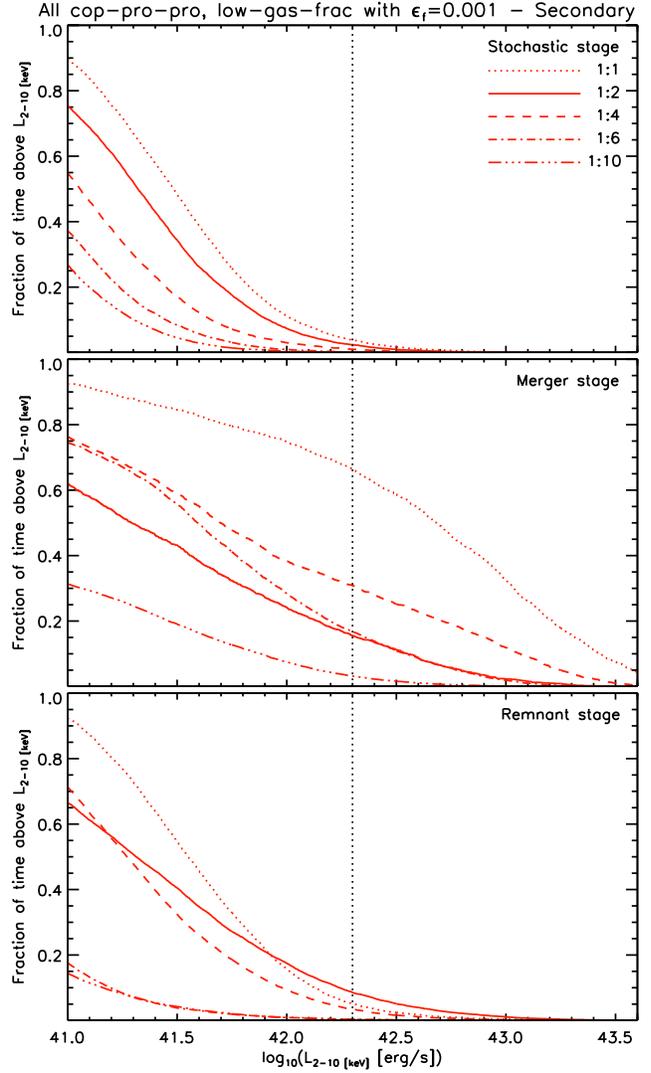}
\vspace{-5pt}
\caption[Fractional cumulative time above a given luminosity -- Coplanar, prograde--prograde mergers with low gas fraction and $\epsilon_f=0.001$ -- Secondary BH]{Fractional cumulative time above a given luminosity -- Coplanar, prograde--prograde mergers with low gas fraction and $\epsilon_f=0.001$ -- Secondary BH. Same as Fig.~\ref{agn2014:fig:integral_agn_comparison_allcoplanar_primary_linlog}, but for the secondary BH.}
\label{agn2014:fig:integral_agn_comparison_allcoplanar_secondary_linlog}
\end{figure}

In order to assess the relative effectiveness of each galaxy and stage in inducing increased BH accretion, it is more convenient to move from the `differential view' of Figs~\ref{agn2014:fig:m4_hr_gf0_3_BHeff0_001_phi000000_five_panels_secondary} and \ref{agn2014:fig:m4_hr_gf0_3_BHeff0_001_phi000000_five_panels_primary} (BH accretion rate as a function of time) to an `integral view', by calculating for how long a given BH is accreting above a given accretion level. In Fig.~\ref{agn2014:fig:m4_hr_gf0_3_BHeff0_001_phi000000_integral_agn_linlog}, we show the fractional cumulative time for which each BH is active above a given luminosity, for the three stages of the merger. In order to compare our results to observations, we do not consider the bolometric luminosity (which is simply proportional to the BH accretion rate), but apply instead a bolometric correction \citep{Hopkins2007} and consider the hard-X-ray luminosity for the 2--10-keV band. During the stochastic and remnant stages, the primary BH is active for longer times than the secondary BH, for almost all relevant luminosity thresholds, even though by a factor of at most a few. This is the case also during the merger stage, if we consider low luminosities [$\log_{10}$($L_{2-10\,\, {\rm keV}}) \lesssim 42.4$], but the situation is inverted for higher luminosities, for which the secondary is active for longer times.

\begin{table*} \centering
\vspace{-3.5pt}
\caption[Cumulative time of AGN activity]{
Cumulative time of AGN activity, per stage, for the primary (t1) and secondary (t2) BH, for all mergers with $\epsilon_f=0.001$ (six major mergers and four minor mergers). To recover the fractional cumulative times f1 and f2 in a given stage, one has to divide t1 and t2 by the stage length. All times are in Gyr. (1) Run number (same as in Table~\ref{agn2014:tab:merger_params}). (2) Time duration of the stochastic stage. (3) Cumulative time, during the stochastic stage, during which the primary BH has $L_{2-10\,\; {\rm keV}} > 2 \times 10^{42}$ erg~s$^{-1}$. (4) Same as column (3), but for the secondary BH. (5) Time duration of the merger stage. (6) Same as column (3), but for the merger stage. (7) Same as column (4), but for the merger stage. (8) Time duration of the remnant stage. (9) Same as column (3), but for the remnant stage. (10) Same as column (4), but for the remnant stage.
\label{agn2014:tab:time}}
\vspace{5pt}
{\small
\begin{tabular*}{0.89\textwidth}{c|ccc|ccc|ccc}
Run & & Stochastic stage & & & Merger \;\;\;\; stage & & & Remnant \, stage & \\
\# & length & t1 & t2 & length & t1 & t2 & length & t1 & t2 \B \\
\hline
01   & 0.805 & 0.144 & 0.030 & 0.173 & 0.078 & 0.115 & 0.529 & 0.046 & 0.027 \T \B \\
02   & 0.849 & 0.039 & 0.021 & 0.206 & 0.145 & 0.032 & 0.849 & 0.059 & 0.073 \B \\
03   & 0.859 & 0.023 & 0.018 & 0.197 & 0.128 & 0.081 & 0.859 & 0.053 & 0.021 \B \\
04   & 0.876 & 0.011 & 0.018 & 0.204 & 0.125 & 0.055 & 0.876 & 0.107 & 0.026 \B \\
05   & 0.839 & 0.039 & 0.004 & 0.168 & 0.107 & 0.071 & 0.839 & 0.088 & 0.045 \B \\
06   & 0.856 & 0.099 & 0.029 & 0.146 & 0.092 & 0.051 & 0.856 & 0.294 & 0.023 \B \\
\hline
07   & 0.973 & 0.017 & 0.010 & 0.263 & 0.088 & 0.081 & 0.973 & 0.047 & 0.033 \T \B \\
08   & 0.983 & 0.014 & 0.004 & 0.390 & 0.074 & 0.071 & 0.983 & 0.028 & 0.001 \B \\
09   & 1.091 & 0.005 & 0.002 & 0.464 & 0.011 & 0.078 & 1.091 & 0.010 & 0.004 \B \\
10   & 1.277 & 0.021 & 0.001 & 1.466 & 0.220 & 0.047 & 1.277 & 0.130 & 0.004 \B \\
\hline
\end{tabular*}
\vspace{5pt}
}
\end{table*}

For an easier comparison, we focus on the value of $L$ usually used to define the lower threshold for AGN activity \citep[e.g.][]{Silverman2011}: $L_{2-10\,\, {\rm keV}} \equiv L_{\rm AGN} \simeq 2 \times 10^{42}$~erg~s$^{-1}$. Using this definition, the primary BH is an AGN for 1.7 per cent of the stochastic stage, 33.4 per cent of the merger stage, and 4.8 per cent of the remnant stage. The secondary BH, on the other hand, is an AGN for 1.0, 30.8, and 3.4 per cent of the stage time, for the stochastic, merger, and remnant stages, respectively. For both BHs, there is a clear increase in fractional time moving from the stochastic stage (when the galaxies can be considered almost in isolation) to the merger stage (when the interaction between the two galaxies is stronger), and a clear decrease moving from the merger stage to the remnant stage (when the dynamically-violent phase has ended and a galactic remnant is being formed). In the next sections, we will show that this trend is valid for all mergers in our suite, but what differs is the degree of luminosity increase/decrease from stage to stage, for each BH in different mergers.


\subsection{Dependence on the initial mass ratio}\label{agn2014:sec:Dependence_on_mass_ratio}

In this section, we show the dependence of our results on the initial mass ratio, by keeping all other variables fixed. We consider all five coplanar, prograde--prograde mergers with low gas fraction and $\epsilon_f=0.001$ in our suite. These have mass ratio 1:1, 1:2 (two major mergers: Runs~1 and 2), 1:4, 1:6 and 1:10 (three minor mergers: Runs~7, 9, and 10)\footnote{In Figs~\ref{agn2014:fig:m2_hr_gf0_3_BHeff0_001_phi000000_five_panels_secondary}--\ref{agn2014:fig:m10_hr_gf0_3_BHeff0_001_phi000000_five_panels_primary} of the Appendix, we show in detail the most relevant quantities during the evolution of two of these mergers (Runs~2 and 10).}.

In Fig.~\ref{agn2014:fig:Mdot_comparison_allcoplanarlowgasfraction_sixpanels}, we compare the BH accretion rate for all the mergers considered in this section. The top panel shows the BH separation for each merger, to highlight the fact that encounters with different initial mass ratios have vastly different merger histories, mostly due to the different dynamical-friction time-scales involved.

The other five panels show the BH accretion rate for both BHs in each merger. As already shown in Section~\ref{agn2014:sec:1to4_merger}, BH accretion is relatively low during the stochastic and remnant stages and achieves its highest values during the merger stage. This is the case for all BHs in all mergers, except for one of the BHs\footnote{By definition, there is no primary or secondary BH in a 1:1 merger, at least initially. For this encounter, the primary and secondary BH designations were given randomly.} in the 1:1 merger, which has an unusually high accretion rate during the stochastic stage, and for the secondary BH in the 1:10 merger, as explained below.

The merger history for all these galactic encounters is similar. Up to the second pericentric passage, that is, during the stochastic stage, all secondary BHs have a relatively low BH accretion rate, usually two orders of magnitude below the Eddington level. SFR is also low, below 0.1 M$_{\odot}$~yr$^{-1}$ in the central 100-pc region around all secondary BHs. Further, the specific angular momentum curves of all central gas shells around the secondary BHs are very flat, indicating that there are no large inflows or outflows in this stage. Immediately after the second pericentric passage, at the onset of the merger stage, the specific angular momentum of all central gas shells around all secondary BHs drops by several orders of magnitude, signalling the occurrence of large-scale gas inflows. These gas inflows have the effect of increasing the central gas mass, SFR, and BH accretion. During the merger stage, we note high peaks of BH accretion rates (with a few cases of Eddington or mildly super-Eddington accretion) for all secondary BHs, concurrent to high peaks of SFR around these BHs (with the central SFR usually almost equalling the entire SFR of the system). The only exception is the merger stage of the 1:10 merger (see also Fig.~\ref{agn2014:fig:m10_hr_gf0_3_BHeff0_001_phi000000_five_panels_secondary} of the Appendix), which can be subdivided into two parts: during the first part, the secondary galaxy and BH experience the merger in the same way of all other encounters. During the second part, the gas of the secondary galaxy is severely ram-pressure stripped by the primary galaxy and the secondary BH becomes devoid of surrounding gas \citep[see also][]{Callegari2011}. This gas will eventually provide the supply to feed the BH in the primary galaxy. Finally, the remnant stage is again similar for all mentioned mergers, with lower values of BH accretion and SFR than in the merger stage, but usually a little higher than in the stochastic stage (this is especially true for the 1:10 merger, where the primary feeds off of the gas stripped from the primary galaxy). During this stage, all specific angular momentum curves are flat again, signalling the end of the dynamically-violent stage of the encounter. We note that BH accretion during the remnant stage is unusually high in the 1:2 merger, almost reaching Eddington levels in a couple of instances (see also Fig.~\ref{agn2014:fig:m2_hr_gf0_3_BHeff0_001_phi000000_five_panels_secondary} of the Appendix). Also, the peculiar periodicity of BH accretion observed in the 1:4 merger seems to be unique to that merger.

As far as the primary galaxy of these mergers is concerned, in minor mergers the specific angular momentum of the central gas around the primary BH shows almost no response during the merger stage. However, in the major mergers, the companion is large enough to affect the primary galaxy significantly, and the specific angular momentum curves around the primary BH have a drop similar to those of the gas around the secondary BH.

These general trends apply to all mergers, but the exact results vary with mass ratio. In the major mergers, the primary BH accretion rate during the merger stage clearly reaches values much higher than during the other two stages because the secondary galaxy significantly affects the dynamics of the primary galaxy. A relatively massive companion (especially in the 1:1 merger) causes stronger tidal torques in the primary galaxy, with subsequent gaseous inflows and ultimately higher BH accretion rates. In the minor mergers, on the other hand, the increase in the activity of the primary BH is much more modest. This is especially true in the 1:6 and 1:10 mergers. The effect of the initial mass ratio on the secondary BH accretion rate is less pronounced, because the smaller galaxy always responds more strongly to the interaction.

A different way to quantify the difference between mass ratios is to calculate the fraction of time a given BH accretion rate has been above a given luminosity, in the same way we calculated it for the 1:4 merger in Fig.~\ref{agn2014:fig:m4_hr_gf0_3_BHeff0_001_phi000000_integral_agn_linlog}. In Figs \ref{agn2014:fig:integral_agn_comparison_allcoplanar_primary_linlog} and \ref{agn2014:fig:integral_agn_comparison_allcoplanar_secondary_linlog}, we show such results for all the mergers in this section, for the primary and secondary BH, respectively (see also Table~\ref{agn2014:tab:time}).

The spread in primary AGN-activity time-fraction is relatively small (except for the 1:1 merger, where one of the BHs accretes at an unusually high rate during the stochastic stage). This is expected, since all these primary galaxies are initially identical and, up to the second pericentric passage, can be considered somewhat in isolation (despite the fact that the galaxies have already undergone one pericentric passage). The spread increases significantly when we consider the merger stage, varying from 2.4 per cent in the 1:6 merger to 70.2 per cent in the 1:2 merger. During the remnant stage, the spread in primary AGN-activity time-fraction decreases. This is because, after the merger stage, what remains is a somewhat quieter remnant galaxy with essentially the same mass (by up to a factor of 50 per cent) for all mergers.

If we look at the secondary BH (again, except for the 1:1 merger) in the stochastic stage, the spread in AGN-activity time-fraction is simply due to the difference in initial galaxy (and BH) mass: the more massive BH accretes more than the least massive BH; recall that Bondi accretion is proportional to the square of the BH mass. This spread does not change much during the encounter, i.e., it is mostly driven by the difference in BH mass.

In all cases, however, we note that the AGN activity time-fraction increases when going from the stochastic stage to the merger stage (by different amounts: major versus minor mergers) and decreases when going from the merger stage to the remnant stage. Usually, the remnant stage has higher time-fractions than the stochastic stage, even though by not much. This probably occurs because the remnant galaxy is not yet a quiet, in-equilibrium system, and because BH masses have increased in the meantime.

\begin{figure}
\centering
\vspace{-10pt}
\includegraphics[width=1.01\columnwidth,angle=0]{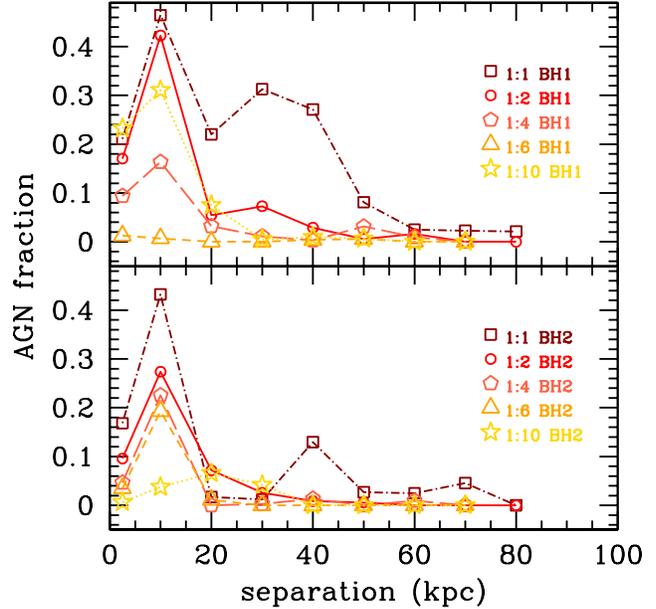}
\vspace{-5pt}
\caption[AGN fraction]{AGN fraction, defined for a threshold hard-X-ray luminosity $L_{\rm AGN} \simeq 2 \times 10^{42}$~erg~s$^{-1}$, as a function of separation, for all mergers with $\epsilon_f=0.001$. {Top panel}: primary BH. {Bottom panel}: secondary BH. Squares: 1:1 merger; circles: all 1:2 mergers; pentagons: all 1:4 mergers; triangles: 1:6 merger; stars: 1:10 merger.}
\label{agn2014:fig:AGNfraction}
\end{figure}

The reason why we have considered the fractional cumulative time instead of the real cumulative time is because we want to compare different stages within a merger, and different mergers, which naturally have different physical durations. By doing this, however, we lose the information of how long is a stage. If we wanted to know the AGN efficiency in a population of mergers, we would also need to know the real times. For this reason, we put in Table~\ref{agn2014:tab:time} the values of the real times (t1 and t2) instead of the fractional times (f1 and f2), so that they can be used for future works. To recover f1 (f2) in a given stage, one has to divide t1 (t2) by the stage length.

\cite{Ellison2011}, \cite{Silverman2011} and \cite{2013MNRAS.435.3627E} study the enhancement in the AGN fraction for galaxies in pairs with respect to isolated galaxies. They all find an enhancement in the AGN fraction in galaxy pairs. \cite{Silverman2011} report an AGN fraction $\sim$10 per cent at separations $<$75~kpc and $\sim$7 per cent at separations $<$150~kpc for a sample at $z<1$; \cite{Lackner_et_al_2014} extend the analysis to close pairs in late merger stages, finding an overall AGN fraction of $\sim$6 per cent at separations $<$10~kpc, while \cite{Ellison2011} report an AGN fraction $\sim$20 per cent at separations $<$10~kpc and $\sim$12 per cent at separations between 10 and 30~kpc, decreasing to $\sim$6 per cent at separations $>$40~kpc for a sample at $z<0.2$.

 Given the different redshift range and selection criteria, it is not trivial to compare our results to their observations. Furthermore, \cite{Ellison2011} use emission lines to identify AGN, a type of diagnostics we cannot model. We therefore propose in the following only a qualitative comparison with observations, and we adopt a fixed luminosity threshold in the hard-X-ray band, similarly to \cite{Silverman2011} as a reference. Fig.~\ref{agn2014:fig:AGNfraction} shows the AGN fraction as a function of separation for our simulated galaxies. We find that the AGN fraction generally increases with decreasing separations, down to $\sim$5~kpc. The drop in the innermost bin occurs because the strongest AGN activity occurs after the second pericentric passage, for a few apo-pericentre oscillations, where the separation between the BHs is between 1 and 10~kpc (recall that the time spent near apocentre is longer than the time spent near pericentre). By the time the BHs remain persistently on sub-kpc scales, the main burst of activity has ended. Compared to observations, we find a lower AGN fraction at large separations, in absolute terms. This happens because we simulate small galaxies with small BHs, and only rarely in the stochastic stage the BHs are above the AGN threshold, as discussed above.

Observationally, the relative enhancement of AGN activity on the primary or secondary BH is still relatively uncertain. For instance, \cite{2013MNRAS.435.2335B} find that in the merging (triple) system NGC~3341 an AGN is triggered only in the smaller galaxy of a minor merger (the mass ratio is 1:25, even less than our most minor merger). \cite{Ellison2011}, however, find that in a sample of Sloan Digital Sky Survey galaxies with a close companion the AGN fraction strongly increases for the massive galaxies in the pairs, while the signal is marginal in the secondary galaxy \citep[see also][]{2007AJ....134..527W}. On the other hand, \cite{2011ApJ...737..101L} find that AGN luminosities and BH accretion rates are higher in the secondary galaxies of their pairs. In our simulations we find that a fixed {\it luminosity} threshold favours detecting activity of the primary, more massive BH. In terms of {\it specific accretion rate}, e.g., Eddington rate, we find instead that the secondary BH is more excited. If we defined the AGN threshold at, e.g., 5 per cent of the Eddington rate, we would find an average AGN fraction of 15 per cent for the secondary BH for separations $<$70~kpc, while the AGN fraction for the primary BH would be $\sim$10 per cent only for separations $<$10~kpc. In the following we expand the discussion on the relative strength of accretion for the two BHs.

\begin{figure}
\centering
\vspace{2.5pt}
\includegraphics[width=0.99\columnwidth,angle=0]{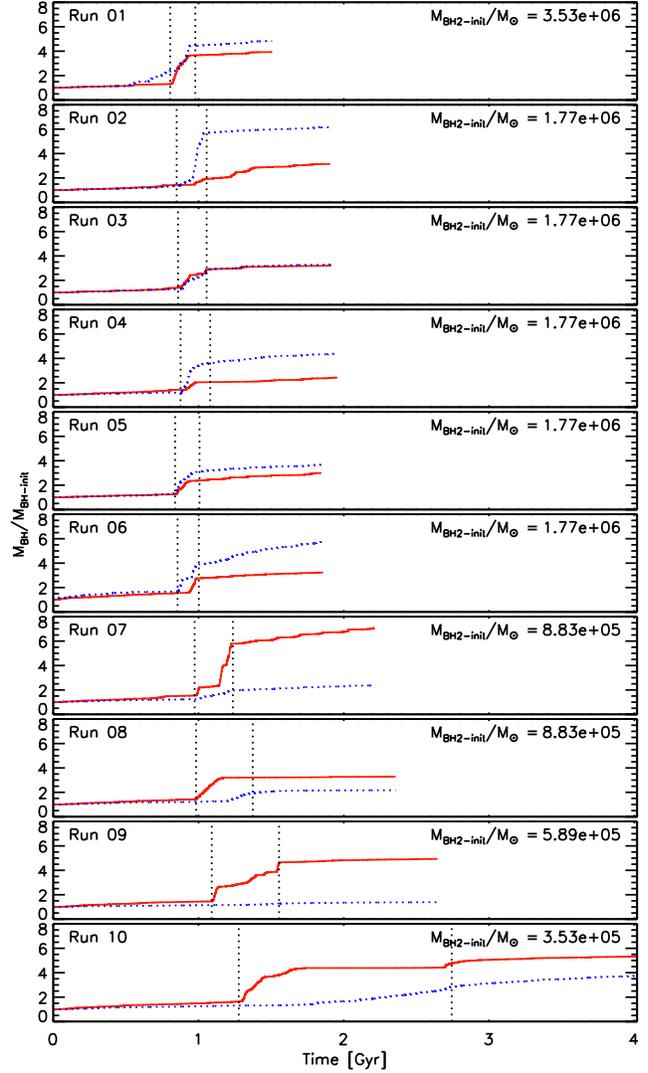}
\vspace{-5pt}
\caption[BH mass, as a function of time, for all mergers with $\epsilon_f=0.001$]{BH mass, as a function of time, for all mergers with $\epsilon_f=0.001$. In all panels: the vertical, dotted, black lines show the separation between the stochastic, the merger, and the remnant stage; the dotted, blue (solid, red) line shows the mass of the primary (secondary) BH, divided by its own value at the beginning of the merger simulation; the initial BH mass of the primary galaxy is always $3.53 \times 10^6$~M$_{\odot}$. The panels are ordered as in Table~\ref{agn2014:tab:merger_params}.}
\label{agn2014:fig:q_comparison_tenpanels}
\end{figure}

In Fig.~\ref{agn2014:fig:q_comparison_tenpanels}, we show the dependence of the mass of each BH with time. Consistent with the fact that BH accretion is usually very low during the stochastic and remnant stages, the mass of each BH grows relatively very little before and after the merger stage, when instead it increases by factors of up to a few. The only large exception is the mass growth of one of the BHs during the stochastic stage of the 1:1 merger, as mentioned before.

\begin{figure}
\centering
\vspace{2pt}
\includegraphics[width=0.99\columnwidth,angle=0]{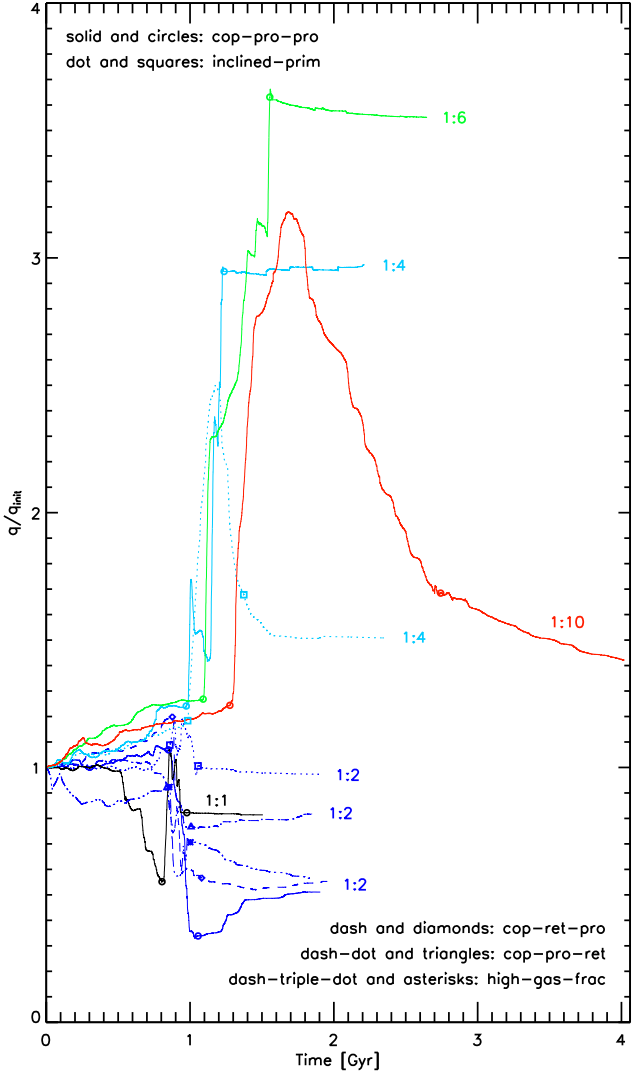}
\vspace{-5pt}
\caption[BH mass ratio, as a function of time, for all mergers with $\epsilon_f=0.001$]{BH mass ratio $q$, as a function of time, for all mergers with $\epsilon_f=0.001$. Each line shows $q$, divided by its own value at the beginning of the merger simulation: 1:1 coplanar, prograde--prograde (solid, black), 1:2 coplanar, prograde--prograde, low-gas-fraction (solid, blue), 1:2 inclined-primary (dotted, blue), 1:2 coplanar, retrograde--prograde (dashed, blue), 1:2 coplanar, prograde--retrograde (dash-dotted, blue), 1:2 coplanar, prograde--prograde, high-gas-fraction (dash-triple-dotted, blue), 1:4 coplanar, prograde--prograde (solid, cyan), 1:4 inclined-primary (dotted, cyan), 1:6 coplanar, prograde--prograde (solid, green), and 1:10 coplanar, prograde--prograde (solid, red) merger. Each line is interrupted by two markers, which indicate the beginning and the end of the merger stage.}
\label{agn2014:fig:q_comparison_singlepanel_1}
\end{figure}

The difference from merger to merger, however, is in how fast the two BHs grow with respect to each other, depending upon the initial mass ratio. In the major mergers (1:1 and 1:2), the primary BH grows more than the secondary BH (the blue lines are above the red lines). In the minor mergers (1:4, 1:6, and 1:10), the opposite occurs, with the smaller BH growing faster than the larger BH (the red lines are above the blue lines).

The evolution of the BH mass ratio, $q\equiv M_{BH2}/M_{BH1}$, can help us understand which BH would grow relatively more during a merger event. In order to better compare the behaviour of these mergers, in Fig.~\ref{agn2014:fig:q_comparison_singlepanel_1}, we show the dependence of the BH mass ratio $q$ with time for all mergers with $\epsilon_f=0.001$ in one panel. The BH mass ratio is somewhat flat during the stochastic and remnant stages and changes significantly during the merger stage (in Fig.~\ref{agn2014:fig:q_comparison_singlepanel_1}, delimited by the two markers on each line), as already inferred from Fig.~\ref{agn2014:fig:q_comparison_tenpanels}, although here we can see that $q$ is not exactly constant during the first and third stage of the encounter. The four minor mergers end up with BH mass ratios larger than their own values at time zero, by factors that vary from $\sim$1.4 (1:10 merger; note, however, that this factor is as high as $\sim$3.2 during the merger stage) to $\sim$3.6 (1:6 merger). The six major mergers, on the other hand, end up with BH mass ratios smaller than their own values at time zero\footnote{We remind the reader that the choice of primary and secondary BH in the 1:1 merger is arbitrary. By inverting the two definitions of BHs in such merger, the final mass ratio is slightly higher than 1, instead of slightly lower than 1. However, since it is still the case that one of the BHs is smaller than the other, the major merger has become a little less major.}, by factors that vary from $\sim$1 to $\sim$2.

\begin{figure*}
\centering
\vspace{2pt}
\includegraphics[width=1.55\columnwidth,angle=90]{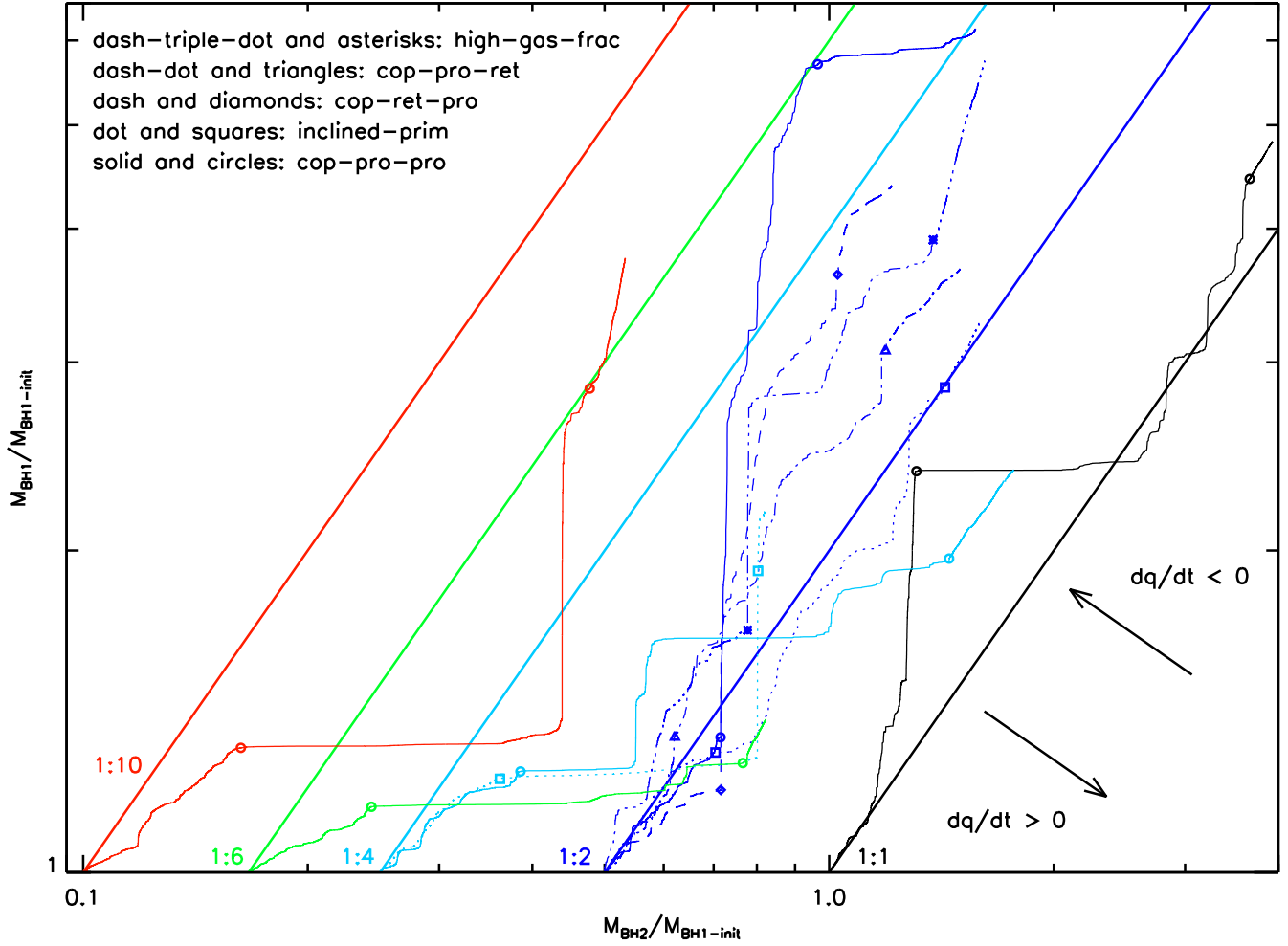}
\vspace{-6pt}
\caption[BH mass phase-diagram, for all mergers with $\epsilon_f=0.001$]{BH mass phase-diagram, for all mergers with $\epsilon_f=0.001$. We show the mass of the primary BH versus the mass of the secondary BH, both divided by the mass of the primary BH at the beginning of the merger simulation: 1:1 coplanar, prograde--prograde (solid, black), 1:2 coplanar, prograde--prograde, low-gas-fraction (solid, blue), 1:2 inclined-primary (dotted, blue), 1:2 coplanar, retrograde--prograde (dashed, blue), 1:2 coplanar, prograde--retrograde (dash-dotted, blue), 1:2 coplanar, prograde--prograde, high-gas-fraction (dash-triple-dotted, blue), 1:4 coplanar, prograde--prograde (solid, cyan), 1:4 inclined-primary (dotted, cyan), 1:6 coplanar, prograde--prograde (solid, green), and 1:10 coplanar, prograde--prograde (solid, red) merger. Each line is interrupted by two markers, which indicate the beginning and the end of the merger stage. The diagonal, thick lines show the lines of constant BH mass ratio: 1:1 (black), 1:2 (blue), 1:4 (cyan), 1:6 (green), and 1:10 (red).}
\label{agn2014:fig:q_comparison_singlepanel_2}
\end{figure*}

In other words, if we look at the BH mass ratio $q$, major mergers tend to become `more minor' ($dq/dt < 0$) and minor mergers tend to become `more major' ($dq/dt > 0$). To better understand the evolution of $q$ during the history of each encounter, in Fig.~\ref{agn2014:fig:q_comparison_singlepanel_2} we show how the BH mass ratio varies, by plotting the mass of one BH as a function of the mass of the other BH. The diagonal, straight lines are the curves of constant $q$. If the two BHs follow this line or a line parallel to it, it means that they are increasing at the same fractional pace [$d(\ln M_{BH})/dt$]. If the slope of the `$M_{BH1}$ versus $M_{BH2}$' curve is lower (higher), it means that the secondary BH is increasing in mass at a greater (smaller) fractional pace than the primary BH. If the `$M_{BH1}$ versus $M_{BH2}$' curve crosses another diagonal straight line, it implies that the BH mass ratio has changed significantly (in particular, if it crosses the cyan line, it means that it went from major to minor, or from minor to major). The BH mass ratio does not depart that much from the lines of constant $q$, during the first and third stage. It is during the merger stage (in Fig.~\ref{agn2014:fig:q_comparison_singlepanel_2}, delimited by the two markers on each line) that $q$ changes significantly. For example, the 1:10 merger started as a minor merger, then briefly became `major' (that is, crossed the cyan diagonal line), then went back to being minor. The 1:4 and 1:6 mergers, on the other hand, quickly enter the major-merger area and never leave it. The 1:2 mergers have the opposite behaviour, and the 1:1 merger more or less stays the same. In the minor mergers, the secondary galaxy is more affected by the encounter, therefore there are stronger gas inflows, more BH accretion and more BH growth. In the major mergers, on the other hand, the secondary galaxy is more resistant to the effects of the merger, therefore both BHs tend to grow in the same way except, being that the primary BH is larger, it grows more, simply because the Bondi accretion formula goes like the square of the BH mass.


\subsection{Dependence on the orbital configuration}\label{agn2014:sec:Dependence_on_orbital_configuration}

In this section, we show the dependence of our results on the initial orbital configuration, by keeping all other variables fixed. We consider all the four 1:2 mergers with low gas fraction and $\epsilon_f=0.001$ of our suite (coplanar, prograde--prograde; inclined-primary; coplanar, retrograde--prograde; and coplanar, prograde--retrograde; Runs~2--5) and, separately, the two 1:4 mergers (coplanar, prograde--prograde; and inclined-primary; Runs~7--8). We recall that the global orbit of the two galaxies is always the same for every merger in our suite: the two galaxies start at a distance equal to the sum of their virial radii and approach each other on a parabolic orbit such that their first pericentric distance is 20 per cent of the virial radius of the larger galaxy.

\begin{figure}
\centering
\vspace{2.5pt}
\includegraphics[width=0.99\columnwidth,angle=0]{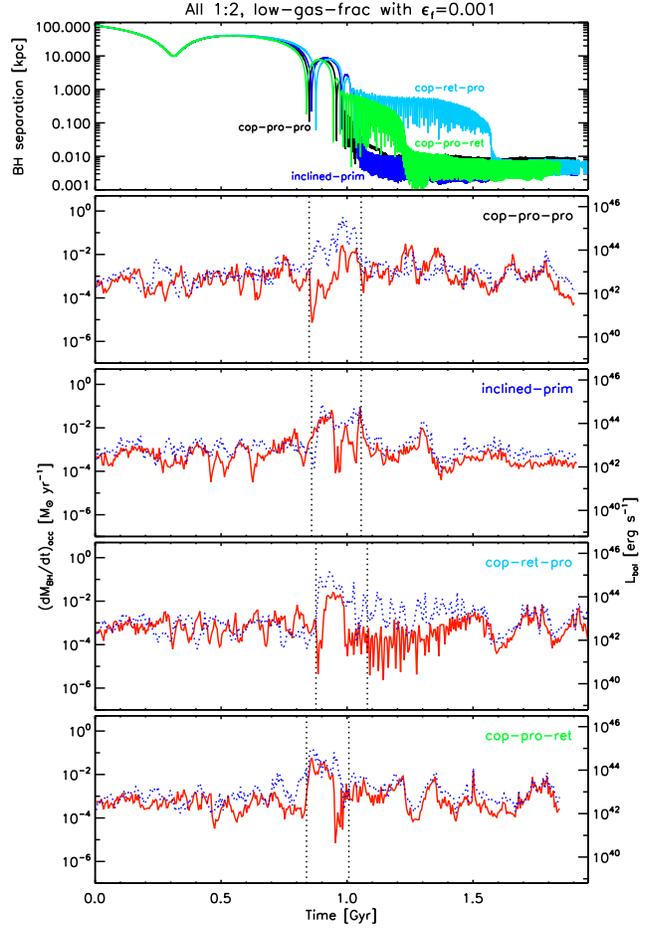}
\vspace{-5pt}
\caption[BH accretion rates for all 1:2 mergers with low gas fraction and $\epsilon_f=0.001$]{BH accretion rates for all 1:2 mergers with low gas fraction and $\epsilon_f=0.001$. In panels 2--5, the vertical, dotted, black lines show the separation between the stochastic, the merger, and the remnant stage. {First panel}: separation between the two BHs, for the prograde--prograde (black), inclined-primary (blue), retrograde--prograde (cyan), and prograde--retrograde (green) merger. {Second panel}: BH accretion rate for the primary (dotted, blue) and secondary (solid, red) BH of the prograde--prograde merger. {Third panel}: same as the second panel, but for the inclined-primary merger. {Fourth panel}: same as the second panel, but for the retrograde--prograde merger. {Fifth panel}: same as the second panel, but for the prograde--retrograde merger.}
\label{agn2014:fig:Mdot_comparison_all1to2lowgasfraction_fivepanels}
\end{figure}

In Fig.~\ref{agn2014:fig:Mdot_comparison_all1to2lowgasfraction_fivepanels}, we compare the BH accretion rate for all the 1:2 mergers considered in this section. The top panel shows the BH separation for each merger, to highlight the fact that encounters with the same initial mass ratio have rather similar merger histories: the beginning and the end of the merger stage happen at very similar times, throughout the four encounters.

\begin{figure}
\centering
\vspace{3.0pt}
\includegraphics[width=0.99\columnwidth,angle=0]{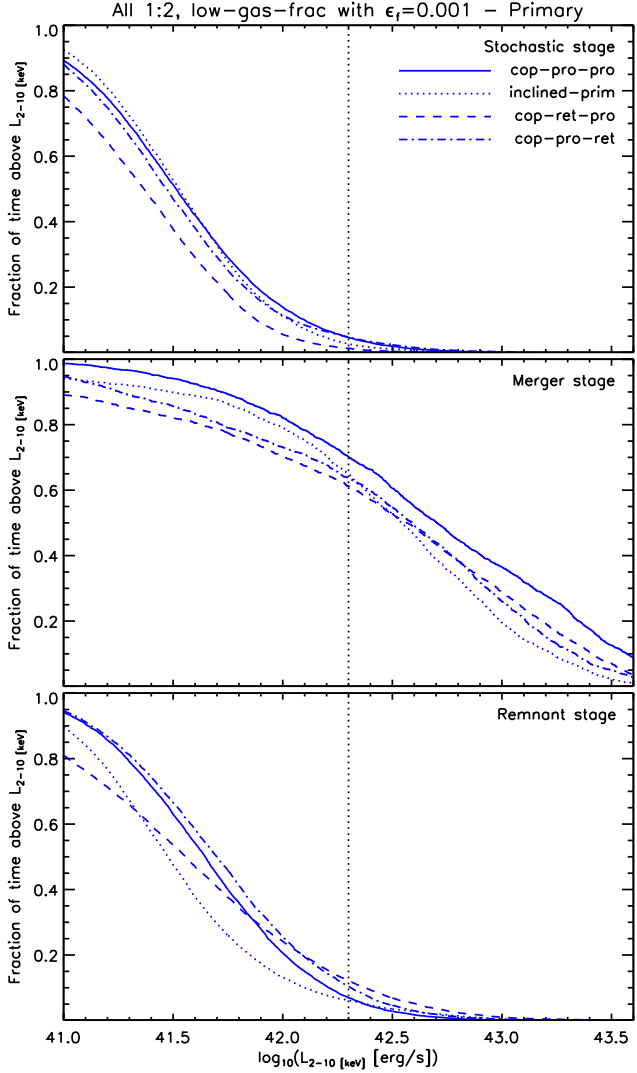}
\vspace{-5pt}
\caption[Fractional cumulative time above a given luminosity -- 1:2 mergers with low gas fraction and $\epsilon_f=0.001$ -- Primary BH]{Fractional cumulative time above a given luminosity -- 1:2 mergers with low gas fraction and $\epsilon_f=0.001$ -- Primary BH. Same as Fig.~\ref{agn2014:fig:m4_hr_gf0_3_BHeff0_001_phi000000_integral_agn_linlog}. {Top panel}: fractional cumulative time for the stochastic stage, for the primary BH of the coplanar, prograde--prograde (solid), inclined-primary (dotted), coplanar, retrograde--prograde (dashed), and coplanar, prograde--retrograde (dash-dotted) merger. {Middle panel}: same as the top panel, but for the merger stage. {Bottom panel}: same as the top panel, but for the remnant stage.}
\label{agn2014:fig:integral_agn_comparison_all1to2_primary_linlog}
\end{figure}

\begin{figure}
\centering
\vspace{3.0pt}
\includegraphics[width=0.99\columnwidth,angle=0]{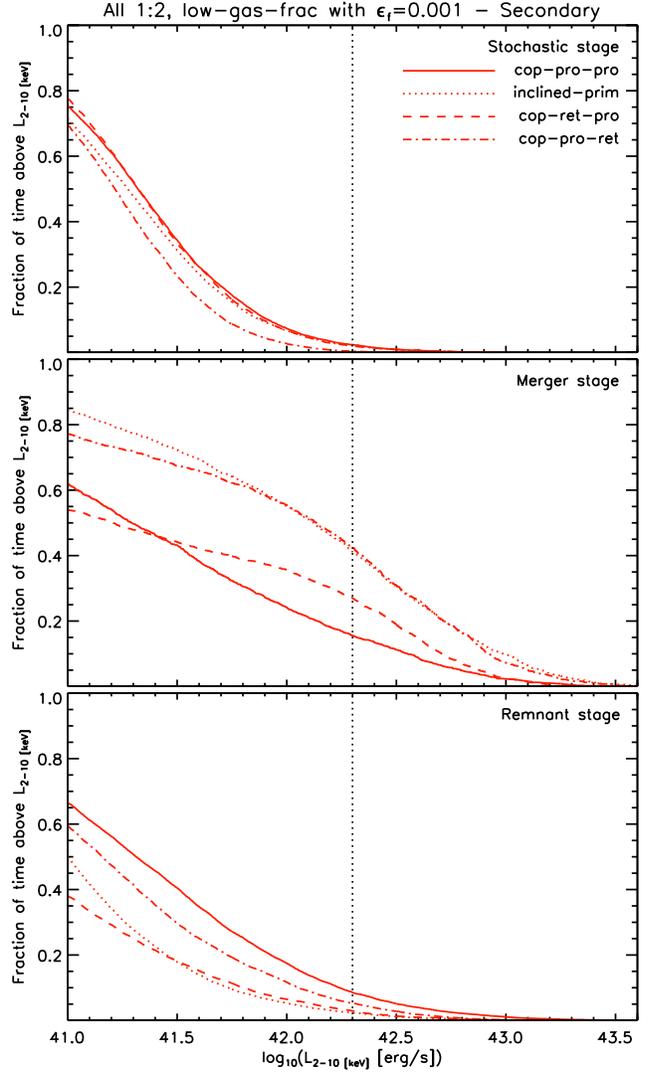}
\vspace{-5pt}
\caption[Fractional cumulative time above a given luminosity -- 1:2 mergers with low gas fraction and $\epsilon_f=0.001$ -- Secondary BH]{Fractional cumulative time above a given luminosity -- 1:2 mergers with low gas fraction and $\epsilon_f=0.001$ -- Secondary BH. Same as Fig.~\ref{agn2014:fig:integral_agn_comparison_all1to2_primary_linlog}, but for the secondary BH.}
\label{agn2014:fig:integral_agn_comparison_all1to2_secondary_linlog}
\end{figure}

The other four panels show the BH accretion rate for both BHs in each merger. The BH accretion rate history is remarkably similar for all mergers. This is even clearer when we look at Figs \ref{agn2014:fig:integral_agn_comparison_all1to2_primary_linlog} and \ref{agn2014:fig:integral_agn_comparison_all1to2_secondary_linlog}, where we show the fractional cumulative time above a given luminosity. We find that, even during the merger stage, there is practically no difference between the encounters, as far as the primary BH is concerned.

\begin{figure}
\centering
\vspace{3.0pt}
\includegraphics[width=0.99\columnwidth,angle=0]{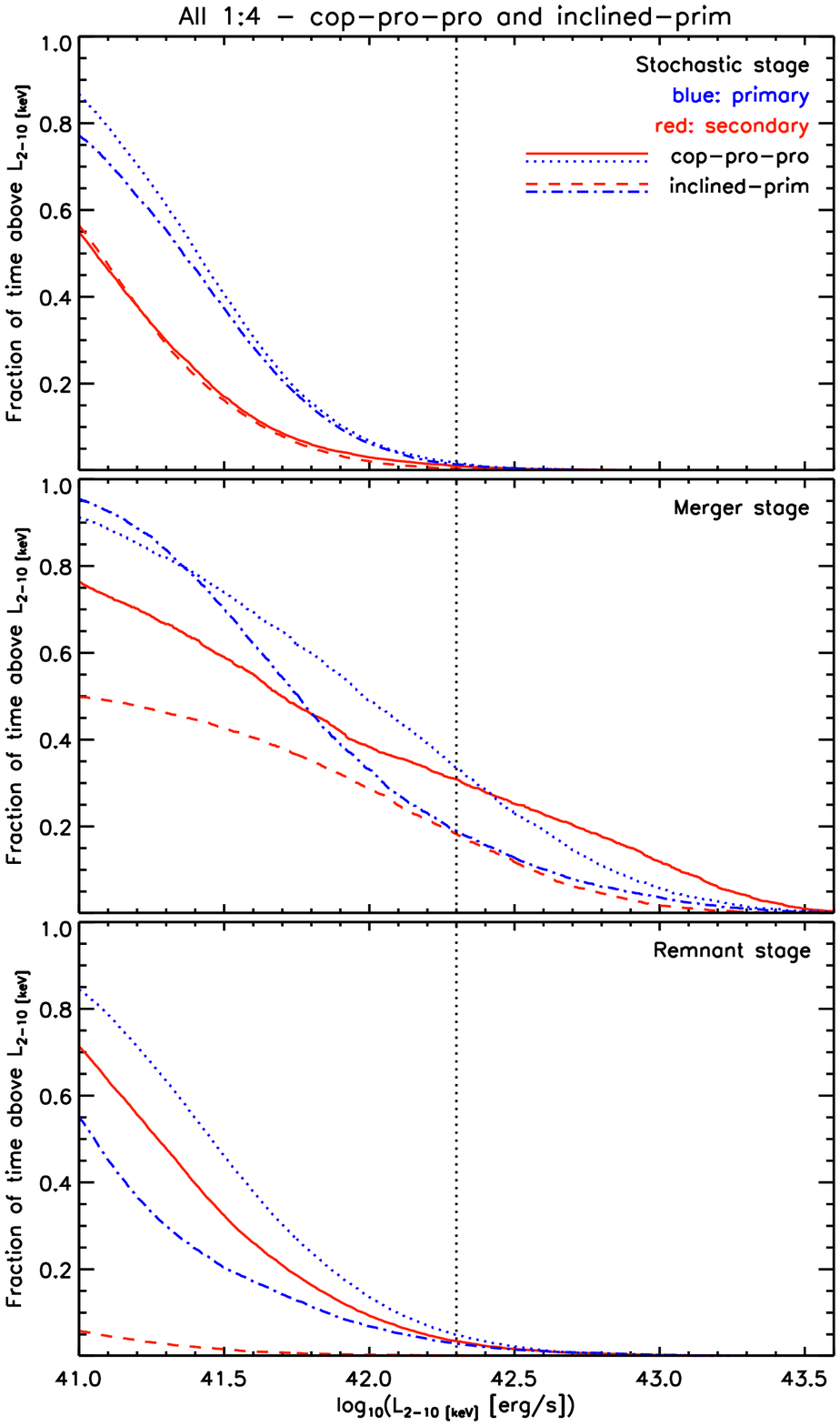}
\vspace{-5pt}
\caption[Fractional cumulative time above a given luminosity -- 1:4 mergers]{Fractional cumulative time above a given luminosity -- 1:4 mergers. Same as Fig.~\ref{agn2014:fig:m4_hr_gf0_3_BHeff0_001_phi000000_integral_agn_linlog}. {Top panel}: fractional cumulative time for the stochastic stage, for the primary (dotted and dash-dotted, blue) and secondary (solid and dashed, red) BH of the coplanar, prograde--prograde (solid and dotted) and inclined-primary (dashed and dash-dotted lines) merger. {Middle panel}: same as the top panel, but for the merger stage. {Bottom panel}: same as the top panel, but for the remnant stage.}
\label{agn2014:fig:integral_agn_comparison_all1to4_linlog}
\end{figure}

In the two 1:4 mergers, the results are similar (see Fig.~\ref{agn2014:fig:integral_agn_comparison_all1to4_linlog}), with the exception that the secondary BH in the inclined-primary merger has almost no AGN activity during the remnant stage. Consistently, the levels of central SFR and gas mass around the secondary BH are much lower than those around the secondary BH of the 1:4 coplanar, prograde--prograde merger.

Finally, a complementary way to view the differences between these mergers is to look at Fig.~\ref{agn2014:fig:q_comparison_singlepanel_2} again. The four major mergers of this section have similar behaviours: in all cases, the final BH mass ratio has decreased from its value at the beginning of the simulation. In other words, all four major mergers have become less `major'. The two minor mergers of this section, on the other hand, have both increased their BH mass ratio, with the 1:4 coplanar, prograde--prograde merger having, at the end of the remnant stage, the highest $q$ of the entire suite of mergers (if we exclude the 1:1 coplanar, prograde--prograde merger). Once again, it seems like the initial mass ratio ($q_{\textnormal{\tiny \textsc{G}}}$) is the most important parameter. However, we also see that, in both groups of major and minor mergers we considered in this section, the inclined mergers tend to keep $q$ relatively closer to its initial value, whereas coplanar mergers tend to change $q$ more (the direction of the change depending on the initial mass ratio). Given the small number of inclined mergers in the suite, we caution the reader that a more thorough study is needed. However, such result is not unreasonable, since in inclined mergers the efficiency of the merger-induced torques is lower than in coplanar mergers (see, e.g., \citealt{Cox2008} for a study of different inclinations in mergers).

\subsection{Dependence on the gas fraction}\label{agn2014:sec:Dependence_on_gas_fraction}

In this section, we show the dependence of our results on the initial gas fraction in the galactic disc, by keeping all other variables fixed. We therefore consider the 1:2 coplanar, prograde--prograde mergers with $\epsilon_f=0.001$ with low gas fraction (30 per cent of the disc mass is in gaseous form; Run~2) and high gas fraction (60 per cent; Run~6).

In terms of merger history, there are not many differences. However, the high-gas-fraction encounter has a merger stage a good 25 per cent shorter than that of its low-gas-fraction counterpart ($\sim$150~Myr versus $\sim$200~Myr).

\begin{figure}
\centering
\vspace{3.0pt}
\includegraphics[width=0.99\columnwidth,angle=0]{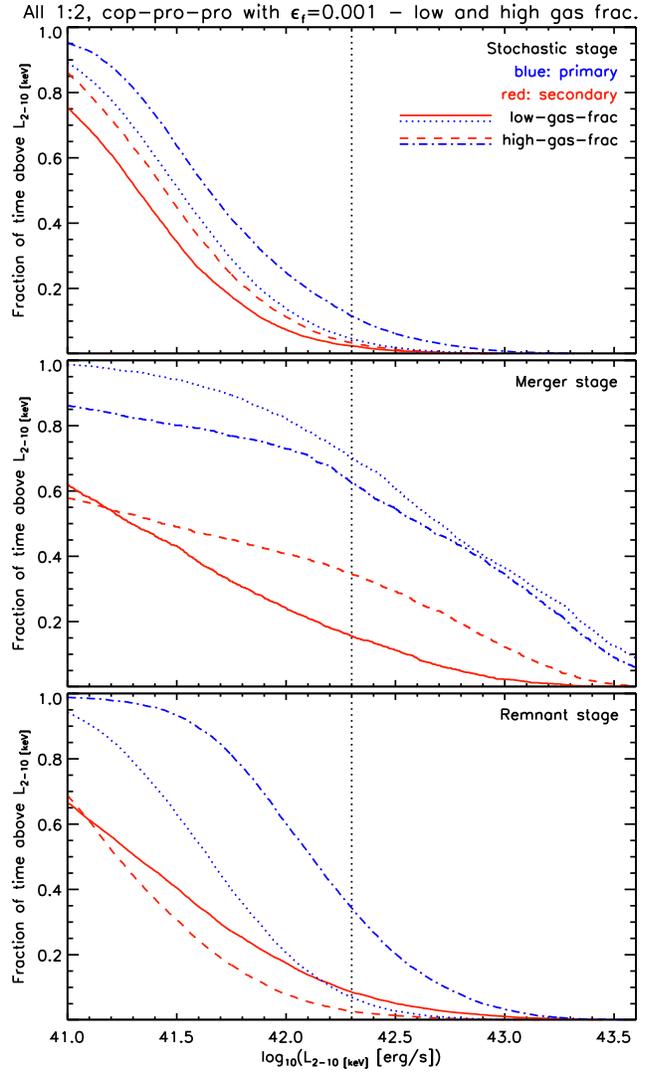}
\vspace{-5pt}
\caption[Fractional cumulative time above a given luminosity -- 1:2 coplanar, prograde--prograde mergers with $\epsilon_f=0.001$ and different gas fractions]{Fractional cumulative time above a given luminosity -- 1:2 coplanar, prograde--prograde mergers with $\epsilon_f=0.001$ and different gas fractions. Same as Fig.~\ref{agn2014:fig:m4_hr_gf0_3_BHeff0_001_phi000000_integral_agn_linlog}. {Top panel}: fractional cumulative time for the stochastic stage, for the primary (dotted and dash-dotted, blue) and secondary (solid and dashed, red) BH of the low-gas-fraction (solid and dotted) and high-gas-fraction (dashed and dash-dotted) merger. {Middle panel}: same as the top panel, but for the merger stage. {Bottom panel}: same as the top panel, but for the remnant stage.}
\label{agn2014:fig:integral_agn_comparison_all1to2coplanar_linlog}
\end{figure}

During the stochastic stage, the BH accretion rate in the high-gas-fraction merger is usually higher than that in the low-gas-fraction merger (see the top panel of Fig.~\ref{agn2014:fig:integral_agn_comparison_all1to2coplanar_linlog}). During the merger stage, both BHs in both mergers experience periods of high activity, with the secondary BH of the high-gas-fraction run being more `active' than that of the low-gas-fraction run, but with the roles inverted (even if barely) for the primary BHs. During the final stage of the encounters, the opposite of the merger stage occurs, with the primary BH being more (less) active than the secondary BH in the high-gas-fraction (low-gas-fraction) merger.

The high-gas-fraction merger is a good example which shows that BH accretion does not necessarily need low levels of specific angular momentum, but instead necessitates the angular momentum to have a negative temporal gradient (recall the discussion in Section~\ref{agn2014:sec:BH_accretion_and_gas_angular_momentum}). Indeed, during the stochastic stage, the specific angular momentum of the central gas of the high-gas-fraction primary galaxy is higher (by a factor of $\sim$2) than that of the low-gas-fraction primary galaxy. Despite this, the primary BH accretion rate in the high-gas-fraction merger is higher than in the low-gas-fraction case.

The unusual levels of specific angular momentum of the high-gas-fraction primary galaxy's central gas are caused by an initial strong peak in BH accretion, due to a combination of higher central gas mass (compared to the low-gas-fraction case) and to an initially slightly lower gas specific angular momentum (compared to the low-gas-fraction case and to the high-gas-fraction secondary galaxy). BHs preferentially accrete gas particles with low specific angular momentum. Additionally, BH feedback, which `pushes' gas particles away, is applied to the gas particle nearest to the BH, which tends to have low specific angular momentum. Higher levels of BH accretion tend, therefore, to increase the specific angular momentum of the central gas.

\subsection{Control runs}\label{agn2014:sec:Control_runs}

In this paper, we have assessed the impact of a few merger quantities (mass ratio, orbital configuration, and gas fraction) on the mass accretion onto and mass growth of BHs. For this reason, we kept the BH parameters the same throughout the analysis (Runs~1--10: Eddington-limited, Bondi accretion and a fixed fraction -- $\epsilon_f=0.001$ -- of the BH accretion energy injected in the nearby gas as thermal energy).

In order to understand the impact of BH physics, we also ran a series of simulations where we changed the BH parameters. See the last three rows in Table~\ref{agn2014:tab:merger_params} for a list of the parameters of the control runs (Runs~C1--C3).

In the first two control runs (Runs~C1 and C2), we assessed the effect of shutting off BH accretion. We ran two versions of the same simulation (1:2 coplanar, prograde--prograde), one with low gas fraction (Run~C1) and the other with high gas fraction (Run~C2), and compared these two control runs with their BH-accreting counterparts (Runs~2 and 6). The two BHs are in the galaxy, but behave only as `giant stars', having no interaction with the other particles in the simulation, other than gravitational.

The lack of BH accretion (and, consequently, of BH feedback) means that (i) the BHs are not `stealing' gas that could be used to form stars and, additionally, that (ii) the gas surrounding the BHs is not being heated and/or pushed away. This implies that the central SFR should be higher in the control runs than in their BH-accreting counterparts. This is indeed the case: SFR in the central 100~pc is higher when BH accretion is shut off. Also, the temporal evolution of the central SFR is much smoother, because there is no BH feedback.

The central ($<$100~pc) gas specific angular momentum curves of the low-gas-fraction merger during the stochastic stage are much less noisy than in the runs with BH accretion. In the stochastic stage of the high-gas-fraction runs, the central gas specific angular momentum around the primary BH is much higher in the BH-accreting run than in its counterpart control run (because BH accretion and feedback tend to deplete the central region of low-specific-angular-momentum gas; see Section~\ref{agn2014:sec:Dependence_on_gas_fraction} for more details).

All other (i.e. non-central) quantities are not affected by the accretion mode of the central BHs. Moreover, from the point of view of the orbital history of the mergers, both mergers have a `normal-length' stochastic stage (0.849~Gyr and 0.854~Gyr, for the low-gas-fraction and high-gas-fraction run, respectively), followed by a `normal-length' (albeit a little on the short side) merger stage (0.158~Gyr and 0.148~Gyr, for the low-gas-fraction and high-gas-fraction run, respectively).

We also ran a simulation (Run~C3: 1:2 coplanar, prograde--prograde, low-gas-fraction) where the BH is accreting but feeds back to the nearby gas a fraction of the BH accretion energy five times higher than in the other runs: $\epsilon_f=0.005$. In this case, the expectation is the opposite to the other control runs. The BH is heating/pushing the gas away much more than in the $\epsilon_f=0.001$ simulation (Run~2). Moreover, an increased BH feedback also means that there is less gas for accretion onto the BH itself.

However, central SFR does not decrease appreciably: the BH feedback efficiency does not have a strong effect on SF (see also Volonteri et al. 2014, submitted). The BH accretion rate, on the other hand, is expectedly lower, by $\sim$one order of magnitude. For this reason, the cumulative time during which each BH is active is also very low. The primary (secondary) BH is active for only 0.002 (0.001), 0.049 (0.008), and 0.013 (0.016)~Gyr, during the stochastic, merger, and remnant stage, respectively. The stochastic (0.849~Gyr) and merger (0.192~Gyr) stages have durations very similar to those of Run~2.

Another direct consequence of the decreased BH accretion is that the BH mass growth is also very small, compared to the encounter with lower BH efficiency: the primary (secondary) BH grows by a factor of 1.7 (1.4) between the beginning of the stochastic stage and the end of the remnant stage, compared to factors of 6.2 and 3.1 for the merger with $\epsilon_f=0.001$. Much of this difference arises during the merger stage, when BH accretion is higher and, consequently, BH feedback has a stronger impact.

Since the BH masses change very little with time, the BH ratio also does not change significantly. However, also in this case, the merger ends up being (slightly) more minor than at the beginning of the simulation, hinting at the fact that the empirical rule for BH mass ratio evolution we found for low BH feedback efficiency ($\epsilon_f=0.001$) might hold even for higher BH feedback efficiency ($\epsilon_f=0.005$).

In summary, changing the BH parameters leaves all galactic quantities (and orbital evolution) unchanged, except for the very central ($<$100~pc) quantities around the BHs, in the expected directions: with no BH accretion, SFR is higher; with more BH feedback, BH accretion is lower.


\section{Conclusions}\label{agn2014:sec:Conclusions}

We present a new suite of high-resolution hydrodynamical simulations of galaxy mergers, where we vary the initial mass ratio (from 1:1 to 1:10), orbital configuration (coplanar, prograde and retrograde, and inclined), and gas fraction in the galactic discs (30 and 60 per cent), focussing on the accretion onto and mass growth of the central BHs, and on the triggering of AGN. Our simulations, with $\sim$10-pc resolution, allow us to show that $100$-pc resolution is not sufficient to resolve the nuclear torques that develop in unequal-mass mergers \citep{VW2014}. This suite of simulations shows that much of the loss of angular momentum that triggers BH accretion occurs below $\sim$50-pc scales, a regime very poorly studied through numerical simulations because of the high resolution required, while at the same time a box of several hundred kpc is needed to capture the complete merger. 

We itemize our findings below.

\begin{enumerate}

\item All encounters in the suite can be subdivided into three clearly distinct stages (\textit{stochastic}, \textit{merger}, and \textit{remnant}), distinguished by the time evolution of the specific angular momentum in several central gas spherical shells around the secondary BH. At the end of the remnant stage the galaxy returns to a quiescent state, similar to that of the early stochastic stage.

\item We find a strong anti-correlation between the specific angular momentum of the central gas and the BH accretion rate. The relevant quantity is not, however, the magnitude of the specific angular momentum, but its temporal gradient. The gas that gets accreted is not necessarily gas with low angular momentum, but the gas that loses angular momentum.

\item We quantify the relative enhancement of BH accretion in the merger stage with respect to the stochastic and remnant stages. While not all AGN activity is merger-driven, the merger stage presents the strongest persistent AGN activity.

\item The initial mass ratio between the galaxies is the parameter that most affects BH accretion and AGN activity in mergers, whereas gas fraction and orbital configuration have very minor effects. 

\item The secondary galaxy always responds strongly to the interaction, almost independently of the mass ratio. The primary galaxies in major mergers are instead vastly more affected than in minor mergers.

\item We compute the evolution of the BH mass ratio during the encounters. In minor mergers, the secondary BH grows faster (fractionally) than the primary BH: \textit{minor mergers tend to become less minor}. In major mergers, the opposite occurs: the primary BH grows faster (fractionally) than the secondary BH: \textit{major mergers tend to become less major}. 

\item We calculate the AGN fraction as a function of separation and find, in broad agreement with observations, that the AGN fraction generally increases with decreasing separations. 

\end{enumerate}

In summary, thanks to the very high spatial ($\sim$10~pc) and temporal ($\sim$1~Myr) resolution of our simulations, we confirm that BH accretion in galaxy mergers is strongly linked to how effective physical processes are in inducing the gas to lose angular momentum and flow towards the centre of the galaxy. This effectiveness does not seem to be very dependent on the initial orbital configuration and/or on the gas fraction, but is strongly dependent on the initial mass ratio. We identified the relative enhancement of merger-driven AGN activity versus the stochastic AGN activity that occurs in dynamically quiescent galaxies, by studying BH accretion in our galaxies before and after the merger started. In short, we find that in the `normal galaxy' phase the BHs are not completely quiescent. The luminosity in this phase, however, seldom exceeds $10^{42}$ erg s$^{-1}$ (depending on the exact bolometric correction adopted) and it is for the most part at sub-Eddington levels.

Moreover, we present detailed analysis on the time evolution of the BH mass ratio, $q$, aiming at understanding which BH would grow the most during a merger event. We find that $q$ changes significantly during the merger, with the direction of this change depending on the initial mass ratio: very unequal BH pairs (with an initial $q \le 0.25$), tend to evolve towards higher $q$, whereas pairs with an initial $q > 0.25$ tend to increase the BH mass contrast. If this trend were confirmed at smaller separations, this would result in a narrow distribution of BH mass ratios expected during the shrinking and coalescence of BH binaries.

\section*{Acknowledgements}
MV acknowledges funding support from NASA, through award ATP NNX10AC84G, from SAO, through award TM1-12007X, from NSF, through award AST 1107675, and from a Marie Curie Career Integration grant (PCIG10-GA-2011-303609). Resources supporting this work were provided by the NASA High-End Computing (HEC) Program through the NASA Advanced Supercomputing (NAS) Division at Ames Research Center, and by TGCC, under the allocations 2013-t2013046955 and 2014-x2014046955 made by GENCI. This research was supported in part by the National Science Foundation under grant no. NSF PHY11-25915, through the Kavli Institute for Theoretical Physics and its program `A Universe of Black Holes'. PRC thanks Sandor Van Wassenhove and Thomas~R. Quinn for their help and useful discussions, and the Institut d'Astrophysique de Paris for hosting him during his visits.

\scalefont{0.94}
\setlength{\bibhang}{1.6em} 
\setlength\labelwidth{0.0em}
\bibliographystyle{mn2e}
\bibliography{mergerdrivenBHaccretion_20141127}

\begin{thebibliography}{70}
\expandafter\ifx\csname natexlab\endcsname\relax\def\natexlab#1{#1}\fi

\bibitem[{{Adelberger} {et~al}\mbox{.}(2005){Adelberger}, {Steidel}, {Pettini},
  {Shapley}, {Reddy}, \& {Erb}}]{adelberger2005b}
{Adelberger} K.~L., {Steidel} C.~C., {Pettini} M., {Shapley} A.~E., {Reddy}
  N.~A., {Erb} D.~K., 2005, \apj, 619, 697

\bibitem[{{Alexander} \& {Hickox}(2012)}]{2012NewAR..56...93A}
{Alexander} D.~M., {Hickox} R.~C., 2012, \nar, 56, 93

\bibitem[{{Barnes} \& {Hernquist}(1992)}]{Barnes92}
{Barnes} J.~E., {Hernquist} L., 1992, \araa, 30, 705

\bibitem[{{Barnes} \& {Hernquist}(1996)}]{BarnesHernquist96}
{Barnes} J.~E., {Hernquist} L., 1996, ApJ, 471, 115

\bibitem[{{Begelman}, {Blandford} \& {Rees}(1980){Begelman}, {Blandford}, \&
  {Rees}}]{BBR1980}
{Begelman} M.~C., {Blandford} R.~D., {Rees} M.~J., 1980, {Nature}, 287, 307

\bibitem[{{Bellovary} {et~al}\mbox{.}(2010){Bellovary}, {Governato}, {Quinn},
  {Wadsley}, {Shen}, \& {Volonteri}}]{Bellovary10}
{Bellovary} J.~M., {Governato} F., {Quinn} T.~R., {Wadsley} J., {Shen} S.,
  {Volonteri} M., 2010, ApJL, 721, L148

\bibitem[{{Benson}(2005)}]{Benson05}
{Benson} A.~J., 2005, MNRAS, 358, 551

\bibitem[{{Bianchi} {et~al}\mbox{.}(2013){Bianchi}, {Piconcelli},
  {P{\'e}rez-Torres}, {Fiore}, {La Franca}, {Mathur}, \&
  {Matt}}]{2013MNRAS.435.2335B}
{Bianchi} S., {Piconcelli} E., {P{\'e}rez-Torres} M.~{\'A}., {Fiore} F., {La
  Franca} F., {Mathur} S., {Matt} G., 2013, \mnras, 435, 2335

\bibitem[{{Callegari} {et~al}\mbox{.}(2011){Callegari}, {Kazantzidis}, {Mayer},
  {Colpi}, {Bellovary}, {Quinn}, \& {Wadsley}}]{Callegari2011}
{Callegari} S., {Kazantzidis} S., {Mayer} L., {Colpi} M., {Bellovary} J.~M.,
  {Quinn} T., {Wadsley} J., 2011, ApJ, 729, 85

\bibitem[{{Callegari} {et~al}\mbox{.}(2009){Callegari}, {Mayer}, {Kazantzidis},
  {Colpi}, {Governato}, {Quinn}, \& {Wadsley}}]{Callegari2009}
{Callegari} S., {Mayer} L., {Kazantzidis} S., {Colpi} M., {Governato} F.,
  {Quinn} T., {Wadsley} J., 2009, ApJL, 696, L89

\bibitem[{{Camm}(1950)}]{Camm_1950}
{Camm} G.~L., 1950, \mnras, 110, 305

\bibitem[{{Chapon}, {Mayer} \& {Teyssier}(2013){Chapon}, {Mayer}, \&
  {Teyssier}}]{Chapon_et_al_2013}
{Chapon} D., {Mayer} L., {Teyssier} R., 2013, \mnras, 429, 3114

\bibitem[{{Cisternas} {et~al}\mbox{.}(2011){Cisternas}, {Jahnke}, {Inskip},
  {Kartaltepe}, {Koekemoer}, {Lisker}, {Robaina}, {Scodeggio}, {Sheth},
  {Trump}, {Andrae}, {Miyaji}, {Lusso}, {Brusa}, {Capak}, {Cappelluti},
  {Civano}, {Ilbert}, {Impey}, {Leauthaud}, {Lilly}, {Salvato}, {Scoville}, \&
  {Taniguchi}}]{Cisternas2011}
{Cisternas} M. {et~al.}, 2011, ApJ, 726, 57

\bibitem[{{Cox} {et~al}\mbox{.}(2008){Cox}, {Jonsson}, {Somerville}, {Primack},
  \& {Dekel}}]{Cox2008}
{Cox} T.~J., {Jonsson} P., {Somerville} R.~S., {Primack} J.~R., {Dekel} A.,
  2008, MNRAS, 384, 386

\bibitem[{{Di Matteo} {et~al}\mbox{.}(2008){Di Matteo}, {Bournaud}, {Martig},
  {Combes}, {Melchior}, \& {Semelin}}]{DiMatteo_et_al_2008}
{Di Matteo} P., {Bournaud} F., {Martig} M., {Combes} F., {Melchior} A.-L.,
  {Semelin} B., 2008, \aap, 492, 31

\bibitem[{{Di Matteo}, {Springel} \& {Hernquist}(2005){Di Matteo}, {Springel},
  \& {Hernquist}}]{DiMatteo2005}
{Di Matteo} T., {Springel} V., {Hernquist} L., 2005, {Nature}, 433, 604

\bibitem[{{Diemer} \& {Kravtsov}(2014)}]{Diemer_Kravtsov_2014}
{Diemer} B., {Kravtsov} A.~V., 2014, preprint (arXiv:1407.4730)

\bibitem[{{Dutton} \& {Macci{\`o}}(2014)}]{Dutton_Maccio_2014}
{Dutton} A.~A., {Macci{\`o}} A.~V., 2014, \mnras, 441, 3359

\bibitem[{{Ellison} {et~al}\mbox{.}(2013){Ellison}, {Mendel}, {Patton}, \&
  {Scudder}}]{2013MNRAS.435.3627E}
{Ellison} S.~L., {Mendel} J.~T., {Patton} D.~R., {Scudder} J.~M., 2013, \mnras,
  435, 3627

\bibitem[{{Ellison} {et~al}\mbox{.}(2011){Ellison}, {Patton}, {Mendel}, \&
  {Scudder}}]{Ellison2011}
{Ellison} S.~L., {Patton} D.~R., {Mendel} J.~T., {Scudder} J.~M., 2011, MNRAS,
  418, 2043

\bibitem[{{Ferrarese} \& {Merritt}(2000)}]{fm00}
{Ferrarese} L., {Merritt} D., 2000, {ApJ}, 539, L9

\bibitem[{{Gabor} \& {Bournaud}(2013)}]{Gabor_Bournaud_2013}
{Gabor} J.~M., {Bournaud} F., 2013, \mnras, 434, 606

\bibitem[{{Gabor} {et~al}\mbox{.}(2009){Gabor}, {Impey}, {Jahnke}, {Simmons},
  {Trump}, {Koekemoer}, {Brusa}, {Cappelluti}, {Schinnerer}, {Smol{\v
  c}i{\'c}}, {Salvato}, {Rhodes}, {Mobasher}, {Capak}, {Massey}, {Leauthaud},
  \& {Scoville}}]{2009ApJ...691..705G}
{Gabor} J.~M. {et~al.}, 2009, \apj, 691, 705

\bibitem[{{Gebhardt} {et~al}\mbox{.}(2000){Gebhardt}, {Bender}, {Bower},
  {Dressler}, {Faber}, {Filippenko}, {Green}, {Grillmair}, {Ho}, {Kormendy},
  {Lauer}, {Magorrian}, {Pinkney}, {Richstone}, \& {Tremaine}}]{Gebhardt2000}
{Gebhardt} K. {et~al.}, 2000, {ApJ}, 539, L13

\bibitem[{{G{\"u}ltekin} {et~al}\mbox{.}(2009){G{\"u}ltekin}, {Richstone},
  {Gebhardt}, {Lauer}, {Tremaine}, {Aller}, {Bender}, {Dressler}, {Faber},
  {Filippenko}, {Green}, {Ho}, {Kormendy}, {Magorrian}, {Pinkney}, \&
  {Siopis}}]{Gultekin2009}
{G{\"u}ltekin} K. {et~al.}, 2009, ApJ, 698, 198

\bibitem[{{Hayward} {et~al}\mbox{.}(2014){Hayward}, {Torrey}, {Springel},
  {Hernquist}, \& {Vogelsberger}}]{Hayward_et_al_2014}
{Hayward} C.~C., {Torrey} P., {Springel} V., {Hernquist} L., {Vogelsberger} M.,
  2014, \mnras, 442, 1992

\bibitem[{{Hernquist}(1990)}]{Hernquist1990}
{Hernquist} L., 1990, ApJ, 356, 359

\bibitem[{{Hopkins} \& {Hernquist}(2006)}]{2006ApJS..166....1H}
{Hopkins} P.~F., {Hernquist} L., 2006, \apjs, 166, 1

\bibitem[{{Hopkins} {et~al}\mbox{.}(2006){Hopkins}, {Hernquist}, {Cox}, {Di
  Matteo}, {Robertson}, \& {Springel}}]{hopkins2006}
{Hopkins} P.~F., {Hernquist} L., {Cox} T.~J., {Di Matteo} T., {Robertson} B.,
  {Springel} V., 2006, {ApJS}, 163, 1

\bibitem[{{Hopkins} \& {Quataert}(2010)}]{HQ2010}
{Hopkins} P.~F., {Quataert} E., 2010, MNRAS, 407, 1529

\bibitem[{{Hopkins}, {Richards} \& {Hernquist}(2007){Hopkins}, {Richards}, \&
  {Hernquist}}]{Hopkins2007}
{Hopkins} P.~F., {Richards} G.~T., {Hernquist} L., 2007, \apj, 654, 731

\bibitem[{{Jogee}(2006)}]{Jogee_2006}
{Jogee} S., 2006, in Lecture Notes in Physics, Berlin Springer Verlag, Vol.
  693, Physics of Active Galactic Nuclei at all Scales, {Alloin} D., ed., p.
  143

\bibitem[{{Johansson}, {Burkert} \& {Naab}(2009){Johansson}, {Burkert}, \&
  {Naab}}]{Johansson2009}
{Johansson} P.~H., {Burkert} A., {Naab} T., 2009, ApJL, 707, L184

\bibitem[{{Karl} {et~al}\mbox{.}(2010){Karl}, {Naab}, {Johansson}, {Kotarba},
  {Boily}, {Renaud}, \& {Theis}}]{Karl_et_al_2010}
{Karl} S.~J., {Naab} T., {Johansson} P.~H., {Kotarba} H., {Boily} C.~M.,
  {Renaud} F., {Theis} C., 2010, \apjl, 715, L88

\bibitem[{{Khochfar} \& {Burkert}(2006)}]{Khochfar2006}
{Khochfar} S., {Burkert} A., 2006, AAP, 445, 403

\bibitem[{{Kim}, {Wise} \& {Abel}(2009){Kim}, {Wise}, \&
  {Abel}}]{Kim_et_al_2009}
{Kim} J.-h., {Wise} J.~H., {Abel} T., 2009, \apjl, 694, L123

\bibitem[{{Kormendy} \& {Richstone}(1995)}]{KormendyRichstone95}
{Kormendy} J., {Richstone} D., 1995, ARAA, 33, 581

\bibitem[{{Kroupa}, {Tout} \& {Gilmore}(1993){Kroupa}, {Tout}, \&
  {Gilmore}}]{Kroupa_IMF}
{Kroupa} P., {Tout} C.~A., {Gilmore} G., 1993, \mnras, 262, 545

\bibitem[{{Lackner} {et~al}\mbox{.}(2014){Lackner}, {Silverman}, {Salvato},
  {Kampczyk}, {Kartaltepe}, {Sanders}, {Capak}, {Civano}, {Ilbert}, {Jahnke},
  {Koekemoer}, {Lee}, {Le Fevre}, {Liu}, {Scoville}, {Sheth}, \&
  {Toft}}]{Lackner_et_al_2014}
{Lackner} C.~N. {et~al.}, 2014, preprint (arXiv:1406.2327)

\bibitem[{{Liu} {et~al}\mbox{.}(2011){Liu}, {Shen}, {Strauss}, \&
  {Hao}}]{2011ApJ...737..101L}
{Liu} X., {Shen} Y., {Strauss} M.~A., {Hao} L., 2011, \apj, 737, 101

\bibitem[{{Lynden-Bell}(1969)}]{LB1969}
{Lynden-Bell} D., 1969, Nature, 223, 690

\bibitem[{{Magorrian} {et~al}\mbox{.}(1998){Magorrian}, {Tremaine},
  {Richstone}, {Bender}, {Bower}, {Dressler}, {Faber}, {Gebhardt}, {Green},
  {Grillmair}, {Kormendy}, \& {Lauer}}]{magorrian1998}
{Magorrian} J. {et~al.}, 1998, {AJ}, 115, 2285

\bibitem[{{Marconi} \& {Hunt}(2003)}]{MarconiHunt2003}
{Marconi} A., {Hunt} L.~K., 2003, ApJL, 589, L21

\bibitem[{{Mayer}(2013)}]{Mayer2013}
{Mayer} L., 2013, Classical and Quantum Gravity, 30, 244008

\bibitem[{{McConnell} \& {Ma}(2013)}]{McConnell2013}
{McConnell} N.~J., {Ma} C.-P., 2013, ApJ, 764, 184

\bibitem[{{Mihos} \& {Hernquist}(1996)}]{MihosHernquist1996}
{Mihos} J.~C., {Hernquist} L., 1996, {ApJ}, 464, 641

\bibitem[{{Milosavljevi{\'c}} \& {Merritt}(2001)}]{milosavljevic2001}
{Milosavljevi{\'c}} M., {Merritt} D., 2001, {ApJ}, 563, 34

\bibitem[{{Navarro}, {Frenk} \& {White}(1996){Navarro}, {Frenk}, \&
  {White}}]{NFW1996}
{Navarro} J.~F., {Frenk} C.~S., {White} S.~D.~M., 1996, ApJ, 462, 563

\bibitem[{{Raiteri}, {Villata} \& {Navarro}(1996){Raiteri}, {Villata}, \&
  {Navarro}}]{Raiteri_et_al_1996}
{Raiteri} C.~M., {Villata} M., {Navarro} J.~F., 1996, \aap, 315, 105

\bibitem[{{Renaud} {et~al}\mbox{.}(2014){Renaud}, {Bournaud}, {Kraljic}, \&
  {Duc}}]{Renaud_et_al_2014}
{Renaud} F., {Bournaud} F., {Kraljic} K., {Duc} P.-A., 2014, \mnras, 442, L33

\bibitem[{{Saitoh} {et~al}\mbox{.}(2011){Saitoh}, {Daisaka}, {Kokubo},
  {Makino}, {Okamoto}, {Tomisaka}, {Wada}, \& {Yoshida}}]{Saitoh_et_al_2011}
{Saitoh} T.~R., {Daisaka} H., {Kokubo} E., {Makino} J., {Okamoto} T.,
  {Tomisaka} K., {Wada} K., {Yoshida} N., 2011, in IAU Symposium, Vol. 270,
  Computational Star Formation, {Alves} J., {Elmegreen} B.~G., {Girart} J.~M.,
  {Trimble} V., eds., pp. 483--486

\bibitem[{{Salpeter}(1964)}]{Salpeter64}
{Salpeter} E.~E., 1964, ApJ, 140, 796

\bibitem[{{Shen}, {Wadsley} \& {Stinson}(2010){Shen}, {Wadsley}, \&
  {Stinson}}]{Shen_Wadsley_Stinson_2010}
{Shen} S., {Wadsley} J., {Stinson} G., 2010, \mnras, 407, 1581

\bibitem[{{Shlosman}, {Begelman} \& {Frank}(1990){Shlosman}, {Begelman}, \&
  {Frank}}]{Shlosman1990}
{Shlosman} I., {Begelman} M.~C., {Frank} J., 1990, Nature, 345, 679

\bibitem[{{Shlosman}, {Frank} \& {Begelman}(1989){Shlosman}, {Frank}, \&
  {Begelman}}]{Shlosman1989}
{Shlosman} I., {Frank} J., {Begelman} M.~C., 1989, Nature, 338, 45

\bibitem[{{Silverman} {et~al}\mbox{.}(2011){Silverman}, {Kampczyk}, {Jahnke},
  {Andrae}, {Lilly}, {Elvis}, {Civano}, {Mainieri}, {Vignali}, {Zamorani},
  {Nair}, {Le F{\`e}vre}, {de Ravel}, {Bardelli}, {Bongiorno}, {Bolzonella},
  {Cappi}, {Caputi}, {Carollo}, {Contini}, {Coppa}, {Cucciati}, {de la Torre},
  {Franzetti}, {Garilli}, {Halliday}, {Hasinger}, {Iovino}, {Knobel},
  {Koekemoer}, {Kova{\v c}}, {Lamareille}, {Le Borgne}, {Le Brun}, {Maier},
  {Mignoli}, {Pello}, {P{\'e}rez-Montero}, {Ricciardelli}, {Peng}, {Scodeggio},
  {Tanaka}, {Tasca}, {Tresse}, {Vergani}, {Zucca}, {Brusa}, {Cappelluti},
  {Comastri}, {Finoguenov}, {Fu}, {Gilli}, {Hao}, {Ho}, \&
  {Salvato}}]{Silverman2011}
{Silverman} J.~D. {et~al.}, 2011, \apj, 743, 2

\bibitem[{{Spitzer}(1942)}]{Spitzer_1942}
{Spitzer}, Jr. L., 1942, \apj, 95, 329

\bibitem[{{Springel}, {Di Matteo} \& {Hernquist}(2005){Springel}, {Di Matteo},
  \& {Hernquist}}]{springel2005b}
{Springel} V., {Di Matteo} T., {Hernquist} L., 2005, {MNRAS}, 361, 776

\bibitem[{{Springel} \& {White}(1999)}]{Springel_White_1999}
{Springel} V., {White} S.~D.~M., 1999, \mnras, 307, 162

\bibitem[{{Stadel}(2001)}]{stadel01}
{Stadel} J.~G., 2001, PhD thesis, AA(UNIVERSITY OF WASHINGTON)

\bibitem[{{Stinson} {et~al}\mbox{.}(2006){Stinson}, {Seth}, {Katz}, {Wadsley},
  {Governato}, \& {Quinn}}]{Stinson2006}
{Stinson} G., {Seth} A., {Katz} N., {Wadsley} J., {Governato} F., {Quinn} T.,
  2006, MNRAS, 373, 1074

\bibitem[{{Tacconi} {et~al}\mbox{.}(2010){Tacconi}, {Genzel}, {Neri}, {Cox},
  {Cooper}, {Shapiro}, {Bolatto}, {Bouch{\'e}}, {Bournaud}, {Burkert},
  {Combes}, {Comerford}, {Davis}, {Schreiber}, {Garcia-Burillo},
  {Gracia-Carpio}, {Lutz}, {Naab}, {Omont}, {Shapley}, {Sternberg}, \&
  {Weiner}}]{Tacconi2010}
{Tacconi} L.~J. {et~al.}, 2010, Nature, 463, 781

\bibitem[{{Teyssier}, {Chapon} \& {Bournaud}(2010){Teyssier}, {Chapon}, \&
  {Bournaud}}]{Teyssier_et_al_2010}
{Teyssier} R., {Chapon} D., {Bournaud} F., 2010, \apjl, 720, L149

\bibitem[{{Van Wassenhove} {et~al}\mbox{.}(2014){Van Wassenhove}, {Capelo},
  {Volonteri}, {Dotti}, {Bellovary}, {Mayer}, \& {Governato}}]{VW2014}
{Van Wassenhove} S., {Capelo} P.~R., {Volonteri} M., {Dotti} M., {Bellovary}
  J.~M., {Mayer} L., {Governato} F., 2014, \mnras, 439, 474

\bibitem[{{Van Wassenhove} {et~al}\mbox{.}(2012){Van Wassenhove}, {Volonteri},
  {Mayer}, {Dotti}, {Bellovary}, \& {Callegari}}]{VW2012}
{Van Wassenhove} S., {Volonteri} M., {Mayer} L., {Dotti} M., {Bellovary} J.,
  {Callegari} S., 2012, ApJL, 748, L7

\bibitem[{{Vito} {et~al}\mbox{.}(2014){Vito}, {Maiolino}, {Santini}, {Brusa},
  {Comastri}, {Cresci}, {Farrah}, {Franceschini}, {Gilli}, {Granato},
  {Gruppioni}, {Lutz}, {Mannucci}, {Pozzi}, {Rosario}, {Scott}, {Viero}, \&
  {Vignali}}]{Vito_et_al_2014}
{Vito} F. {et~al.}, 2014, \mnras, 441, 1059

\bibitem[{{Vitvitska} {et~al}\mbox{.}(2002){Vitvitska}, {Klypin}, {Kravtsov},
  {Wechsler}, {Primack}, \& {Bullock}}]{Vitvitska_et_al_2002}
{Vitvitska} M., {Klypin} A.~A., {Kravtsov} A.~V., {Wechsler} R.~H., {Primack}
  J.~R., {Bullock} J.~S., 2002, \apj, 581, 799

\bibitem[{{Wadsley}, {Stadel} \& {Quinn}(2004){Wadsley}, {Stadel}, \&
  {Quinn}}]{gasoline}
{Wadsley} J.~W., {Stadel} J., {Quinn} T., 2004, New Astronomy, 9, 137

\bibitem[{{Woods} \& {Geller}(2007)}]{2007AJ....134..527W}
{Woods} D.~F., {Geller} M.~J., 2007, \aj, 134, 527

\bibitem[{{Younger} {et~al}\mbox{.}(2008){Younger}, {Hopkins}, {Cox}, \&
  {Hernquist}}]{Younger2008}
{Younger} J.~D., {Hopkins} P.~F., {Cox} T.~J., {Hernquist} L., 2008, ApJ, 686,
  815

\end{thebibliography}
\normalsize

\appendix

\section{Detailed evolution of select mergers}

\begin{figure}
\centering
\vspace{2.5pt}
\includegraphics[width=0.99\columnwidth,angle=0]{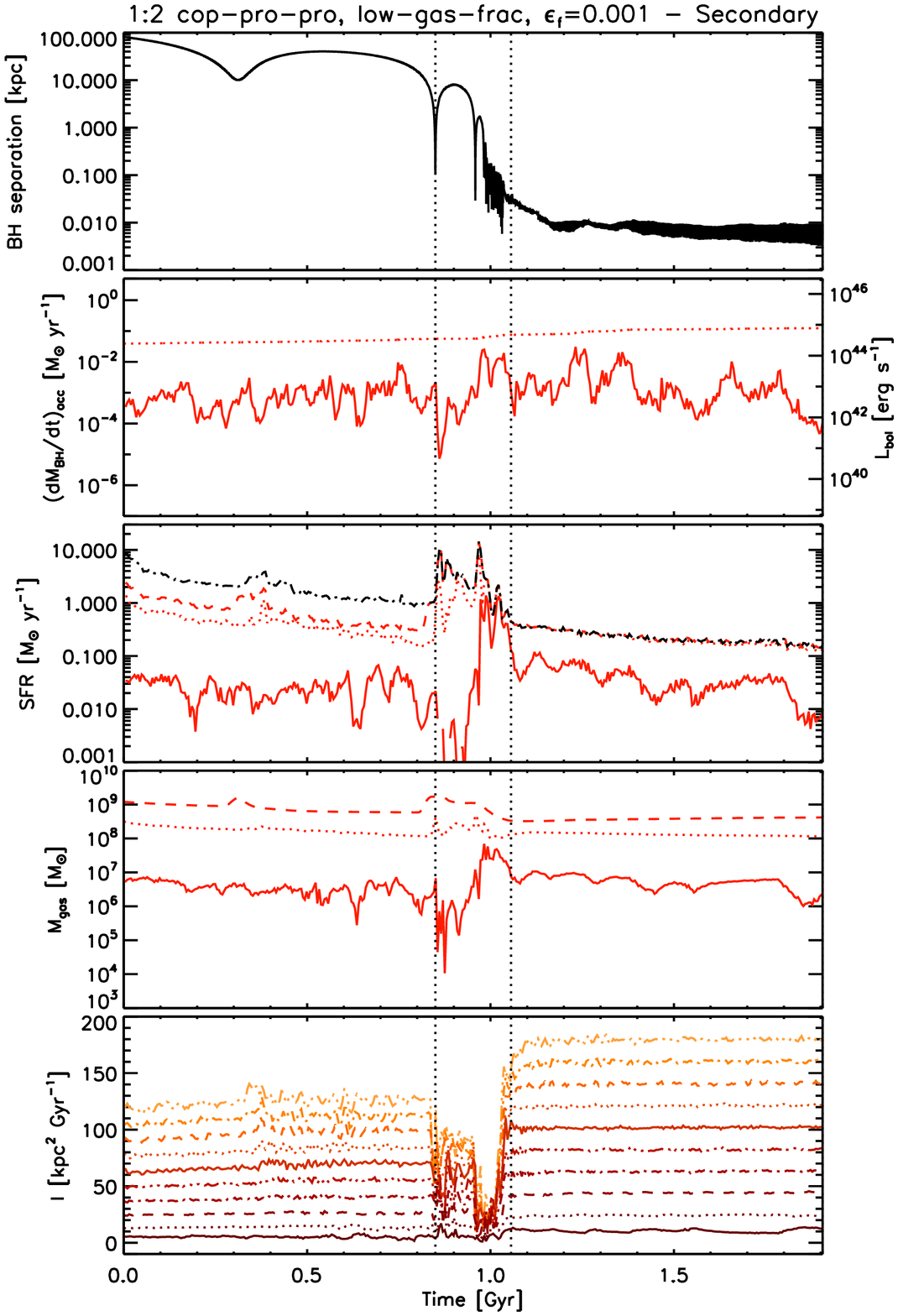}
\vspace{-5pt}
\caption[Temporal evolution of the 1:2 coplanar, prograde--prograde merger with low gas fraction and $\epsilon_f=0.001$ -- Secondary BH]{Temporal evolution of the 1:2 coplanar, prograde--prograde merger with low gas fraction and $\epsilon_f=0.001$ -- Main quantities of and around the secondary BH. Same as Fig.~\ref{agn2014:fig:m4_hr_gf0_3_BHeff0_001_phi000000_five_panels_secondary}.}
\label{agn2014:fig:m2_hr_gf0_3_BHeff0_001_phi000000_five_panels_secondary}
\end{figure}

\begin{figure}
\centering
\vspace{2.5pt}
\includegraphics[width=0.99\columnwidth,angle=0]{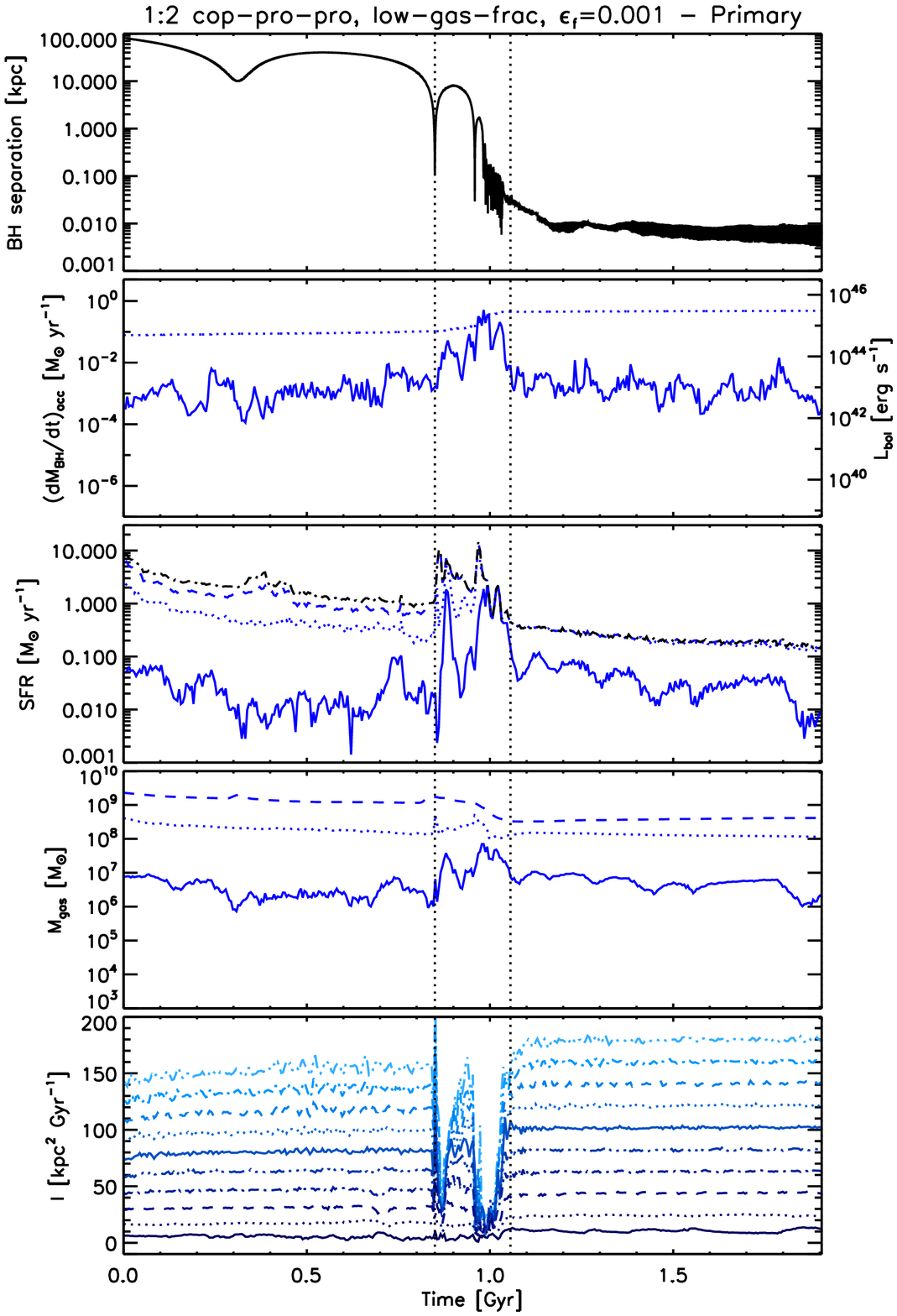}
\vspace{-5pt}
\caption[Temporal evolution of the 1:2 coplanar, prograde--prograde merger with low gas fraction and $\epsilon_f=0.001$ -- Primary BH]{Temporal evolution of the 1:2 coplanar, prograde--prograde merger with low gas fraction and $\epsilon_f=0.001$ -- Main quantities of and around the primary BH. Same as Fig.~\ref{agn2014:fig:m4_hr_gf0_3_BHeff0_001_phi000000_five_panels_primary}.}
\label{agn2014:fig:m2_hr_gf0_3_BHeff0_001_phi000000_five_panels_primary}
\end{figure}

\begin{figure}
\centering
\vspace{2.5pt}
\includegraphics[width=0.99\columnwidth,angle=0]{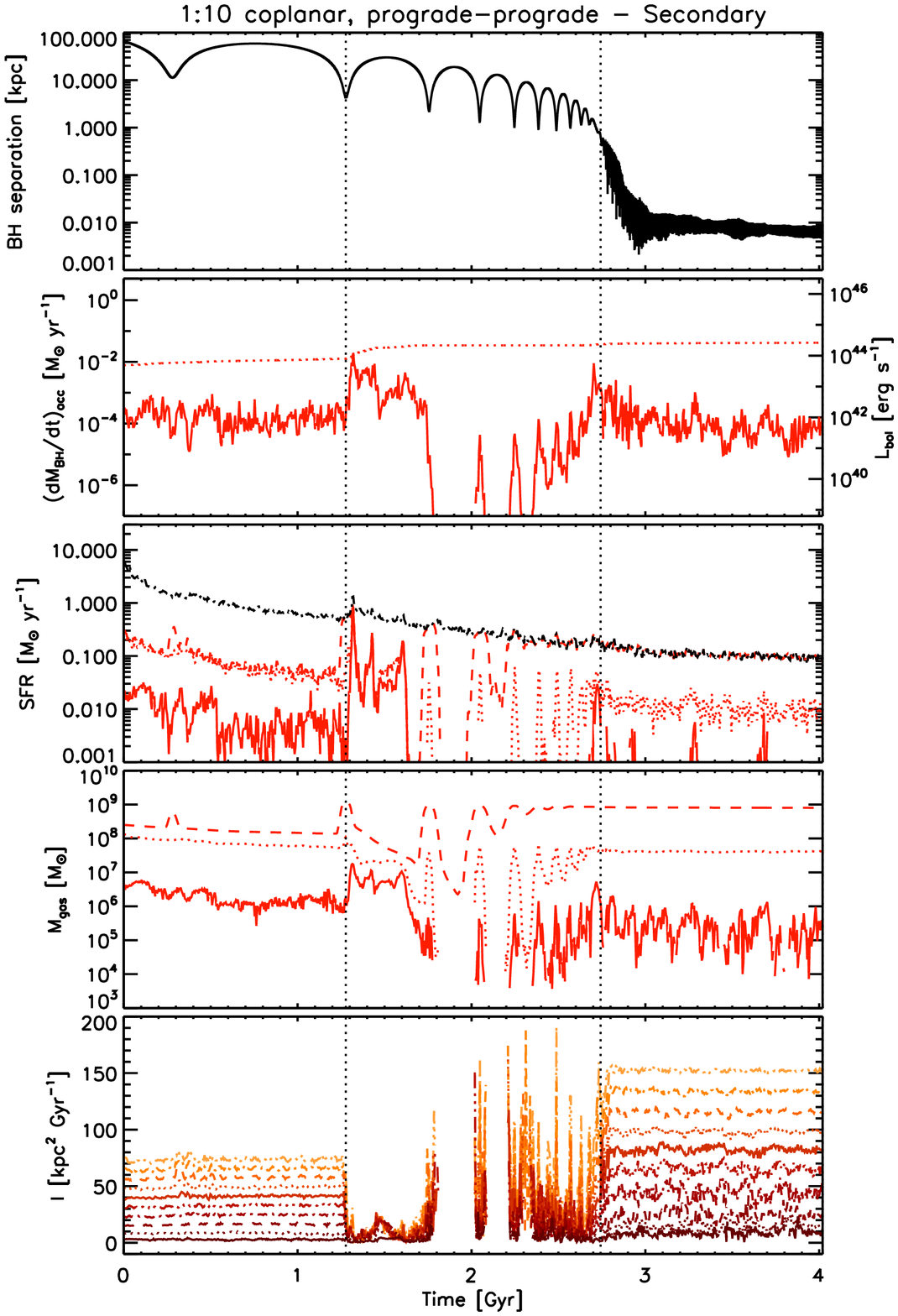}
\vspace{-5pt}
\caption[Temporal evolution of the 1:10 coplanar, prograde--prograde merger -- Secondary BH]{Temporal evolution of the 1:10 coplanar, prograde--prograde merger -- Main quantities of and around the secondary BH. Same as Fig.~\ref{agn2014:fig:m4_hr_gf0_3_BHeff0_001_phi000000_five_panels_secondary}.}
\label{agn2014:fig:m10_hr_gf0_3_BHeff0_001_phi000000_five_panels_secondary}
\end{figure}

\begin{figure}
\centering
\vspace{2.5pt}
\includegraphics[width=0.99\columnwidth,angle=0]{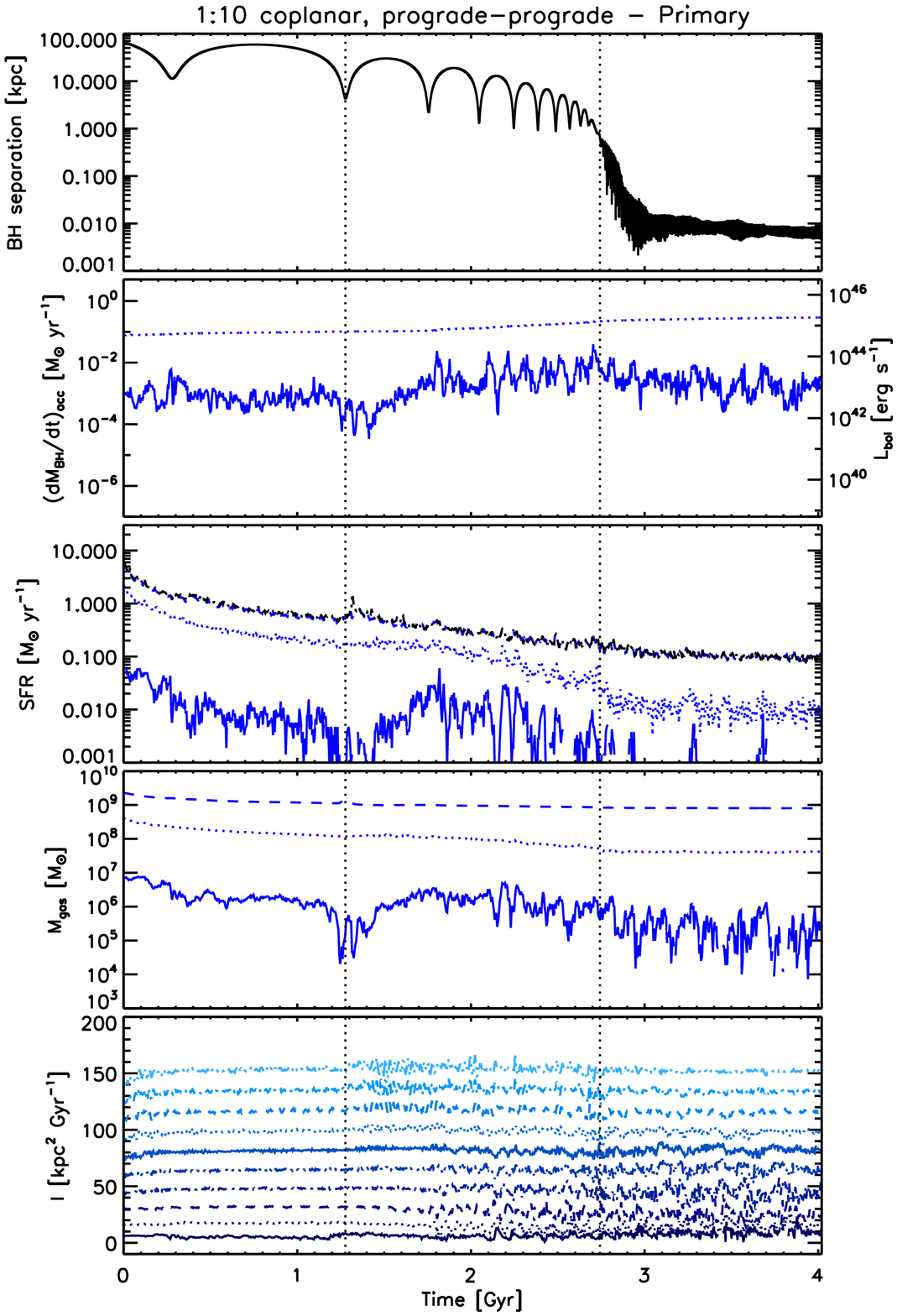}
\vspace{-5pt}
\caption[Temporal evolution of the 1:10 coplanar, prograde--prograde merger -- Primary BH]{Temporal evolution of the 1:10 coplanar, prograde--prograde merger -- Main quantities of and around the primary BH. Same as Fig.~\ref{agn2014:fig:m4_hr_gf0_3_BHeff0_001_phi000000_five_panels_primary}.}
\label{agn2014:fig:m10_hr_gf0_3_BHeff0_001_phi000000_five_panels_primary}
\end{figure}

In Figs~\ref{agn2014:fig:m2_hr_gf0_3_BHeff0_001_phi000000_five_panels_secondary} and \ref{agn2014:fig:m2_hr_gf0_3_BHeff0_001_phi000000_five_panels_primary}, we show in detail the evolution of the 1:2 coplanar, prograde--prograde merger with low gas fraction and $\epsilon_f=0.001$, for the main quantities of and around the secondary and primary BH, respectively. In Figs~\ref{agn2014:fig:m10_hr_gf0_3_BHeff0_001_phi000000_five_panels_secondary} and \ref{agn2014:fig:m10_hr_gf0_3_BHeff0_001_phi000000_five_panels_primary}, we do the same for the 1:10 coplanar, prograde--prograde merger. We thus chose an example of major merger and one of minor merger to show in detail the effect of the initial mass ratio (see Section~\ref{agn2014:sec:Dependence_on_mass_ratio}).

Due to the vastly different dynamical friction time-scales, the durations of both the stochastic and merger stages are longer in the 1:10 than in the 1:2 merger (even though the initial distance between the two galaxies is smaller in the minor merger -- recall that $R_{\rm init}$ is equal to the sum of the initial virial radii of the merging galaxies). During the remnant stage, the two BHs quickly find themselves within $\sim$10~pc from each other, even though the approach is slower in the minor merger case.

The primary galaxy is affected by the encounter with the secondary galaxy more in the major merger case: during the merger stage, the primary BH accretion, together with the central SFR and central gas mass around the primary BH, increases by a few orders of magnitude. At the same time, the central gas specific angular momentum around the primary BH drops significantly. In the minor merger case, on the other hand, the primary galaxy is barely affected, as it can be especially seen from the curves of the central gas specific angular momentum around the primary BH, which do not vary for the entire duration of the encounter. The increase in primary BH accretion (and central SFR and gas mass around the primary BH) during the second part of the merger stage is mostly due to the fact that the secondary galaxy has been completely disrupted of its gas and the primary galaxy has `stolen' it.

The secondary galaxy is instead very much affected by the encounter in both the minor and major cases. In both encounters, secondary BH accretion and central SFR and gas mass around the secondary BH increase by several orders of magnitude during the merger stage, while the central gas specific angular momentum around the secondary BH decreases significantly. The main difference between the two encounters is that, at a certain point during the merger stage of the minor merger, the secondary gaseous disc is completely disrupted by the primary, whereas in the major merger the secondary disc survives the close encounter. This has also a direct effect on the evolution of the BH mass ratio, especially during the merger stage.

\begin{figure}
\centering
\vspace{2.5pt}
\includegraphics[width=0.914\columnwidth,angle=0]{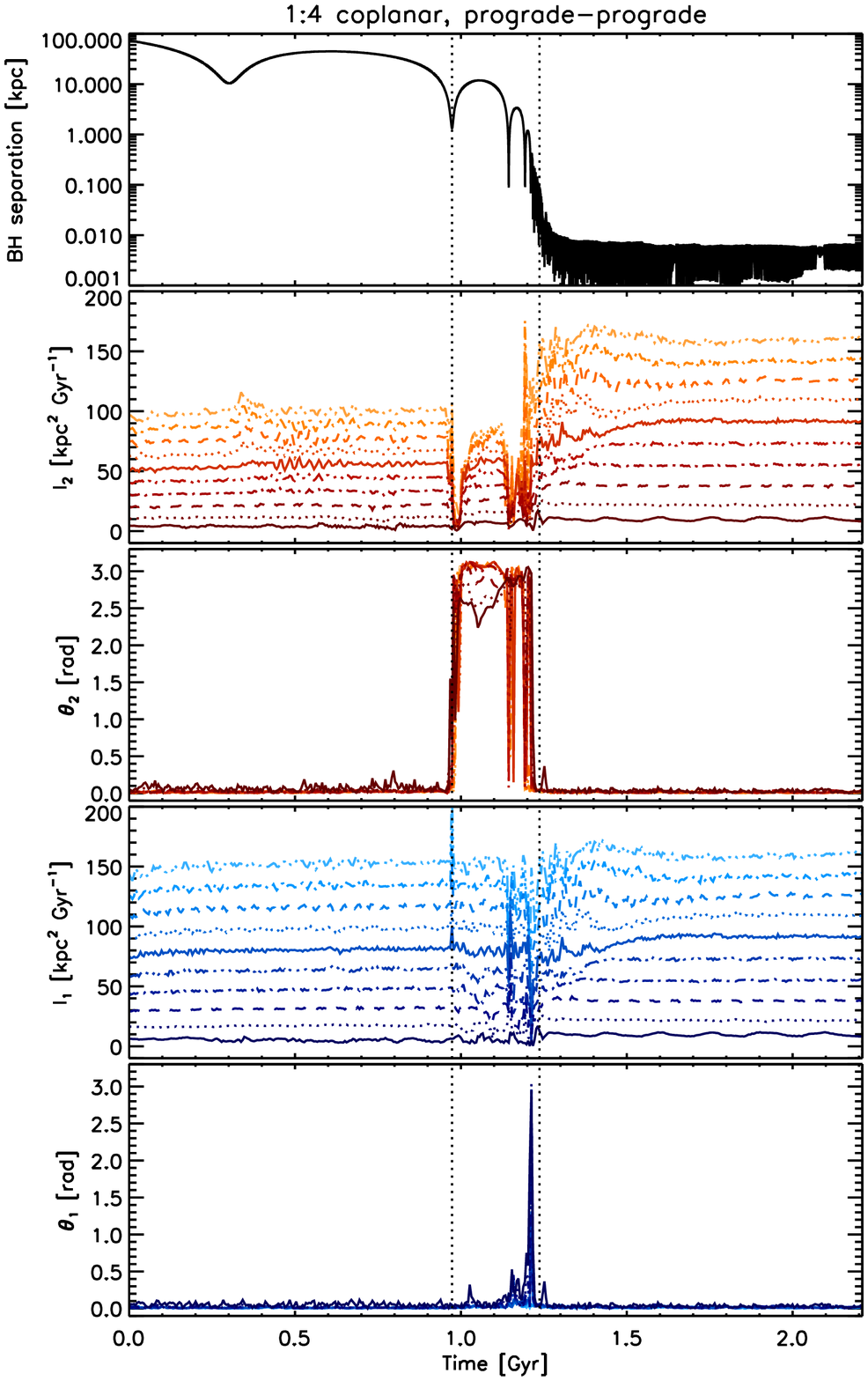}
\vspace{5pt}
\caption[Temporal evolution of the gas specific angular momentum -- 1:4 coplanar, prograde--prograde merger]{Temporal evolution of the gas specific angular momentum -- 1:4 coplanar, prograde--prograde merger. In all panels, the vertical, dotted, black lines show the separation between the stochastic, the merger, and the remnant stage. {First panel}: separation between the two BHs. {Second panel}: gas specific angular momentum magnitude in concentric shells around the local centre of mass near the secondary BH: 0--100 (solid, red), 100--200 (dotted, blue), 200--300 (dashed, red), 300--400 (dash-dotted, blue), 400--500 (dash-triple-dotted, red), 500--600 (solid, blue), 600--700 (dotted, red), 700--800 (dashed, blue), 800--900 (dash-dotted, red), 900-1000 (dash-triple-dotted, blue) pc. {Third panel}: same as the second panel, but for the polar angle of the specific angular momentum vector. {Fourth panel}: same as the second panel, but for the gas specific angular momentum magnitude around the primary BH. {Fifth panel}: same as the fourth panel, but for the polar angle.}
\label{agn2014:fig:m4_hr_gf0_3_BHeff0_001_phi000000_angular_momentum_panels}
\end{figure}

In Fig.~\ref{agn2014:fig:m4_hr_gf0_3_BHeff0_001_phi000000_angular_momentum_panels}, we show the angular momentum flip, introduced in Section~\ref{agn2014:sec:1to4_merger} as a complementary method to distinguish the merger stage from the stochastic and remnant stages, for the 1:4 coplanar, prograde--prograde merger. At the onset of the merger stage, the polar angle of the angular momentum of the gas in the central shells around the secondary BH sharply changes by $\sim$180 degrees: the central gaseous disc now rotates in the opposite direction as before, effectively changing the orbit of the encounter from a prograde--prograde to a prograde--retrograde merger \citep[as also seen in some of the mergers in][]{VW2014}. A few hundred Myr later, at a time close to the beginning of the remnant stage, the central gas undergoes a `counter-flip', when the polar angle of its angular momentum suddenly changes by another $\sim$180 degrees. This last change is likely due to the fact that the two galactic gaseous discs are now overlapping in space: being gas collisional, two counter-rotating discs cannot coexist, and the larger (i.e. primary) disc forces the smaller (i.e. secondary) disc to change its internal rotation. These two sharp changes are common to all central shells within 1~kpc from the secondary BH and affect only the gas. Stars (except of course for newly formed stellar particles, which keep the angular momentum of the `flipped' gas particles they originated from) are not affected by the flip, hinting at the fact that this phenomenon is not related to gravitational torques due to merger dynamics (which would have affected the stars as well), but is probably related to some gas process (such as ram-pressure from the primary disc). We postpone the detailed study of this newly found phenomenon to a future work, where we aim to study in more detail the dynamics of the encounters themselves.

\end{document}